\documentclass[11pt,a4paper]{article}
\pdfoutput=1
\usepackage{jheppub}
\usepackage{amsmath}
\usepackage{epsfig}
\usepackage{amssymb}
\usepackage{graphics}
\usepackage[active]{srcltx}
\usepackage{epstopdf}
\usepackage{shuffle}
\usepackage[utf8]{inputenc}

\newcommand{\bea}{\begin{eqnarray}}
\newcommand{\eea}{\end{eqnarray}}
\newcommand{\bean}{\begin{eqnarray*}}
\newcommand{\eean}{\end{eqnarray*}}
\newcommand{\nn}{\nonumber \\}

\def\W #1{\widetilde{#1}}
\def\WH #1{\widehat{#1}}

\def\eref#1{(\ref{#1})}

\def\a{{\alpha}}

\def\b{{\beta}}

\def\eps{\epsilon}

\def\Label#1{\label{#1}%
  \smash{\hbox to0pt{\raise1ex\hbox{\tiny[#1]}\hss}}}

\def\YM{{\tiny\mbox{YM}}}
\def\G{{\tiny\mbox{G}}}

\def\YMs{{\tiny\mbox{YMs}}}
\def\CHY{{\tiny \mbox{CHY}}}
\def\PT{{\mbox{PT}}}
\def\EYM{{\tiny \mbox{EYM}}}
\def\Pf{{\mbox{Pf}}}

\newcommand{\Sl}{\sum\limits}

\title{Expansion of  Einstein-Yang-Mills Amplitude }
\author[a]{Chih-Hao Fu}
\author[b]{Yi-Jian Du\footnote{The correspondence author.}}
\author[c]{ Rijun Huang}
\author[d,e]{and Bo Feng\footnote{The unusual ordering of authors instead of the
standard alphabet ordering is for young researchers to get proper recognition of
contributions under the current out-dated practice in China. }}

\affiliation[a]{School of Physics and Information Technology,
Shaanxi Normal University,\\
No.620 West Chang'an Avenue, Xi'an 710119, P.R. China.}
\affiliation[b]{Center for Theoretical Physics, School of Physics and Technology,
Wuhan University, \\
No.299 Bayi Road, Wuhan 430072, P.R. China.}
\affiliation[c]{Department of Physics and Institute of Theoretical Physics, Nanjing Normal University,\\
 No.1 Wenyuan Road, Nanjing 210046, P. R. China.}
\affiliation[d]{Zhejiang Institute of Modern Physics, Department of Physics,
 Zhejiang University,\\
 No.38 Zheda Road, Hangzhou 310027, P.R. China.}
\affiliation[e]{Center of Mathematical Science,
  Zhejiang University,\\
  No.38 Zheda Road, Hangzhou 310027, P.R. China.}

\emailAdd{chihhaofu@snnu.edu.cn}
\emailAdd{yijian.du@whu.edu.cn}
\emailAdd{huang@nbi.dk}
\emailAdd{fengbo@zju.edu.cn}
\date{\today}
\abstract{In this paper, we study from various perspectives
the expansion of {tree level} single trace Einstein-Yang-Mills amplitudes into linear combination of color-ordered Yang-Mills amplitudes. By applying the gauge invariance principle, a programable recursive construction is {devised to expand} EYM amplitude with arbitrary number of gravitons into EYM amplitudes with {fewer gravitons}. {Based on this recursive technique we write down} the complete expansion of any single trace EYM amplitude in the basis of color-order Yang-Mills amplitude. As a byproduct, an algorithm for constructing { a} polynomial form of { the } BCJ numerator for Yang-Mills amplitudes { is also outlined in this paper}. { In addition}, by applying BCFW recursion relation we show how to arrive at the same EYM amplitude expansion from the on-shell { perspective}. And  we examine the EYM expansion {using} KLT relations and { show how to  evaluate} the expansion coefficients efficiently.

}

\begin{document}
\maketitle \flushbottom

\section{Introduction}
\label{SecIntroduction}

A fairly non-trivial relation between Einstein-Yang-Mills (EYM)
amplitude and pure Yang-Mills amplitudes was proposed in
\cite{Stieberger:2016lng} recently, where the amplitude of $n$
gluons coupled to a single graviton is expanded as linear sum of
$(n+1)$-point pure gluon amplitudes in an elegant formulation, { which is}
different from the earlier proposed relations {that} express
$n$-gluon $m$-graviton amplitudes by $(n+2m)$-gluon amplitudes
\cite{Chen:2009tr,Stieberger:2009hq,Chen:2010sr,Chen:2010ct,Stieberger:2014cea}.
It is now widely known that, the non-trivial relations among
amplitudes are important both in the practical evaluation and the
analytical {study}, while the $U(1)$-relation,
Kleiss-Kuijf(KK)relation \cite{Kleiss:1988ne} and especially the
Bern-Carrasco-Johansson(BCJ) relations \cite{Bern:2008qj} among
amplitudes of the same field theory have received considerable
investigations in the past few years, and {inspired} the
color-kinematics duality for gravity and Yang-Mills amplitudes
\cite{Bern:2010yg,Bern:2010ue}.
{ As an analogous scenario, where
amplitudes of two originally seemingly unrelated theories take part in novel identity,
recall that the
}
 famous Kawai-Lewellen-Tye
(KLT) relation \cite{Kawai:1985xq} was proposed quite a long time
ago, which formulates a closed string amplitude as products of two
open string amplitudes, and in the field theory limit it expands a
pure gravity amplitude as bi-linear sum of Yang-Mills amplitudes.
The newly proposed linear EYM amplitude relation was also inspired
by the study of string theory, where monodromy relations for mixed
closed-open string amplitudes, previously been applied to the study
of BCJ relations
\cite{BjerrumBohr:2009rd,Stieberger:2009hq,Stieberger:2015vya}, has
been considered.

Because of its compact and simple nature, 
{ a substantial research interests has been drawn to the
study of EYM amplitude relations and to its generalizations}
\cite{Nandan:2016pya,delaCruz:2016gnm,Schlotterer:2016cxa,Du:2016wkt,Nandan:2016ohb,He:2016mzd}\footnote{Remark that in paper \cite{Adamo:2015gia}, a formula for single trace EYM amplitudes in four dimension for arbitrary many gravitons is provided, although not mentioning the amplitude relations.}.
{ In particular} most of the discussions are based on the Cachazo-He-Yuan (CHY)
formulation
\cite{Cachazo:2013gna,Cachazo:2013hca,Cachazo:2013iea,Cachazo:2014nsa,Cachazo:2014xea},
by genuinely reformulating the CHY-integrand in an appropriate form.
{Notably,} explicit expressions for EYM amplitude relations with arbitrary
number of gluons coupled {to} up to three gravitons {were} provided in
\cite{Nandan:2016pya}. The technique for reformulating the
CHY-integrands in these papers {developed into} a systematic explanation in
\cite{Bjerrum-Bohr:2016axv}, and it is revealed therein that the
cross-ratio identity and other off-shell identities of integrands
\cite{Bjerrum-Bohr:2016juj,Cardona:2016gon} are  crucial tools
for deforming CHY-integrands into alternative forms corresponding to
different field theories. These powerful tools benefit from the
integration rule method
\cite{Baadsgaard:2015voa,Baadsgaard:2015ifa,Baadsgaard:2015hia,Huang:2016zzb}
{ developed for the purpose of} evaluating CHY-integrand without referring to the
scattering equations. The {idea} of integration rule and
cross-ratio identity method {was to decompose} arbitrary
CHY-integrand {using} cross-ratio identities {into} those
corresponding to
cubic-scalar Feynman diagrams dressed with kinematic factors. By
carefully organizing terms
one can identify the resulting CHY-integrands as amplitudes of
certain field theories, hence the amplitude relations, as was done
in \cite{Nandan:2016pya,delaCruz:2016gnm}. In fact, there is more
about EYM amplitude relations from the perspective of CHY-framework.
Starting from CHY-integrand of a theory, it is always possible to
reformulate it to another form by cross-ratio and other off-shell
relations, for instance the Yang-Mills-scalar (YMs) amplitude can be
expanded as linear sum of bi-adjoint cubic-scalar amplitudes. We
shall discuss this later in this paper.

{ As it is very often, on-shell technique
can prove to be a powerful tool for the purpose of understanding
non-trivial amplitude relations  within field theory framework.
One such example is the} on-shell proof of BCJ
relations
%
\cite{Feng:2010my,Chen:2011jxa}.
The {central idea} is to deduce {physical identities} only from  general
principles
{ such as}
locality, unitarity
{and} gauge
invariance.
{ This is particularly true with the advent of}
Britto-Cachazo-Feng-Witten(BCFW) on-shell recursion relation
\cite{Britto:2004ap,Britto:2005fq},
{ which utilizes the first two.}
In most cases, the BCFW recursion relation computes the
amplitude in a way such that only contributions from finite local
single poles are summed over, which requires a vanishing behavior in
the boundary of BCFW complex parameter plane. This is exactly the
case for BCJ relations of Yang-Mills amplitudes. However, for
generic situations, the amplitude as a rational function of BCFW
parameter $z$ is not vanishing in $z\to \infty$ and the boundary
contributions can not be avoid. This is a problem one would meet
when applying BCFW recursions to the EYM amplitude relations, and
such subtlety 
complicate{s} the on-shell proof of EYM amplitude
relations. The evaluation of boundary contributions is {generically a difficult problem},
but many methods have been proposed to deal with it.
{ Noteworthily}
systematic algorithm has also been proposed recently
\cite{Benincasa:2011kn,Feng:2014pia,Jin:2014qya,Feng:2015qna,Jin:2015pua,
Cheung:2015cba,Cheung:2015ota} {so that at least
in principle} it is {indeed} possible to
systematically study the EYM amplitude relations {using} BCFW recursion
relations.
{ On the other hand it is also known that very often
gauge invariance can become a very handy tool in constraining the specific analytic form of the scattering amplitude.}
Recent {progresses} have pushed the gauge
invariance principle forward and indicate that, the gauge invariance
along with cubic graph expansion are enough to determine the
amplitudes
\cite{Barreiro:2013dpa,Boels:2016xhc,Berg:2016fui,Arkani-Hamed:2016rak,Rodina:2016mbk,Rodina:2016jyz}.
{In} a less but still quite {challenging} situation, we claim that the
gauge invariance {should} uniquely determine the EYM amplitude
relations, and from which we can explicitly write down the expansion
for EYM amplitude with arbitrary number of gravitons.

As the number of gravitons increases and that of gluons decreases,
in the extremal limit we would come to the amplitude with pure
gravitons. This is the important problem of expanding gravity
amplitude as pure Yang-Mills amplitudes. Furthermore, with the
philosophy of decomposing CHY-integrands, the same argument applies
to the Yang-Mills amplitudes which would be expanded as pure
bi-adjoint cubic-scalar amplitudes. This is exactly the cubic-graph
expansion of Yang-Mills amplitude which makes the color-kinematics
duality manifest \cite{Bjerrum-Bohr:2016axv}. The EYM amplitude
relation combined with CHY-integrand, more specifically the Pfaffian
expansion, would produce the non-trivial expansion for Yang-Mills
amplitude as cubic-scalar graphs, as well as expansion for gravity
amplitude as pure Yang-Mills amplitudes and eventually the
cubic-scalar graphs. This provides a way of computing the BCJ
numerators, which is usually considered to be very
difficult \cite{Mafra:2011kj,Monteiro:2011pc,Broedel:2011pd,BjerrumBohr:2012mg,Fu:2012uy,Boels:2012sy,Monteiro:2013rya,Fu:2014pya,Naculich:2014rta,Chiodaroli:2014xia,Chiodaroli:2015rdg,Du:2016tbc,Carrasco:2016ygv,Berg:2016fui}.
When KLT relation is in action, the EYM amplitude relation can be
connected to the BCJ numerator problem. We will learn more about
this in later sections.

In this paper we
{ examine}
the EYM amplitude relations
{ from the perspectives of}
CHY-formulation, BCFW
on-shell recursion, KLT relation, {and through the contruction of} BCJ numerators.
This paper is organized as follows. In \S
\ref{secCHY}, we present the general theoretical playground of
non-trivial amplitude relations from the CHY-formulation, and
explain the expansion of amplitudes as the expansion of Pfaffian of
CHY-integrand. In \S \ref{secGauge}, we facilitate the principle of
gauge invariance to determine the EYM amplitude relations for gluons
coupled to arbitrary number of gravitons. In \S\ref{secBCJ4YM}, we
generalize the EYM amplitude relations to pure Yang-Mills amplitudes
and apply the non-trivial relation to the computation of BCJ
numerators. In \S\ref{secBCFW}, we provide the on-shell proof of
some EYM amplitude relations by BCFW recursion relations. In
\S\ref{secKLT} we study in the language of KLT relations. Conclusion
is presented in \S\ref{SecConclusion} and some useful backgrounds
are summarized in the Appendix.

\section{Amplitude relations from the perspective of CHY-formulation}
\label{secCHY}

The non-trivial relation revealed recently between EYM amplitudes
and pure Yang-Mills amplitudes
\cite{Stieberger:2016lng,Nandan:2016pya,delaCruz:2016gnm} has an
intuitive interpretation in the CHY-framework. In fact, the
CHY-formulation tells more beyond the EYM amplitudes. In the
CHY-formula, it is the so called CHY-integrand $\mathcal{I}^\CHY$
that describes specific field theories. The CHY-integrand is an
uniform weight-$4$ rational function of $n$ complex variables $z_i$
for $n$-point scattering system, i.e., with the $1/z_i^4$ scaling
behavior in the $z_i\to \infty$ limit.

For almost all known theories, the weight-4 CHY-integrand can be
factorized as two weight-2 ingredients, formally written as
\bea \mathcal{I}^{\CHY}=\mathcal{I}_L\times
\mathcal{I}_R~.~~~\label{gen-1}\eea
Let us then define two new weight-4 CHY integrands as follows
\bea \mathcal{I}^\CHY_L(\alpha):=\mathcal{I}_L\times
\PT(\alpha)~~~,~~~\mathcal{I}^\CHY_R(\beta):=\mathcal{I}_R\times
\PT(\beta)~,~~~\label{gen-2}\eea
where $\PT(\alpha)$ is the Parke-Taylor factor
\bea \PT(\alpha):={1\over (z_{\alpha_1}-z_{\alpha_2})\cdots
(z_{\alpha_{n-1}}-z_{\alpha_n})(z_{\alpha_n}-z_{\alpha_1})}~.~~~\eea
Supposing the two CHY-integrands
$\mathcal{I}^\CHY_L,\mathcal{I}^\CHY_R$ also describe certain
physical meaningful field theories and produce the corresponding
color-ordered amplitudes $A_L(\alpha),A_R(\beta)$ after
CHY-evaluation, then by CHY-construction
\cite{Cachazo:2013gna,Cachazo:2013hca,Cachazo:2013iea} we could
arrive at the following generalized KLT relation,
\bea A=\sum_{\sigma,\widetilde{\sigma}\in
S_{n-3}}A_L(n-1,n,\sigma,1)\mathcal{S}[\sigma|\widetilde{\sigma}]A_R(1,\widetilde{\sigma},n-1,n)~,~~~\label{gen-3}\eea
where $A$ is the amplitude of specific field theory determined by the theories of $A_L$, $A_R$, while $S_n$ denotes permutations on $n$ elements and
$\mathcal{S}[\sigma|\widetilde{\sigma}]$ is some kinematic kernel.
The summation is over $S_{n-3}$ permutations of sets
$\{2,\ldots,n-2\}$, depending on our choice of legs
$k_1,k_{n-1},k_n$ being fixed.

The expression (\ref{gen-3}) denotes a general expansion for the
original amplitude $A$ defined by CHY-integrand (\ref{gen-1}). If
for a specific $\widetilde{\sigma}$ ordering, we sum over all
$S_{n-3}$ permutations of $\sigma$ and define the result as
\bea \mathcal{C}(\widetilde{\sigma}):=\sum_{\sigma\in
S_{n-3}}A_L(n-1,n,\sigma,1)\mathcal{S}[\sigma|\widetilde{\sigma}]~,~~\label{gen-4}\eea
then the original amplitude can be expressed as
\bea A=\sum_{\widetilde{\sigma}\in
S_{n-3}}\mathcal{C}(\widetilde{\sigma})A_R(1,\widetilde{\sigma},n-1,n)~,~~~\label{gen-5}\eea
where $\mathcal{C}(\widetilde{\sigma})$ serves as the expansion
coefficients. Similarly, if for a specific $\sigma$ ordering we sum
over all $S_{n-3}$ permutations of $\widetilde{\sigma}$ and define
the summation as
\bea \widetilde{C}(\sigma):=\sum_{\widetilde{\sigma}\in
S_{n-3}}\mathcal{S}[\sigma|\widetilde{\sigma}]A_R(1,\widetilde{\sigma},n-1,n)~,~~~\label{gen-6}\eea
then the original amplitude can be expanded as
\bea A=\sum_{\sigma\in
S_{n-3}}\widetilde{C}(\sigma)A_L(n-1,n,\sigma,1)~.~~~\label{gen-7}\eea
The expressions \eref{gen-5} and \eref{gen-7} have provided two
different expansions of the original theory. There are several
general remarks regarding the expansion in above,
\begin{itemize}

\item Firstly, the expansion is into a chosen  $(n-3)!$ BCJ basis, and the corresponding expansion coefficients $C(\W
\sigma)$ and $\W C(\sigma)$ would also be unique. However, as we
will discuss soon, sometimes it is better to expand the original
amplitude into the $(n-2)!$ KK basis. Because of the BCJ
relations among color-ordered partial amplitudes, the expansion
coefficients in the KK basis will not be unique and depend on
the {\sl generalized gauge choice} in the BCJ sense.

\item Secondly, with the amplitude expansion formula in hand, the next
is to compute the expansion coefficients. For this purpose,
there are several approaches. The first approach is to use the
definitions \eref{gen-4} and \eref{gen-6} directly. However, in
general it is very hard to evaluate the summation for generic
$n$-point situation, and only in certain special case a direct
evaluation is possible, which we shall explain later. The second
approach seeds back to the expression \eref{gen-1}, and the
major idea is to expand the weight-2 ingredients $\mathcal{I}_L$
or $ \mathcal{I}_R$ into the $\PT(\alpha)$ factor of $n$
elements. In fact, this is the approach followed in
\cite{Nandan:2016pya,delaCruz:2016gnm}. The expansion can be
systematically achieved by successively applying cross-ratio
identities to the CHY-integrands, where in each step a gauge
choice should be taken in the cross-ratio identity. In general,
such expansion leads to a result with $(n-1)!$ cyclic basis.
Then one can use the KK relation to rewrite it into the $(n-2)!$
KK basis. As already mentioned, the gauge dependence remains in
the expansion coefficients at each step, and it would disappear
only after using the BCJ relations to rewrite all into $(n-3)!$
BCJ basis.

Besides the above two direct evaluation methods for expansion
coefficients, there are also some indirect ways. For example,
one can propose some ansatz for the expansion coefficients, then
prove and generalize it by on-shell recursion relations. One can
also use some general considerations, for instance the gauge
invariance or the soft behavior, to determine the coefficients
\cite{Arkani-Hamed:2016rak,Rodina:2016mbk,Rodina:2016jyz}.

In this paper, we will investigate the expansion from these
different views.

\item Thirdly, although in most theories, the CHY-integrand is given by products of
two weight-2 ingredients as \eref{gen-1}, for some theories the
CHY-integrand is defined by the product of four weight-1
ingredients. So there are various combinations of them to form
weight-2 parts. In other words, there are possibilities to have
more than two expansions given in \eref{gen-5} and \eref{gen-7}.
It would be interesting to survey the consequence of different
combinations for these theories.

\end{itemize}

After above general discussions, now we focus on our major topic in
this paper, i.e., the single trace part of EYM theory, whose
CHY-integrand is defined as
\bea \mathcal{I}^{\EYM}_{r,s}(\alpha)=\PT_r(\alpha)\Pf~ \Psi_s\times
\Pf'~\Psi_n~,~~~\label{EYM-single-trace}\eea
for scattering system of $r$ gluons and $s$ gravitons with $r+s=n$,
and $\alpha=\{\alpha_1,\ldots,\alpha_r\}$. We can define two new
CHY-integrands as
\bea
\mathcal{I}^\CHY_{L}(\a|\W\sigma):=\PT_r(\alpha)\Pf~\Psi_s\times
\PT_n(\W\sigma)~~~,~~~\mathcal{I}^{\CHY}_{R}(\b):=\Pf'~\Psi_n\times
\PT_n(\sigma)~,~~~\label{EYM-single-trace-1}\eea
where $\sigma,\W\sigma\in S_n$. It is easy to tell that the
$\mathcal{I}^{\CHY}_L$ is the CHY-integrand of
Yang-Mills-scalar(YMs) theory and $\mathcal{I}^{\CHY}_R$ is the
CHY-integrand of Yang-Mills theory. Correspondingly, the amplitude
$A_L$ is the color-ordered YMs amplitude $A^{\YMs}_{r,s}$ with $r$
scalars and $s$ gluons, which has two trace structures associated with the two PT-factors,  while the amplitude $A_R$ is color-ordered
Yang-Mills amplitude $A^{\YM}_n$. One thing to emphasize is that the
scalar carries two  groups (one gauge group and one flavor group)
and has bi-adjoint scalar-cubic interactions.

An immediate consequence from (\ref{gen-3}) reads
\bea A^{\EYM}_{r,s}(\a)&=&\sum_{\sigma,\widetilde{\sigma}\in
S_{n-3}}A^{\YM}_n(n-1,n,\sigma,1)\mathcal{S}[\sigma|\widetilde{\sigma}]A^{\YMs}_{r,s}(\a|1,\widetilde{\sigma},n-1,n)\nonumber\\
&=&\sum_{\sigma\in
S_{n-3}}\widetilde{C}(\a|\sigma)A^{\YM}_n(n-1,n,\sigma,1)~,~~~\label{EYM-by-YM-exp-1}\eea
with
\bea \widetilde{C}(\a|\sigma)=\sum_{\widetilde{\sigma}\in
S_{n-3}}\mathcal{S}[\sigma|\widetilde{\sigma}]A_{r,s}^{\YMs}(\a|1,\widetilde{\sigma},n-1,n)~.~~~\label{EYM-by-YM-exp-1-1}\eea
The expansion (\ref{EYM-by-YM-exp-1}) is into the BCJ basis with
$(n-3)!$ independent Yang-Mills amplitudes. However, as it will be
clear soon, an expansion into $(n-2)!$ KK basis is more favorable,
and we would present it here as
\bea A_{r,s}^{\EYM}(\a)=\sum_{\sigma\in
S_{n-2}}\widetilde{C}'(\a|\sigma)A_n^{\YM}(n,\sigma,1)~,~~~\label{gen-mine-1}
\eea
with the expansion coefficients
\bea \widetilde{C}'(\a|\sigma)=\lim_{k_n^2\to 0}{1\over k_n^2}
\sum_{\widetilde{\sigma}\in
S_{n-2}}\mathcal{S}[\sigma|\widetilde{\sigma}]A_{r,s}^{\YMs}(\a|1,\widetilde{\sigma},n)~.~~~\eea

The expansion coefficient in \eref{gen-mine-1} is the desired
quantity we want to compute in this paper. As we have discussed in
previous paragraph, these coefficients are determined by only one
weight-2 ingredient in the CHY-integrand in \eref{gen-1}. This means
that while keeping the same weight-2 ingredient, we have the freedom
to change the other weight-2 ingredient. As an implication of such
modification, we could work out the expansion for different field
theories but with the same expansion coefficients. This freedom
could simplify our investigation of expansion coefficients. For
example, in the context of EYM amplitude as Yang-Mills amplitudes,
we can change the $\Pf'~\Psi_n$ in (\ref{EYM-single-trace}) as
$\PT_n(\rho)$. The resulting CHY-integrand
\bea \mathcal{I}^{\YMs}_{r,s}(\alpha|\rho)=\PT_r(\alpha)\Pf~
\Psi_s\times \PT_n(\rho)~~~~\label{YMs-single-trace}\eea
describes a Yang-Mills-scalar amplitude with $r$ scalars and $s$
gluons, and the weight-2 ingredients are now
$\mathcal{I}_L=\PT_r(\alpha)\Pf~ \Psi_s$ and
$\mathcal{I}_R=\PT_n(\rho)$. With the same philosophy as in
(\ref{EYM-single-trace-1}), we can define two new CHY-integrands as
\bea \mathcal{I}^\CHY_{L}:=\PT_r(\alpha)\Pf~\Psi_s\times
\PT_n(\gamma)~~~,~~~\mathcal{I}^{\CHY}_{R}:=\PT_n(\rho)\times
\PT_n(\beta)~.~~~\label{YMs-single-trace-1}\eea
Hence we arrive at the following expansion
\bea A^{\YMs}_{r,s}(\a|\rho)&=&\sum_{\sigma\in
S_{n-3}}\widetilde{C}(\a|\sigma)A^{\phi^3}_n(\rho|n-1,n,\sigma,1)~,~~~\label{EYM-by-YM-exp-2}\eea
with the same expansion coefficients as in
(\ref{EYM-by-YM-exp-1-1}). This non-trivial relation expresses any
single trace color-ordered amplitude of Yang-Mills-scalar theory as
linear combination of color-ordered amplitude of bi-adjoint scalar
$\phi^3$ theory.

After studying the expansion of single trace part of EYM theory to
YM theory, we will briefly discuss the expansion of gravity theory
to YM theory. The CHY-integrand of gravity theory is
\bea \mathcal{I}^{\G}_{r,s}(\alpha)=\Pf'~\Psi_n\times
\Pf'~\Psi_n~.~~~\label{GR-1}\eea
If expanding the reduced Pfaffian
\bea \Pf'~\Psi_n = \sum_{\sigma \in S_{n-2}} n(1,\sigma,n)~
\PT_n(1,\alpha,n)~~~\label{GR-2}\eea
by cross-ratio identities, we will get
\bea A^{\G} = \sum_{\sigma \in S_{n-2}} n(1,\sigma,n)~ A^{\YM}
(1,\sigma,n)~~~\label{GR-3}\eea
by \eref{GR-1}. As already pointed out in papers
\cite{Kiermaier,BjerrumBohr:2010hn,Cachazo:2013iea,
Fu:2012uy,Fu:2013qna, Du:2013sha, Fu:2014pya},
the coefficients
$n(1,\sigma,n)$ in the expansion \eref{GR-2}(hence also the one in the expansion \eref{GR-3}) is nothing but the DDM
basis for the BCJ numerator of YM amplitude. While in the expansion (\ref{gen-7}), i.e., $A=\sum_{\sigma\in S_{n-3}}\widetilde{C}(\sigma)A_L(n-1,n,\sigma,1)$, suppose we can rewrite the $(n-3)!$ BCJ basis into $(n-2)!$ KK basis, then identifying the resulting formula with the one given by (\ref{GR-3}), and equaling the expansion coefficients of the same KK basis, we will get the BCJ numerator $n(1,\sigma,n)$ as linear combination of $\widetilde{C}(\sigma)$. Thus here we provided a new way of computing the BCJ numerators via the computation of amplitude expansion (\ref{gen-7}). Although in (\ref{GR-3}) we have taken gravity amplitude as example, the same consideration can be applied to large number of theories, and the BCJ numerators of those theories can also be identified as the expansion coefficients after rewriting (\ref{gen-7}) into KK basis.

With the above general framework for studying the non-trivial
relations among different theories, we will start our exploration
from next section. As we will see, these relations encode many
surprising structures and connect many important topics, such as the
construction of BCJ numerators mentioned above, the boundary
contribution of amplitudes under BCFW on-shell recursion relation,
the DDM chain and BCJ relations, etc,.

\section{The gauge invariance determines the amplitude relations}
\label{secGauge}

A physical amplitude should be gauge invariant, i.e., vanishes on
the condition $\epsilon_i\to k_i$. If considering the amplitudes
with gravitons and expressing the graviton polarization states as
products of two Yang-Mills polarization states $\epsilon^{\pm}_i
\epsilon^{\pm}_i$, then it should also vanish by setting one of the
polarization vector as $k_i$. The gauge invariance is an important
property of amplitude, and of course it is also valid in the
amplitude relations. As we have mentioned in previous section, there
are different approaches to study expansion coefficients. In this
section, we will demonstrate how to use the gauge invariance to fix
coefficients, which is the same spirit  spelled out in
\cite{Arkani-Hamed:2016rak,Rodina:2016mbk,Rodina:2016jyz}.

\subsection{With single graviton}

To motivate our discussion, let us start with the single trace EYM
amplitude with one graviton with the known expansion
\bea A^{\EYM}_{n,1}(1,2,\ldots,n;p)=\sum_{\shuffle} (\epsilon_p\cdot
Y_p)
A_{n+1}^{\YM}(1,\{2,\ldots,n-1\}\shuffle\{p\},n)~,~~~\label{EYM1gra}\eea
where the summation $\shuffle$ is over all shuffles $\sigma\shuffle
\widetilde{\sigma}$,  i.e.,  all permutation sets of $\sigma\cup
\widetilde{\sigma}$ while preserving the ordering of each
$\sigma,\widetilde{\sigma}$. The color-ordering of external legs in $A_{n+1}^{\YM}$ has cyclic invariance. However if we conventionally fix the leg $1$ in the first position, then every external leg could have a definite position in the color-ordering. In this convention, we can define $Y_p$(also $X_p$ in the following paragraph) as the sum of momenta of all the gluons at the left hand side (LHS) of leg $p$, given the definite color-ordering of color-ordered YM amplitudes.

A clarification of the definition $Y_p$ is needed here for the
future usage. In the $(n+1)$-point pure Yang-Mills amplitude, the
gluon legs has two different correspondents in the EYM amplitude,
i.e., The gluon legs $k_i$, $i=1,\ldots, n$ are also gluons in EYM
amplitude while gluon leg $p$ is originally graviton leg in EYM
amplitude. $Y_p$ is specifically defined as the sum of momenta {\bf
in the gluon subset} of EYM amplitude at the LHS of $p$, while we also
define another quantity $X_p$ as the sum of all momenta at the LHS of $p$
{\bf no matter it is in the gluon subset or graviton subset} of EYM
amplitude. $X_p, Y_p$ would be different when considering EYM
amplitude with more than one gravitons, but in the current
discussion they are the same.

Let us now return to the relation (\ref{EYM1gra}). In the LHS, imposing any gauge condition $\epsilon_i\to k_i$ for
gluon legs would vanish the EYM amplitude, while any Yang-Mills
amplitudes in the right hand side(RHS) also vanish under the same
gauge condition. For the graviton polarization
$\epsilon_p^{\pm}\epsilon_p^{\pm}$, setting either $\epsilon_p\to p$
would vanish the EYM amplitude. In the RHS, the graviton
polarization is distributed in two places: one is in the Yang-Mills
amplitude and the other, in the expansion coefficients. The vanish
of RHS for the former case is trivial, while for the latter case, it
vanishes due to the fundamental BCJ relations
\bea \sum_{\shuffle} (p\cdot Y_p)A^{\YM}_{n+1}(1,\{2,\ldots,
n-1\}\shuffle\{p\},n)=0~.~~~\eea
This consequence is rather interesting. For the non-trivial relation
(\ref{EYM1gra}) to be true and the gauge invariance be not violated,
we eventually end up with BCJ relations. On the other hand, if we
assume $A^{\EYM}_{n,1}$ can be expanded as linear combination of
Yang-Mills amplitudes in the KK basis for convenience, and the
expansion coefficients should be certain sum of $\epsilon_p\cdot
k_i$ to compensate the extra graviton polarization,
\bea A^{\EYM}_{n,1}(1,\ldots,n;p)= \sum_{\alpha\in
S_{n-1}}(\epsilon_p\cdot
x_p)A^{\YM}_{n+1}(1,\alpha_2,\ldots,\alpha_{n-1},\alpha_p,n)~,~~~\eea
then $\sum_{\alpha}(p\cdot
x_p)A^{\YM}_{n+1}(1,\alpha_2,\ldots,\alpha_{n-1},\alpha_p,n)$ should
be in the BCJ relation form!

The lesson learned from the EYM amplitude with one graviton suggests
that, while expressing EYM amplitudes as linear combination of
Yang-Mills amplitudes, the gauge invariance strongly constraints the
possible form of expansion coefficients. This property motivates
us to find the expansion of the single trace multi-graviton EYM
amplitude with more than one graviton
\bea A_{n,m}^{\EYM}(1,\ldots,n;h_1,\ldots,h_m)~,~~~\eea
by gauge invariance, i.e., we want coefficients to  make the
expression to zero under each gauge condition $\epsilon_{h_i}\to
h_i$.

In order to construct the non-trivial relations for EYM amplitudes
with generic number of gravitons, we start with the following two
ansatz,
\begin{itemize}
  \item {\bf Ansatz 1:} The single trace EYM amplitude $A^{\EYM}_{n,m}$ with $m$
  gravitons can always be expanded into EYM amplitudes $A^{\EYM}_{n+m-m',m'}$
  with $m'<m$ gravitons.

  \item {\bf Ansatz 2:} When an EYM amplitude $A^{\EYM}_{n,m}$ is expanded
  into pure Yang-Mills amplitudes, the terms whose expansion
  coefficients contains only $\epsilon\cdot k$ but no
  $\epsilon\cdot \epsilon$ takes the form\footnote{We have taken
  the convention that, for an EYM amplitude we choice the sign
  of these terms to be $(+)$. It would be possible that for
  results in other conventions, for instance the CHY results in
   recent literatures, there could  be a  sign difference.},
  \bea \sum_{\alpha\in S_m}
  \sum_{\shuffle}(\epsilon_{h_{\alpha_{1}}}\cdot
  X_{h_{\alpha_1}})\cdots (\epsilon_{h_{\alpha_m}}\cdot
  X_{h_{\alpha_m}})A_{n+m}^{\YM}(1,\{1,2,\ldots,n\}\shuffle\{\alpha_{h_1},\ldots,\alpha_{h_m}\},n)~.~~~\label{key-observation}\eea
  \end{itemize}
These two ansatz come from lower-point known results. The first ansatz is in fact a general statement saying that a recursive construction for EYM amplitude expansion exists. While the second ansatz is presented in an explicit expression which has an obvious BCJ-like relation form. The validation of ansatz 2 in fact can be verified by BCFW recursions. In the expression (\ref{key-observation}), we emphasize again that $X_{h_i}$ is defined to be the sum of all momenta in the LHS of leg $h_i$, no matter those legs representing gluons or gravitons in the EYM amplitude.

Bearing in mind that any EYM amplitude expansion relations should follow the above mentioned two ansatz, we are now ready to extend the introductory one graviton example to EYM amplitudes with arbitrary number of gravitons. However, before presenting the general algorithm, let us familiarize ourselves by studying EYM
amplitudes with two and three gravitons.

\subsection{Expressing $n$-gluon two-graviton EYM amplitudes as Yang-Mills amplitudes}

The algorithm for producing general EYM amplitude relations is to expand $A^{\EYM}_{n,m}$ as $A^{\EYM}_{n+1,m-1}$ {\sl successively} until $A^{\EYM}_{n+m,0}\equiv A^{\YM}_{m+n}$. Note that the gravitons are colorless, and it has no color-ordering in EYM amplitudes. But in our construction of EYM amplitude relation by gauge invariance principle, it is necessary to specify a graviton in each step of expansion $A^{\EYM}_{n,m}\to A^{\EYM}_{n+1,m-1}$, which in the $A^{\EYM}_{n,m}$ amplitude it denotes a graviton but in the $A^{\EYM}_{n+1,m-1}$ amplitude it denotes a gluon, such that we can apply gauge invariance principle with that graviton. Furthermore, it requires us to select one arbitrary graviton to start with.

Now let us outline the arguments that lead to the correct expansion of EYM amplitude with two gravitons $A_{n,2}^{\EYM}(1,2,\ldots,n;p,q)$. In the first step, let us specify the graviton $h_p$, and following the {\bf Ansatz 1} let us propose the following terms that {\sl would} contribute to the expansion of $A^{\EYM}_{n,2}$,
\bea T_1=\sum_{\shuffle} (\epsilon_p\cdot
Y_p)A^{\EYM}_{n+1,1}(1,\{2,\ldots,n-1\}\shuffle\{p\},n;q)~.~~~\label{Y-2g-T1}\eea
In fact, these proposed terms (\ref{Y-2g-T1}) are reminiscent of the expression \eref{EYM1gra} for expanding the single trace EYM amplitude with one graviton. This is not yet the complete expansion expression for $A^{\EYM}_{n,2}$, but we will explain soon after how to recover the remaining terms. Let us investigate the gauge invariance of gravitons $h_p$ and $h_q$ for the proposed terms (\ref{Y-2g-T1}). The gauge invariance for $h_q$ is obvious since $A_{n+1,1}^{\EYM}(\cdots;q)$ vanishes under $\epsilon_q\to q$ gauge condition. However, $T_1$ is not gauge invariant under $\epsilon_p\to p$ due to the expansion coefficients $\epsilon_p\cdot Y_p$, and there are some missing terms in order to produce the complete expansion for $A^{\EYM}_{n,2}$. Let us proceed to expand the $A^{\EYM}_{n+1,1}$ in $T_1$ with the known result (\ref{EYM1gra}), which gives
\bea T_1=\sum_{\shuffle_1} (\epsilon_p\cdot Y_p)\sum_{\shuffle_2}
(\epsilon_q\cdot X_q)
A^{\YM}_{n+2}(1,(\{2,\ldots,n-1\}\shuffle_1\{p\})\shuffle_2\{q\},n)~.~~~\label{Y-2g-T1-0}\eea
Note that in the permutation shuffle $\{2,\ldots,n-1\}\shuffle\{p\}\shuffle\{q\}$, the position of leg $q$ would be either at the LHS of $p$ or RHS of $p$. But the leg $p$ denotes a graviton in $A^{\EYM}_{n,2}$. So the expansion coefficient is $\epsilon_q\cdot X_q$ but not $\epsilon_q\cdot Y_q$(remind that $X_q$ is the sum of all momenta in the LHS of $q$, while $Y_q$ is the sum of all momenta in the LHS of $q$ excluded the leg $p$, which means that if $p$ is at the RHS of $q$, $X_q=Y_q$, but if $p$ is at the LHS of $q$, $X_q=Y_q+p$).

From the {\bf Ansatz 2} (\ref{key-observation}), we know that in the $A^{\EYM}_{n,2}$ expansion, the correct terms with coefficients $(\epsilon_q\cdot \bullet)(\epsilon_q\cdot \bullet)$ must be
\bea \sum_{\shuffle_1} \sum_{\shuffle_2} (\epsilon_p\cdot
X_p)(\epsilon_q\cdot
X_q)A^{\YM}_{n+2}(1,\{2,\ldots,n-1\}\shuffle_1\{p\}\shuffle_2\{q\},n)~.~~~\label{twoGraEK}\eea
Comparing $T_1$ (\ref{Y-2g-T1-0}) with the correct result (\ref{twoGraEK}), it is easy to see that for those terms with $p$ in the LHS of $q$, $Y_p=X_p$ so that the corresponding terms in $T_1$ and (\ref{twoGraEK}) are the same. While for those terms with $p$ in the RHS of $q$, we have $Y_p+q=X_p$. So in order to reproduce the correct result (\ref{twoGraEK}), we should add another contribution
\bea T_2=\sum_{\shuffle} (\epsilon_p\cdot q)(\epsilon_q\cdot
X_q)A_{n+2}^{\YM}(1,\{2,\ldots,n-1\}\shuffle\{q,p\},n)~~~~\eea
such that
\bea T_1+T_2&=&\sum_{\shuffle} (\epsilon_p\cdot
Y_p)A^{\EYM}_{n+1,1}(1,\{2,\ldots,n-1\}\shuffle\{p\},n;q)\nonumber\\
&&+\sum_{\shuffle} (\epsilon_p\cdot q)(\epsilon_q\cdot
X_q)A_{n+2}^{\YM}(1,\{2,\ldots,n-1\}\shuffle\{q,p\},n)~~~~\label{T1T2}\eea
is exactly equivalent to the correct result (\ref{twoGraEK}). Remind that {\bf Ansatz 2} gives correct answer for contributing terms without $(\epsilon\cdot \epsilon)$ coefficients for EYM amplitude expansion, and up to now, we have reformulated the correct result as (\ref{T1T2}) which guarantees an easy generalization.

Of course, (\ref{T1T2}) is still not yet the complete expansion for $A^{\EYM}_{n,2}$, since those terms with coefficient $(\epsilon_p\cdot \epsilon_q)$ are still missing. Let us propose that the complete expansion takes the form
\bea A^{\EYM}_{n,2}(1,2,\ldots,n;p,q)=T_1+T_2+(\epsilon_p\cdot
\epsilon_q)T_3~,~~~\eea
and the remaining task is to determine $T_3$. It can be determined either by gauge condition $\epsilon_p\to p$ or by $\epsilon_q\to q$, however the latter is much more convenient since $T_1$ is already manifestly gauge invariant for leg $q$. Setting $\epsilon_q\to q$, we have
\bea 0=\Big(T_1+T_2+(\epsilon_p\cdot
\epsilon_q)T_3\Big)\Big|_{\epsilon_q\to q}=(\epsilon_p\cdot
q)\Big(\sum_{\shuffle} (q\cdot
X_q)A_{n+2}^{\YM}(1,\{2,\ldots,n-1\}\shuffle\{q,p\},n)+T_3\Big)~,~~~\eea
which has a solution
\bea T_3=-\sum_{\shuffle} (q\cdot
X_q)A_{n+2}^{\YM}(1,\{2,\ldots,n-1\}\shuffle\{q,p\},n)~.~~~\eea
Hence we get the non-trivial relation for EYM amplitude with two
gravitons as
\bea A^{\EYM}_{n,2}(1,2,\ldots,n;p,q)&=&\sum_{\shuffle}
(\epsilon_p\cdot
Y_p)A^{\EYM}_{n+1,1}(1,\{2,\ldots,n-1\}\shuffle\{p\},n;q)\nonumber\\
&&+\sum_{\shuffle} (\epsilon_p\cdot q)(\epsilon_q\cdot
X_q)A_{n+2}^{\YM}(1,\{2,\ldots,n-1\}\shuffle\{q,p\},n)\nonumber\\
&&-\sum_{\shuffle} (\epsilon_p\cdot \epsilon_q)(q\cdot
X_q)A_{n+2}^{\YM}(1,\{2,\ldots,n-1\}\shuffle\{q,p\},n)~.~~~\label{recurEYM2gra-1}\eea
If we define the tensor
\bea F_q^{\mu\nu}:=q^\mu\epsilon_q^{\nu}-\epsilon_q^\mu
q^{\nu}~,~~~\eea
the above EYM amplitude expansion can be reformulated in a more compact form as
\bea A^{\EYM}_{n,2}(1,2,\ldots,n;p,q)&=&\sum_{\shuffle}
(\epsilon_p\cdot
Y_p)A^{\EYM}_{n+1,1}(1,\{2,\ldots,n-1\}\shuffle\{p\},n;q)\nonumber\\
&&+\sum_{\shuffle} (\epsilon_p\cdot F_q\cdot
X_q)A_{n+2}^{\YM}(1,\{2,\ldots,n-1\}\shuffle\{q,p\},n)~.~~~\label{G2-manifest-1}\eea

From expression \eref{G2-manifest-1} we can infer some important
features. Firstly, for terms in the first line, leg $p$ denotes a gluon and leg $q$ denotes a graviton, while for terms in the second line, leg $q$ denotes gluon instead of graviton. This difference leads to the difference of expansion coefficient, such that the $Y_p$ factor in the first line has been replaced by the factor $F_q\cdot X_q$. Or we can say a factor $F_q$ is inserted. As
we would see shortly after, this is a general pattern for EYM amplitudes
involving more gravitons.

Secondly, in the expression \eref{G2-manifest-1}, the gauge
invariance for leg $q$ is manifest, while gauge
invariance for leg $p$ is not manifest and requires further checking. Although it can be checked directly, we will follow
another approach. Note that the whole result should be symmetric
under switching $p\leftrightarrow q$. For the terms with kinematic
factors $(\eps\cdot k)(\eps\cdot k)$, this symmetry is manifest
since it is given by \eref{twoGraEK}. For the terms with kinematic
factors $(\eps_p\cdot \eps_q)$, the result is not manifestly
symmetric. In order to shown the symmetry, we need to use the
generalized BCJ relation. Let us divide the set $\{2,...,n-1\}$ into
two ordered subsets $\a=\{a_1,...,\a_m\}$ and $\b=\{\b_1,...,\b_t\}$
such that $m+t=n-2$, then the general BCJ relation is given by
\cite{BjerrumBohr:2009rd,Chen:2011jxa}
\bea \sum_{\shuffle} (\sum_{i=1}^t k_{\b_i}\cdot X_{\b_i}  ) A(1,
\a\shuffle \b,n)=0~,~~~\label{generalized-BCJ} \eea
where the first summation is over all shuffles, and $X_{\b_i}$ is
the summation of all momenta of legs at the LHS of leg $\b_i$. Using
\eref{generalized-BCJ} with $\b=\{q,p\}$ it is easy to see that
\bea &&(\eps_q\cdot \eps_p)\sum_{\shuffle}(q\cdot X_q)A(1,
\{2,\ldots,n-1\}\shuffle \{q,p\},n)\nonumber\\
&&~~~~~~~~~~~~~~ =  -(\eps_q\cdot \eps_p)\sum_{\shuffle}(p\cdot
X_p)A(1, \{2,\ldots,n-1\}\shuffle \{q,p\},n)~.~~~\eea
Next, we use the general BCJ relation \eref{generalized-BCJ} with
the choice $\b=\{p\}$ (i.e., the fundamental BCJ relation) to reach
\bea (\eps_q\cdot \eps_p)\sum_{\shuffle}(p\cdot X_p)A(1,
\{2,...,n-1\}\shuffle \{p,q\},n)~.~~~\eea
Hence the symmetry of legs $q,p$ for the terms with factor
$(\eps_p\cdot \eps_q)$ is apparent. Since the gauge invariance for
leg $q$ is satisfied, by the symmetry, the gauge invariance for leg
$p$ is also satisfied.

The above discussion allows a systematical generalization to EYM
amplitude with any number of gravitons. Before doing so, let us
introduce a new quantity which would be useful in later discussions.
Assuming the gravitons have been split into two subsets
$\alpha,\beta$, where $\alpha$ is the ordered length-$m_1$ set in
the gluon side and $\beta$ is a length-$m_2$ set in the graviton
side whose ordering is not relevant, we define
\bea \mathcal{T}[\alpha|\beta]=\sum_{\shuffle}
\Big(\prod_{i=1}^{m_1-1}\epsilon_{\alpha_{i}}\cdot
k_{\alpha_{i+1}}\Big)(\epsilon_{\alpha_{m_1}}\cdot
Y_{\alpha_{m_1}})A^{\EYM}_{n+m_1,m_2}(1,\{2,\ldots,n-1\}\shuffle
\{\alpha_{m_1},\ldots,\alpha_{1}\},n;\beta)~.~~~\label{defT}\eea
For example, if $\alpha=\{p\}, \beta=\{q\}$, we have $m_1=m_2=1$, and
\bea \mathcal{T}[\{p\}|\{q\}]=\sum_{\shuffle} (\epsilon_p\cdot Y_p)A^{\EYM}_{n+1,1}(1,\{2,\ldots,n-1\}\shuffle\{p\},n;q)~,~~~\eea
which is in fact the first line of (\ref{recurEYM2gra-1}). While if $\alpha=\{q,p\},\beta=\emptyset$, $m_1=2,m_2=0$, and we have
\bea \mathcal{T}[\{q,p\}|\emptyset]=\sum_{\shuffle}(\epsilon_p\cdot q)(\epsilon_q\cdot Y_q)A^{\YM}_{n+2}(1,\{2,\ldots,n-1\}\shuffle\{q,p\},n)~,~~~\eea
which reproduces the second line of (\ref{recurEYM2gra-1}).

\subsection{Expressing $n$-gluon three-graviton EYM amplitudes as Yang-Mills amplitudes}

Now let us explore the details by the EYM amplitude with three
gravitons $A^{\EYM}_{n,3}(1,\ldots,n;p,q,r)$. Our purpose is to
construct a recursive algorithm for EYM amplitude expansion which is
manifestly gauge invariant in each step for  gravitons (except the
initial one), and the terms without $(\epsilon_{h_i}\cdot
\epsilon_{h_j})$ factor matches the {\bf Ansatz 2}
(\ref{key-observation}). In the current case it is
\bea \sum_{\shuffle_1}\sum_{\shuffle_2}\sum_{\shuffle_3}
(\epsilon_{p}\cdot X_{p})(\epsilon_{q}\cdot X_{q})(\epsilon_{r}\cdot
X_{r})A^{\YM}_{n+3}(1,\{2,\ldots,n-1\}\shuffle_1\{p\}\shuffle_2\{q\}\shuffle_3\{r\},n)~.~~~\label{ThreeGraEK}\eea
Again, the starting point is specifying an arbitrary graviton for expanding $A^{\EYM}_{n,3}\to A^{\EYM}_{n+1,2}$ and without lose of generality we choice $p$. The proposed contributing terms are
\bea \mathcal{T}[\{p\}|\{q,r\}]=\sum_{\shuffle}(\epsilon_p\cdot
Y_p)A^{\EYM}_{n+1,2}(1,\{2,\ldots,n-1\}\shuffle\{p\},n;q,r)~.~~~\eea
Note that $q,r$ are manifestly gauge invariant in
$\mathcal{T}[\{p\}|\{q,r\}]$, and legs $q, r$ denote gravitons.

To match the correct result (\ref{ThreeGraEK}), we need to add terms where
leg $q$ or $r$ is at the LHS of $p$. This means that we need to add terms $A^{\EYM}_{n+2,1}$ where leg $q$ or leg $r$ now denotes gluon and its position is at the LHS of leg $p$. For $A^{\EYM}_{n+2,1}$ terms where leg $p,q$ are gluons but leg $r$ is graviton, the added term should be
\bea \mathcal{T}[\{q,p\}|\{r\}]=\sum_{\shuffle}(\epsilon_p\cdot
q)(\epsilon_q\cdot
Y_q)A^{\EYM}_{n+2,1}(1,\{2,\ldots,n-1\}\shuffle\{q,p\},n;r)~.~~~\label{threeGreL2}\eea
These terms introduce the missing terms for
$\mathcal{T}[\{p\}|\{q,r\}]$ in order to match the result
(\ref{ThreeGraEK}), however the gauge invariance for $q$ is still broken.
In order to keep gauge invariance for legs $q,r$ at every step, we
should further modify (\ref{threeGreL2}) by adding terms with
$(\epsilon_p\cdot \epsilon_q)$ coefficients, and the resulting terms should not alter the matching with result
(\ref{ThreeGraEK}). From experiences gained in the previous subsection, we can propose the following modification
\bea \mathcal{G}[\{q,p\}|\{r\}]&=&\sum_{\shuffle}(\epsilon_p\cdot
q)(\epsilon_q\cdot
Y_q)A^{\EYM}_{n+2,1}(1,\{2,\ldots,n-1\}\shuffle\{q,p\},n;r)\nonumber\\
&&-\sum_{\shuffle}(\epsilon_p\cdot \epsilon_q)(q\cdot
Y_q)A^{\EYM}_{n+2,1}(1,\{2,\ldots,n-1\}\shuffle\{q,p\},n;r)\nonumber\\
&=&\sum_{\shuffle}(\epsilon_p\cdot F_q\cdot
Y_q)A^{\EYM}_{n+2,1}(1,\{2,\ldots,n-1\}\shuffle\{q,p\},n;r)~,~~~\label{threeGreL2-1}\eea
which are manifestly gauge invariant for $q,r$.

Similarly, for $A^{\EYM}_{n+2,1}$ terms where legs $p,r$ are gluons but leg $q$ is graviton, the proposed gauge invariant term should be
\bea \mathcal{G}[\{r,p\}|\{q\}]=\sum_{\shuffle} (\epsilon_p\cdot
F_r\cdot
Y_r)A^{\EYM}_{n+2,1}(1,\{2,\ldots,n-1\}\shuffle\{r,p\},n;q)~.~~~\label{threeGraL2-2}\eea
Emphasize again that the above proposals are based on the gauge invariant principle, the {\bf Ansatz 1} and the {\bf Ansatz 2}.

Notice that in (\ref{threeGreL2-1}) and (\ref{threeGraL2-2}), we
have $(\epsilon_q\cdot Y_q)$ or $(\epsilon_r\cdot Y_r)$ instead of
$(\epsilon_{q,r}\cdot X_{q,r})$. So in order to arrive at a complete matching with result
(\ref{ThreeGraEK}), we should further add $A^{\EYM}_{n+3,0}$ terms where all $p,q,r$ are gluon legs. For
$\mathcal{G}[\{q,p\}|\{r\}]$, the {\bf Ansatz 1} guides us to propose additional terms as %
\bea \sum_{\shuffle}(\epsilon_p\cdot F_q\cdot r)(\epsilon_r\cdot
Y_r)A_{n+3}^{\EYM}(1,\{2,\ldots,n-1\}\shuffle\{r,q,p\},n)~.~~~\label{threeGraL3-1}\eea
However, these terms are not gauge invariant for leg $r$, and according to gauge invariance principle we need to modify (\ref{threeGraL3-1}) as
\bea \mathcal{G}[\{r,q,p\}|\emptyset]&=&\sum_{\shuffle}
(\epsilon_p\cdot F_q\cdot r)(\epsilon_r\cdot
Y_r)A_{n+3}^{\YM}(1,\{2,\ldots,n-1\}\shuffle\{r,q,p\},n)\nonumber\\
&&-\sum_{\shuffle} (\epsilon_p\cdot F_q\cdot \epsilon_r)(r\cdot
Y_r)A_{n+3}^{\YM}(1,\{2,\ldots,n-1\}\shuffle\{r,q,p\},n)\nonumber\\
&=&\sum_{\shuffle} (\epsilon_p\cdot F_q\cdot F_r\cdot
Y_r)A_{n+3}^{\YM}(1,\{2,\ldots,n-1\}\shuffle\{r,q,p\},n)~,~~~\label{threeGraL3-2}\eea
which reproduces the correct result (\ref{ThreeGraEK}) yet is gauge invariant manifestly. The $(\epsilon_p\cdot F_q\cdot \epsilon_r)$ coefficient in the second
line of (\ref{threeGraL3-2}) is
\bea (\epsilon_p\cdot F_q\cdot \epsilon_r)=(\epsilon_p\cdot
q)(\epsilon_q\cdot \epsilon_r)-(\epsilon_p\cdot \epsilon_q)(q\cdot
\epsilon_r)~,~~~\eea
so we can see clearly that, the second line of (\ref{threeGraL3-2})
only introduces terms with $(\epsilon_{h_i}\cdot \epsilon_{h_j})$
factor which will not contribute to the (\ref{ThreeGraEK}) terms.

Similarly, for $\mathcal{G}[\{r,p\}|\{q\}]$, we need to add the following gauge
invariant terms
\bea
\mathcal{G}[\{q,r,p\}|\emptyset]=\sum_{\shuffle}(\epsilon_p\cdot
F_r\cdot F_q\cdot
Y_q)A_{n+3}^{\YM}(1,\{2,\ldots,n-1\}\shuffle\{q,r,p\},n)~.~~~\eea
Summarizing all parts together, we have
\bea
A^{\EYM}_{n,3}(1,\ldots,n;p,q,r)&=&\mathcal{T}[\{p\}|\{q,r\}]+\mathcal{G}[\{q,p\}|\{r\}]+\mathcal{G}[\{r,p\}|\{q\}]+\mathcal{G}[\{r,q,p\}|\emptyset]+\mathcal{G}[\{q,r,p\}|\emptyset]\nonumber\\
&=&\sum_{\shuffle} (\epsilon_p\cdot
Y_p)A^{\EYM}_{n+1,2}(1,\{2,\ldots,n-1\}\shuffle\{p\},n;q,r)\nonumber\\
&&+\sum_{\shuffle}(\epsilon_p\cdot F_q\cdot
Y_q)A^{\EYM}_{n+2,1}(1,\{2,\ldots,n-1\}\shuffle\{q,p\},n;r)\nonumber\\
&&+\sum_{\shuffle} (\epsilon_p\cdot F_r\cdot
Y_r)A^{\EYM}_{n+2,1}(1,\{2,\ldots,n-1\}\shuffle\{r,p\},n;q)\nonumber\\
&&+\sum_{\shuffle} (\epsilon_p\cdot F_q\cdot F_r\cdot
Y_r)A_{n+3}^{\YM}(1,\{2,\ldots,n-1\}\shuffle\{r,q,p\},n)\nonumber\\
&&+\sum_{\shuffle}(\epsilon_p\cdot F_r\cdot F_q\cdot
Y_q)A_{n+3}^{\YM}(1,\{2,\ldots,n-1\}\shuffle\{q,r,p\},n)~.~~~\label{threeGraRe}\eea
Expression (\ref{threeGraRe}) has demonstrated the recursive
construction pattern more clearly,
i.e., expanding the EYM amplitude successively and keep the gauge invariance in each step by introducing additional terms. The starting point is to specify an arbitrary graviton and propose the contributing terms $\mathcal{T}[\{p\}|\{q,r\}]$, which are terms of $A^{\EYM}_{n+1,2}$. It reproduces a part of the correct result (\ref{ThreeGraEK}) from {\bf Ansatz 2}, and another part would come from $A^{\EYM}_{n+2,1}$ terms. Specifying graviton $q$, we can propose the contributing terms $\mathcal{G}[\{q,p\}|\{r\}]$, deduced from gauge invariance principle, {\bf Ansatz 1} and the matching of {\bf Ansatz 2}. While specifying graviton $r$, we can propose $\mathcal{G}[\{r,p\}|\{q\}]$. The remaining part could be proposed by specifying the last graviton, which gives $\mathcal{G}[\{r,q,p\}|\emptyset]$,
$\mathcal{G}[\{q,r,p\}|\emptyset]$. The correctness of terms without
$(\epsilon_{h_i}\cdot \epsilon_{h_j})$ is guaranteed by
construction, while the terms with $(\epsilon_{h_i}\cdot
\epsilon_{h_j})$ factor are determined by gauge invariance in each
step. The gauge invariance for $q,r$ is then manifest at each term,
except for the leg $p$. It is also easy to see that, in each step when leg $h_i$ in the amplitude denotes a gluon while in the previous step it denotes a graviton, the corresponding gauge invariant term is just to insert a $F_{h_i}$ into the kinematic
factor in an appropriate position. It corresponds to replacing
$k_{h_i}^{\mu}\epsilon_{h_i}^{\nu}$ as $F_{h_i}^{\mu\nu}$. So
similar to the definition of $\mathcal{T}[\alpha|\beta]$ in
(\ref{defT}), we can define a new quantity
\bea
\mathcal{G}[\alpha|\beta]=\sum_{\shuffle}(\epsilon_{{\alpha_{1}}}\cdot
F_{\alpha_{2}}\cdot F_{\alpha_3}\cdots F_{\alpha_{m_1}}\cdot
Y_{\alpha_{m_1}}) A^{\EYM}_{n+m_1,m_2}(1,\{2,\ldots,n-1\}\shuffle
\{\alpha_{m_1},\ldots,\alpha_{1}\},n;\beta)~.~~~\label{defG}\eea
Note that when $m_1=1$,
$\mathcal{T}[\alpha|\beta]=\mathcal{G}[\alpha|\beta]$.

Before  presenting the algorithm for general EYM amplitude
relations, we give a remark on the gauge invariance of $p$. It is
not apparent, but one can show the full $S_3$ symmetry among three
gravitons after using various BCJ relations. Hence the gauge
invariance of leg $p$ should be indicated by the symmetry.

\subsection{A constructive algorithm for producing general EYM amplitude relations}

The basic idea of constructive algorithm for producing general EYM amplitude relations $A^{\EYM}_{n,m}$ is to write down the contributing terms $A^{\EYM}_{n+1,m-1},A^{\EYM}_{n+2,m-2},\ldots,A^{\EYM}_{n+m,0}$ successively, and the explicit expression corresponding to $A^{\EYM}_{n+m',m-m'}$ relies on $A^{\EYM}_{n+m'-1,m-m'+1}$ recursively. Briefly speaking, provided we have written down the contribution of $A^{\EYM}_{n-m_2,m_2}$, where the gravitons in this EYM amplitude are labeled as $\beta=\{\beta_1,\ldots,\beta_{m_2}\}$. Then by specifying a graviton, say $h_{\beta_i}$, $$\beta_i\in
\beta=\{\beta_1,\ldots,\beta_{m_2}\}~,~~~$$
we can directly write down a gauge invariant contributing term $A^{\EYM}_{n-m_2+1,m_2-1}$ as
$\mathcal{G}[\{\beta_i\}\cup \alpha|\beta/\{\beta_i\}]$, whose coefficients are obtained
by replacing $Y_{\alpha_{m_1}}\to F_{{\beta_i}}\cdot Y_{\beta_i}$ in the coefficients of $A^{\EYM}_{n-m_2,m_2}$.

Now let us describe the algorithm for generic EYM amplitude
relations. For the EYM amplitude with $m$ gravitons
\bea A^{\EYM}_{n,m}(1,2,\ldots,n;h_1,h_2,\ldots,h_m)~,~~~\eea
\begin{itemize}
  \item {\bf Step 1:} Specify arbitrary one graviton, say $h_1$, and record the contribution
  \bea \mathcal{G}[\{h_1\}|\{h_2,\ldots,h_m\}]~.~~~\eea
  \item {\bf Step 2:} From the previous step, specify one graviton $h'_2\in \{h_2,\ldots,h_m\}$, and record the corresponding contribution
  \bea
  \mathcal{G}[\{h'_2,h_1\}|\{h_2,\ldots,h_m\}/\{h'_2\}]~~,~~h'_2\in
  \{h_2,\ldots,h_m\}~.~~~\eea
  \item {\bf Step 3:} For each $\mathcal{G}$ in the previous step,
  specify one graviton $h'_3\in \{h_2,\ldots,h_m\}/\{h'_2\}$
  and record the corresponding contribution
  \bea
  \mathcal{G}[\{h'_3,h'_2,h_1\}|\{h_2,\ldots,h_m\}/\{h'_2,h'_3\}]~~,~~h'_2\in
  \{h_2,\ldots,h_m\}~~,~~h'_3\in
  \{h_2,\ldots,h_m\}/\{h'_2\}~.~~~\eea
  \item $\cdots\cdots$
  \item {\bf Step m:} For each $\mathcal{G}$ in the previous step,
  specify one graviton
  $h'_m=\{h_2,\ldots,h_m\}/\{h'_2,\ldots,h'_{m-1}\}$
  and record the contribution
  \bea
  \mathcal{G}[\{h'_m,h'_{m-1},\ldots,h'_2,h_1\}|\emptyset]~~,~~h'_i\in
  \{h_2,\ldots,h_m\}/\{h'_2,\ldots,h'_{i-1}\}~~\mbox{for}~~i=2,\ldots,
  m~.~~~\eea
\end{itemize}
Summing over all the recorded contributions, we get the relation for
generic EYM amplitude expansion as
\bea
&&A^{\EYM}_{n,m}(1,2,\ldots,n;h_1,h_2,\ldots,h_m)=\sum_{|H|=0}^{m-1}\sum_{H\subset
\{h_2,\ldots,h_m\}}\sum_{\sigma\in
S_{|H|}}\mathcal{G}[\{\sigma_H,h_1\}|\{h_2,\ldots,h_m/H\}]~,~~~\label{generalEYM}\eea
where $H$ is a subset of $\{h_2,\ldots,h_m\}$, and $|H|$ is the
length of set $H$. Explicitly writing down, we have
\bea
&&A^{\EYM}_{n,m}(1,2,\ldots,n;h_1,h_2,\ldots,h_m)\nonumber\\
&=&\mathcal{G}[\{h_1\}|\{h_2,\ldots,h_m\}]+\Big(\sum_{h'_2\in\{h_2,\ldots,h_m\}}
\mathcal{G}[\{h'_2,h_1\}|\{h_2,\ldots,h_m\}/\{h'_2\}]\Big)\nonumber\\
&&+\cdots
+\Big(\sum_{\{h'_2,\ldots,h'_{k}\}\subset\{h_2,\ldots,h_m\}}\sum_{\sigma\in
S_{k-1}}
\mathcal{G}[\{\sigma_{h'_2},\ldots,\sigma_{h'_k},h_1\}|\{h_2,\ldots,h_m\}/\{h'_2,\ldots,h'_k\}]\Big)\nonumber\\
&&+\cdots+\Big(\sum_{\sigma\in
S_{m-1}}\mathcal{G}[\{\sigma_{h'_2},\ldots,\sigma_{h'_m},h_1\}|\emptyset]\Big)~.~~~\label{Final-recursive}
\eea
Since $\mathcal{G}[\alpha|\beta]$ is well-defined in (\ref{defG}),
the explicit expression for (\ref{generalEYM}) can be readily
written down. Note that relation (\ref{generalEYM}) expands an EYM
amplitude with $m$ gravitons as linear combination of EYM amplitudes
with $m'<m$ gravitons and Yang-Mills amplitudes. In this expression (\ref{Final-recursive}),
the gauge invariance is manifest for $(m-1)$ gravitons
$\{h_2,...,h_m\}$, since by construction, each contributing term $\mathcal{G}[\alpha|\beta]$ that building up the expansion relation is gauge invariant for $\{h_2,...,h_m\}$. For the leg $h_1$, the gauge invariance is not
manifest. However, as argued in
\cite{Arkani-Hamed:2016rak,Rodina:2016mbk,Rodina:2016jyz}, for
$n$-point Yang-Mills amplitudes, manifest gauge invariance for
$(n-1)$ points will be enough to guarantee the correctness of the
result, so the gauge invariance of the $n$-th point. We believe the
same conclusion can be made for EYM theory by similar argument. If
we buy this argument, result \eref{generalEYM} must be the right
expression.

As a demonstration, let us briefly present the non-trivial relations
for EYM amplitude with four gravitons
$A^{\EYM}_{n,4}(1,\ldots,n;h_1,h_2,h_3,h_4)$. The contributions in each step are abbreviated as follows,
\bea ~[\{h_1\}|\{h_2, h_3, h_4\}]\to \left\{
\begin{array}{c} [\{h_2,h_1\}|\{h_3, h_4\}]\to \left\{
\begin{array}{c} [\{h_3,h_2,h_1\}|\{ h_4\}] \to [\{h_4,h_3,h_2,h_1\}|\{ \emptyset\}]\\  ~\\
~[\{h_4,h_2, h_1\}|\{h_3  \}]  \to [\{h_3, h_4,h_2, h_1\}|\{\emptyset  \}]
\end{array}\right.\\
\\
~[\{h_3, h_1\}|\{h_2,  h_4\}]\to \left\{
\begin{array}{c} ~[\{h_2,h_3, h_1\}|\{  h_4\}] \to [\{h_4,h_2,h_3, h_1\}|\{ \emptyset\}]\\~\\
~[\{h_4,h_3, h_1\}|\{h_2\}] \to [\{h_2,h_4,h_3, h_1\}|\{\emptyset\}]\end{array} \right.\\
\\
~[\{h_4, h_1\}|\{h_2, h_3\}]  \to \left\{
\begin{array}{c}  ~[\{h_2,h_4, h_1\}|\{h_3\}]\to [\{h_3,h_2,h_4, h_1\}|\{\emptyset\}]\\
~\\~[\{h_3,h_4, h_1\}|\{h_2\}]\to [\{h_2,h_3,h_4, h_1\}|\{\emptyset\}]
\end{array} \right.
\end{array}\right.~.~~~\nonumber\eea
The first vertical line corresponds to the contributions of $A^{\EYM}_{n+1,3}$, where we have specified leg $h_1$ as the gluon leg in $A^{\EYM}_{n+1,3}$ amplitude. The second vertical line corresponds to the contributions of $A^{\EYM}_{n+2,2}$, and seen from the first vertical line, we can specify either $h_2,h_3$ or $h_4$ as gluon leg in $A^{\EYM}_{n+2,2}$ amplitudes. Thus we get three contributions. The third vertical line corresponds to the contributions of $A^{\EYM}_{n+3,1}$, deduced from the second vertical line by specifying a graviton leg in $A^{\EYM}_{n+2,2}$ as gluon leg in $A^{\EYM}_{n+3,1}$. And so arrives at the fourth vertical line. Each one in the above table corresponds to a gauge invariant term $\mathcal{G}$ defined in (\ref{defG}), and summing over all contributions we get the expansion for EYM amplitude with four gravitons,
\bea
&&A^{\EYM}_{n,4}(1,\ldots,n;h_1,h_2,h_3,h_4)\\
&=&\sum_{\shuffle}(\epsilon_{h_1}\cdot
Y_{h_1})A^{\EYM}_{n+1,3}(1,\{2,\ldots,n-1\}\shuffle\{h_1\},n;h_2,h_3,h_4)\nonumber\\
&&+\sum_{i=2,3,4}\sum_{\shuffle}(\epsilon_{h_1}\cdot F_{h_i}\cdot
Y_{h_i})A^{\EYM}_{n+2,2}(1,\{2,\ldots,n-1\}\shuffle\{h_i,h_1\},n;\{h_2,h_3,h_4\}/\{h_i\})\nonumber\\
&&+\sum_{2\leq i,j\leq 4\atop i\neq
j}\sum_{\shuffle}(\epsilon_{h_1}\cdot F_{\sigma_{h_i}}\cdot
F_{\sigma_{h_j}}\cdot
Y_{\sigma_{h_j}})A^{\EYM}_{n+3,1}(1,\{2,\ldots,n-1\}\shuffle\{\sigma_{h_j},\sigma_{h_i},h_1\},n;\{h_2,h_3,h_4\}/\{h_i,h_j\})\nonumber\\
&&+\sum_{\sigma\in S_3}\sum_{\shuffle} (\epsilon_{h_1}\cdot
F_{\sigma_{h_2}}\cdot F_{\sigma_{h_3}}\cdot F_{\sigma_{h_4}}\cdot
Y_{\sigma_{h_4}})A^{\YM}_{n+4}(1,\{2,\ldots,n-1\}\shuffle\{\sigma_{h_2},\sigma_{h_3},\sigma_{h_4},h_1\},n)~.~~~\nonumber\eea
%

\subsection{Expanding to pure Yang-Mills amplitudes: ordered splitting formula}

The recursive construction given in   (\ref{generalEYM}) is easy to
implement and  one can eventually get an expansion with pure
Yang-Mills amplitudes. In this subsection, we will present the
related discussion.

To familiarize ourselves with this problem, let us start with some
examples. The first example is the one with two gravitons. After
substituting \eref{EYM1gra} into the first term of
\eref{G2-manifest-1}, we get
\bea
A^{\EYM}_{n,2}(1,2,\ldots,n;p,q)&=&\sum_{\shuffle_1,~\shuffle_2}
(\epsilon_p\cdot Y_p)(\epsilon_q\cdot
X_q)A^{\YM}_{n+2}(1,\{2,\ldots,n-1\}\shuffle_1\{p\}\shuffle_2\{q\},n)\nonumber\\
&&+\sum_{\shuffle} (\epsilon_p\cdot F_q\cdot
X_q)A_{n+2}^{\YM}(1,\{2,\ldots,n-1\}\shuffle\{q,p\},n)~,~~~\label{G2-YM}\eea
where it is crucial to use $X_q$ instead of $Y_q$ in the first term
of the expansion, since to the leg $q$, leg $p$ is actually a gluon.
Although the expression \eref{G2-YM} is very suggestive, the pattern
is still not clear, so we go ahead to the examples with three
gravitons \eref{threeGraRe}. Doing similar manipulations, we arrive
at
\bea & & A^{\EYM}_{n,3}(1,\ldots,n;p,q,r)\label{G3-YM}\\
&=&\sum_{\shuffle} (\epsilon_p\cdot Y_p)(\eps_q\cdot
Z_q)(\eps_r\cdot Z_r)A^{\YM}_{n+3}(1,\{2,\ldots,n-1\}\shuffle\{p\}
\shuffle\{q\}\shuffle\{r\},n)\nonumber\\
& & + \sum_{\shuffle} (\epsilon_p\cdot Y_p)(\eps_q\cdot F_r\cdot
Z_r)A^{\YM}_{n+3}(1,\{2,\ldots,n-1\}\shuffle\{p\}\shuffle\{r,q\}
,n)\nn
&&+\sum_{\shuffle}(\epsilon_p\cdot F_q\cdot
Y_q)(\eps_r\cdot Z_r)A^{\YM}_{n+3}(1,\{2,\ldots,n-1\}\shuffle\{q,p\}\shuffle\{r\},n)\nonumber\\
&&+\sum_{\shuffle} (\epsilon_p\cdot F_r\cdot
Y_r)(\eps_q\cdot Z_q)A^{\YM}_{n+3}(1,\{2,\ldots,n-1\}\shuffle\{r,p\}\shuffle\{q\},n)\nonumber\\
&&+\sum_{\shuffle} (\epsilon_p\cdot F_q\cdot F_r\cdot
Y_r)A_{n+3}^{\YM}(1,\{2,\ldots,n-1\}\shuffle\{r,q,p\},n)\nonumber\\
&&+\sum_{\shuffle}(\epsilon_p\cdot F_r\cdot F_q\cdot
Y_q)A_{n+3}^{\YM}(1,\{2,\ldots,n-1\}\shuffle\{q,r,p\},n)~.~~~\nonumber\eea
Some explanations for \eref{G3-YM} are in order. Firstly, when
expanding $A^{\EYM}_{n+1,2}(1,\{2,\ldots,n-1\}\shuffle\{p\},n;q,r)$
we need to specify a graviton leg which would be the gluon leg in $A^{\EYM}_{n+2,1}$, and our choice is leg $q$. Secondly, we have defined a new
notation $Z_{h_i}$. To define $Z_{h_i}$, we shall introduce a new
concept, i.e., the {\sl ordered splitting of $m$ elements}. To
define the ordered splitting, we must first define an ordering of
$m$ elements, for example, $h_1\prec h_2\prec \cdots \prec h_m$ (we
will call it {\sl ordered gauge} ). Once the {\sl ordered gauge} is
fixed, the ordered splitting is then defined by the following {\sl
ordered set} of subsets $\{\a_1,\ldots,\a_t\}$ satisfying following
conditions,
\begin{itemize}

\item Each subset $\a_i\subset\{h_1,\ldots,h_m\}$ is ordered,

\item Join$[\{\a_1, \ldots \a_t\}]=\{h_1,\ldots,h_m\}$,

\item Denoting $R_{\a_i}$ as the last element of the ordered subset
$\a_i$ (or named the pivot), then $R_{\a_1}\prec R_{\a_2}\prec
\cdots \prec R_{\a_t}$ according to the {\sl ordered gauge}(it
defines the ordering of subset $\a_i$ in the set
$\{a_1,\ldots,a_t\}$),

\item In each subset, all other elements must be larger than $R_{\a_i}$
according to the {\sl ordered gauge}. However, there is no
ordering requirement for all other elements.

\end{itemize}
To better understand the definition of ordered splitting, we take
the set $\{p,q,r\}$ with {\sl ordered gauge} $p\prec q\prec r$ as an
example to write down all ordered splitting,
\begin{itemize}

\item With only one subset, we can have two cases: $\{r,q,p\}$ and
$\{q,r,p\}$,

\item With two subsets, we can have three cases: $\{\{p\},\{r,q\}\}$, $\{\{r,p\},\{q\}\}$
and $\{\{q,p\},\{r\}\}$,

\item With three subsets, we have only one case
$\{\{p\},\{q\},\{r\}\}$.

\end{itemize}

Now let us define the notation $Z_{h_i}$. It is easy to notice that,
the six lines in \eref{G3-YM} are one-to-one mapped to the six
ordered splitting of $\{p,q,r\}$ with {\sl ordered gauge} $p\prec
q\prec r$. The $Z_{h_i}$ is the sum of momenta of legs satisfying
the following two conditions: (1) legs at the LHS of the leg
$h_i$ in the color-ordered Yang-Mills amplitudes, (2) legs at the LHS
of the label chain defined by the ordered splitting. The label chain
for a given ordered splitting is the ordered set
$\{1,2,\ldots,n-1,\alpha_1,\ldots,\alpha_t,n\}$. For instance, for
the ordered splitting $\{\{p\},\{q\},\{r\}\}$ in the first line of
(\ref{G3-YM}), the label chain is $\{1,2,\ldots, n-1,p,q,r,n\}$, and
for $\{\{p\},\{r,q\}\}$ in the second line of (\ref{G3-YM}), the
label chain is $\{1,2,\ldots,n-1,p,r,q,n\}$.

With the understanding of $Z_{h_i}$, it is easy to see that all
$Y_{h_i}$ appearing in \eref{G3-YM} is equal to $Z_{h_i}$, so we can
rewrite \eref{G3-YM} as
\bea & & A^{\EYM}_{n,3}(1,\ldots,n;p,q,r)\label{G3-YM-1}\\
&=&\sum_{\shuffle} (\epsilon_p\cdot Z_p)(\eps_q\cdot
Z_q)(\eps_r\cdot Z_r)A^{\YM}_{n+3}(1,\{2,\ldots,n-1\}\shuffle\{p\}
\shuffle\{q\}\shuffle\{r\},n)\nonumber\\
& & + \sum_{\shuffle} (\epsilon_p\cdot Z_p)(\eps_q\cdot F_r\cdot
Z_r)A^{\YM}_{n+3}(1,\{2,\ldots,n-1\}\shuffle\{p\}\shuffle\{r,q\}
,n)\nn
&&+\sum_{\shuffle}(\epsilon_p\cdot F_q\cdot
Z_q)(\eps_r\cdot Z_r)A^{\YM}_{n+3}(1,\{2,\ldots,n-1\}\shuffle\{q,p\}\shuffle\{r\},n)\nonumber\\
&&+\sum_{\shuffle} (\epsilon_p\cdot F_r\cdot
Z_r)(\eps_q\cdot Z_q)A^{\YM}_{n+3}(1,\{2,\ldots,n-1\}\shuffle\{r,p\}\shuffle\{q\},n)\nonumber\\
&&+\sum_{\shuffle} (\epsilon_p\cdot F_q\cdot F_r\cdot
Z_r)A_{n+3}^{\YM}(1,\{2,\ldots,n-1\}\shuffle\{r,q,p\},n)\nonumber\\
&&+\sum_{\shuffle}(\epsilon_p\cdot F_r\cdot F_q\cdot
Z_q)A_{n+3}^{\YM}(1,\{2,\ldots,n-1\}\shuffle\{q,r,p\},n)~.~~~\nonumber\eea
We have numerically checked the above relation, by comparing with $A_{n,3}^{\EYM}$ directly evaluated with the CHY definition and found agreements in the lower-point examples up to $A^{\EYM}_{3,3}$. { In addition, when expanding the amplitude $A^{\EYM}_{n,3}(1,\ldots,n;p,q,r)$ into terms of pure Yang-Mills ones by (\ref{G3-YM-1}), (\ref{G2-YM}) and (\ref{EYM1gra}) and considering BCJ relations, we obtain the same result proposed in \cite{Nandan:2016pya}.}

With the above result \eref{G3-YM-1}, it is ready to outline the
rule for generalizing (\ref{G3-YM}) for arbitrary number of
gravitons,
\begin{itemize}

\item Decide an {\sl ordered gauge} a priori, and write down all possible ordered
splitting.

\item For each ordered splitting $\{\a_1,\a_2,\ldots,\a_t\}$, write
down a factor $(\eps_{e_1}\cdot F_{e_2}\cdots
F_{e_{|\alpha_i|-1}}\cdot F_{e_{|\alpha_i|}}\cdot
Z_{e_{|\alpha_i|}})$ for each subset
$\alpha_i=\{e_{|\alpha_i|},\ldots,e_2,e_1\}$, and product the
factors for all $\a_i's$ in the ordered splitting. This is the
desired coefficients for the color-ordered amplitudes with
color-ordering defined by the corresponding ordered splitting.

\item Sum over the results from all possible ordered splitting, and
we get the desired expansion of EYM amplitudes into pure
Yang-Mills amplitudes.

\end{itemize}
A final remark says that, because of the choice of {\sl ordered
gauge} a priori, in the expansion the gauge invariance of gravitons
is not as obvious as the one given by the recursive expansion in the
previous subsection.

\subsection{Expanding to pure Yang-Mills amplitudes: KK basis formula}

The formula \eref{G3-YM-1} and its generalizations provide an
expansion of EYM amplitudes into Yang-Mills amplitudes in the
framework of ordered splitting. However, to get the explicit
expression for BCJ numerators, we need an expansion based on KK
basis. From the framework of ordered splitting to the KK basis is
somehow straightforward, and the only problem is to identify all the
corresponding ordered splitting that can produce the desired KK
basis. More explicitly, we need to reconstruct the ordered splitting
from a given color-ordering in KK basis. It is easy to propose such
algorithm, and we would like to demonstrate with a six-graviton
example.

Assuming the {\sl ordered gauge} is $h_1\prec h_2\prec h_3\prec h_4
\prec h_5 \prec h_6$, and the color-ordering in KK basis is
$\{\ldots h_5\ldots h_4\ldots h_1\ldots h_2\ldots h_6\ldots
h_3\ldots\}$, we reconstruct the coefficient of KK basis as follows,
\begin{itemize}

\item Start with the lowest element in the {\sl ordered gauge}, here
$h_1$, and write down all possible subsets $\a_i\subset
\{h_1,\ldots,h_6\}$ which contains $h_1$ as its last element
respecting the given color-ordering in KK basis. Since $h_5,
h_4$ are at the LHS of $h_1$, we can write down four subsets,
\bea
\a_{h_1,1}=\{h_1\}~~,~~\a_{h_1,2}=\{h_5,h_1\}~~,~~\a_{h_1,3}=\{h_4,h_1\}~~,~~
\a_{h_1,4}=\{h_5,h_4,h_1\}~.~~~\eea

\item For each $\alpha_{h_1,i}$, we then drop its element in
$\{h_1,\ldots,h_6\}$ and re-do the step one for the remaining
elements $\{h_1,\ldots,h_6\}/\{\a_{h_1,i}\}$ with the lowest
element therein. For instance, for $\a_{h_1,2}=\{h_5,h_1\}$, its
complement regarding $\{h_1,\ldots,h_6\}$ is $\{h_2, h_3, h_4,
h_6\}$, and the lowest element is $h_2$. Now the color-ordering
$\{\ldots h_5\ldots h_4\ldots h_1\ldots h_2\ldots h_6\ldots
h_3\ldots\}$ becomes $\{\ldots h_4\ldots h_2\ldots h_6\ldots
h_3\}$ after dropping elements in $\a_{h_1,2}$. Repeating the
step one, we get two possible subsets $\a_{h_2,1}=\{h_2\}$ and
$\a_{h_2,2}=\{h_4, h_2\}$.

\item  Repeat until we get the complete ordered splitting.

\end{itemize}
For our current example, we can write the recursive construction
starting from $\a_{h_1,4}$ as follows,
\bea \a_{h_1,4}=\{h_5,h_4,h_1\} & \to & \a_{h_2,1}=\{h_2\} \to
\left\{
\begin{array}{lll}
\a_{h_3,1}= \{h_3\} & \to & \a_{h_6,1}=\{h_6\}~.~~~\\
\a_{h_3,2}= \{h_6,h_3\}~.~~~ &  & \\
\end{array} \right.~~~~\label{recon-ordred-split} \eea
This gives two ordered splitting.

After generating all possible ordered splitting regarding the given
color-ordering in KK basis, we can readily write down its
coefficient in KK basis as the sum of those from the ordered
splitting. However, we remark that, the definition of $Z_{h_i}$ relies on the ordered splitting, so the explicit expression of $Z_{h_i}$ in different ordered splitting is different. Thus we need to be careful when summing over results of all possible ordered splitting.

Finally, let us present an complete example with three gravitons to
demonstrate the algorithm, which is shown as follows,
\begin{itemize}

\item Choose the {\sl ordered gauge} as $p\prec q\prec r$.

\item For the color-ordering $\{\ldots p\ldots q\ldots r\ldots \}$ in the KK basis, the
only possible ordered splitting is $\{\{p\}, \{q\},\{r\}\}$.
Thus its coefficient is
\bea (\epsilon_p\cdot Z_p)(\eps_q\cdot Z_q)(\eps_r\cdot
Z_r)|_{\{\{p\}, \{q\},\{r\}\}}~,~~~\eea
where for $Z_{h_i}$ to be clearly defined, we have explicitly
write down the ordered splitting for reference.

\item For the color-ordering $\{\ldots p\ldots r\ldots q\ldots \}$ in the KK basis, the
only possible ordered splitting are $\{\{p\}, \{q\},\{r\}\}$ and
$\{\{p\}, \{r,q\}\}$. Thus its coefficient is
\bea (\epsilon_p\cdot Z_p)(\eps_q\cdot Z_q)(\eps_r\cdot
Z_r)|_{\{\{p\}, \{q\},\{r\}\}}+ (\epsilon_p\cdot
Z_p)(\eps_q\cdot F_r\cdot Z_r)|_{\{\{p\}, \{r,q\}\}}~.~~~\eea

\item For the color-ordering $\{\ldots q\ldots p\ldots r\ldots \}$ in the KK basis, the
only possible ordered splitting are $\{\{p\}, \{q\},\{r\}\}$ and
$\{\{q,p\},\{r\}\}$. Thus its coefficient is
\bea (\epsilon_p\cdot Z_p)(\eps_q\cdot Z_q)(\eps_r\cdot
Z_r)|_{\{\{p\}, \{q\},\{r\}\}}+ (\epsilon_p\cdot F_q\cdot
Z_q)(\eps_r\cdot Z_r)|_{\{\{q,p\},\{r\}\}}~.~~~\eea

\item For the color-ordering $\{\ldots r\ldots p\ldots q\ldots \}$ in the KK basis, the
only possible ordered splitting are $\{\{p\}, \{q\},\{r\}\}$
$\{\{p\}, \{r, q\}\}$ and $\{\{r,p\},\{q\}\}$. Thus the
coefficient is
\bea &&(\epsilon_p\cdot Z_p)(\eps_q\cdot Z_q)(\eps_r\cdot
Z_r)|_{\{\{p\}, \{q\},\{r\}\}}\nonumber\\
&&~~~~~~~~~~~~~~~~~~~~~+ (\epsilon_p\cdot Z_p)(\eps_q\cdot
F_r\cdot Z_r)|_{\{\{p\}, \{r,q\}\}}+ (\epsilon_p\cdot F_r\cdot
Z_r)(\eps_q\cdot Z_q)|_{\{\{r,p\},\{q\}\}}~.~~~\eea
As we have emphasized, $Z_q$ has different meaning in the first
and third terms. It will contain $k_r$ in the third term, but
not in the first term.

\item For the color-ordering $\{\ldots q\ldots r\ldots p\ldots \}$ in the KK basis, the
only possible ordered splitting are $\{\{p\}, \{q\},\{r\}\}$,
$\{\{q,p\},\{r\}\}$, $\{\{r,p\},\{q\}\}$ and $\{q,r,p\}$. Thus
its coefficient is
\bea & & (\epsilon_p\cdot Z_p)(\eps_q\cdot Z_q)(\eps_r\cdot
Z_r)|_{\{\{p\}, \{q\},\{r\}\}}+ (\epsilon_p\cdot F_q\cdot
Z_q)(\eps_r\cdot Z_r)|_{\{\{q,p\},\{r\}\}}\nonumber\\
&&~~~~~~~~~~~~~~~~~~~~~+ (\epsilon_p\cdot F_r\cdot
Z_r)(\eps_q\cdot Z_q)|_{\{\{r,p\},\{q\}\}} +(\epsilon_p\cdot
F_r\cdot F_q\cdot Z_q)|_{\{q,r,p\}}~.~~~\eea

\item For the color-ordering $\{\ldots r\ldots q\ldots p\ldots \}$ in the KK basis, the
only possible ordered splitting are $\{\{p\}, \{q\},\{r\}\}$,
$\{\{q,p\},\{r\}\}$, $\{\{r,p\},\{q\}\}$, $\{\{p\},\{r,q\}\}$
and $\{r,q,p\}$. Thus its coefficient is
\bea & & (\epsilon_p\cdot Z_p)(\eps_q\cdot Z_q)(\eps_r\cdot
Z_r)|_{\{\{p\}, \{q\},\{r\}\}}+ (\epsilon_p\cdot F_q\cdot
Z_q)(\eps_r\cdot Z_r)|_{\{\{q,p\},\{r\}\}}\nonumber\\
&&+ (\epsilon_p\cdot F_r\cdot Z_r)(\eps_q\cdot
Z_q)|_{\{\{r,p\},\{q\}\}}+ (\epsilon_p\cdot Z_p)(\eps_q\cdot
F_r\cdot Z_r)|_{\{\{p\}, \{r,q\}\}} +(\epsilon_p\cdot F_q\cdot
F_r\cdot Z_r)|_{\{r,q,p\}}~.~~~\eea

\end{itemize}
The above example clearly shows how to reproduce the EYM amplitude expansion from the ordered splitting formula to the KK basis. Although for EYM amplitudes with many gravitons the procedure would be quite involved, but with the help of computer, it would not be a serious problem. And the expansion in KK basis is ideal for the study of BCJ numerators, which we will mention in the next section.

\section{The BCJ numerator of Yang-Mills theory}
\label{secBCJ4YM}

In this section, we will apply the story of EYM theory to pure
Yang-Mills theory, to provide an algorithmic construction of BCJ
numerators for Yang-Mills amplitudes. Let us start from some
necessary backgrounds. In paper \cite{Lam:2016tlk}, an expansion of
Pfaffian is introduced as,
 \bea {\rm Pf} ~{\Psi}=(-1)^{{1\over 2}n(n+1)}\sum_{ p\in
S_n}(-)^{\rm {sign}(
p)}\Psi_I\Psi_J\cdots\Psi_K~,~~~\label{pfsign}\eea
where the sum is over permutations $S_n$, and $I,J,\ldots,K$ are
closed cycles from splitting of the permutation. The factor $\Psi_I$
is defined as
\bea  \Psi_I={F_I\over  z_I}={{1\over 2}{\rm Tr}(F_{ i_1}\cdot F_{
i_2}\cdots F_{ i_{|I|}})\over   z_{ i_1 i_2}  z_{ i_2 i_3}\cdots z_{
i_{|I|} i_1}}~~~,~~~F_{ i}^{\mu\nu}=k_{ i}^\mu\eps^{\nu}_{
i}-\eps^{\nu}_{ i}k^{\mu}_{ i}~~~~\label{cycle}\eea
for $I$ containing more than one $z_i's$, and
 \bea
 \Psi_{( i)}=C_{ i i}=-\sum_{ j\not=
i}{\eps_i\cdot k_j\over  z_{ i j}}~.~~~\label{single-cycle}\eea
for $I$ containing only one $z_i$, which is explicitly the diagonal
term of $C$ matrix. Although the definition \eref{pfsign} is
proposed for the full $2n\times 2n$ Pfaffian matrix, it is also
valid for the sub-matrix of Pfaffian. In the case of Yang-Mills
theory, we are interested in the reduced Pfaffian,
\bea {\rm Pf}'~\Psi^{\lambda\nu}_{\lambda\nu}=-2^{n-3}{\sum_{p\in S_n}}'(-)^{{\rm
sign}(p)}{W_I\over z_I}\Psi_J \cdots \Psi_K~,~~~\label{pfprime}\eea
where the $\lambda$ and $\nu$-th rows and columns in $\Psi$ have been removed. The prime in the $\sum$ indicates that the sum is taken over all $p\in S_n$ permutation such that $\nu$ is changed into $\lambda$, which we call constraint permutation. The constraint permutation $p$ has been decomposed to closed
cycles $I,J,\ldots,K$. In this paper, we take $1,n$ as the gauge
choice, so the constraint permutation is the closed cycle $I=(1 i_2
i_3\cdots i_{|I|} n)$, and
\bea W_I=\eps_1 \cdot F_{i_2}\cdot F_{i_3}\cdots
F_{i_{|I|-1}}\cdot\eps_n~,~~~\label{W}\eea

With above gauge choice, we can decompose the constraint permutation
sum in \eref{pfprime} as three summations. The first summation is
the summation of splitting $\{2,3,\ldots, n-1\}$ into a gluon subset
$A$ and a graviton subset $B$, while both subsets could be empty.
The second summation is the permutation over gluon subset $A$. The
third summation is the permutation over graviton subset $B$. This
gives \footnote{Naively, there will be a sign factor $(-)^{{1\over
2} n_B(n_B+1)}$ if using the result \eref{pfsign}. However,
comparing with our explicit example and the argument using the BCFW
on-shell recursion relation, it seems this sign factor should not be
there.}
\bea {\rm Pf}'\Psi & = & -2^{n-3}\sum_{\rm split}\left[\sum_{\rho\in
S_A}(-)^{{\rm sign}(\rho)}{W_{(1\rho(A) n)}\over z_{(1\rho(A)
n)}}\right] ~\left[\sum_{\W\rho\in S_B}(-)^{{\rm
sign}(\W\rho)}\Psi_J ... \Psi_K\right]\nn
& = & - 2^{n-3}\sum_{\rm split}\left[\sum_{\rho\in S_A}(-)^{{\rm
sign}(\rho)}{W_{(1\rho(A) n)}\over z_{(1\rho(A) n)}}\right] ~{\rm
Pf}\Psi_B\nn
& = & - 2^{n-3}\sum_{\rm split}\sum_{\rho\in S_A}(-)^{{\rm
sign}(\rho)}W_{(1\rho(A) n)}\left\{{1\over z_{(1\rho(A) n)}} ~{\rm
Pf}\Psi_B\right\}~,~~~\label{pfprime-new}\eea
where in the second line we have used \eref{pfsign}.

If we multiply (\ref{pfprime-new}) by another copy of reduced Pfaffian, then the expression in the curly bracket in the third line of (\ref{pfprime-new}) can be identified as CHY-integrand of single trace EYM amplitude with gluon legs $\{1,A,n\}$ and graviton legs $B$. Thus the above
relation between reduced Pfaffian and Pfaffians for fewer points,
exactly establishes the same relation between Einstein gravity and
EYM/Yang-Mills amplitudes. As already pointed out in \S \ref{secCHY}, the relation between gravity and YM amplitudes shares the same kinematic coefficients for relations between Yang-Mills amplitude and cubic-scalar amplitudes. Since we have already worked out the EYM amplitude relations in the previous section, we can use the known results for EYM amplitude to directly write down the basis of BCJ numerators for Yang-Mills theory, without going through again the expansion of Yang-Mills amplitudes into cubic-scalar amplitudes.

Now we present the algorithm for finding the BCJ numerator of one KK
basis with the color-ordered $A^{\YM}(1, h_2, h_3,\ldots,h_{n-1},
n)$. The key idea is again to reconstruct related ordered splitting
in \eref{pfprime-new} from the given ordering
$\{h_2,\ldots,h_{n-1}\}$,
\begin{itemize}

\item Firstly, we reconstruct the $\rho(A)$. The length of subset
$\rho(A)$ could be values from 0 to $n-2$, so we can explicitly
get
\bea n_{\rho(A)}=0: &~~~& \rho(A)=\emptyset~,~~~\nn
n_{\rho(A)}=1: &~~~& \rho(A)=\{h_i\}~~~,~~~i=2,\ldots,n-1~,~~~
\nn
n_{\rho(A)}=2: &~~~& \rho(A)=\{h_i,h_j\}~~~,~~~2\leq i< j\leq
n-1~,~~~ \nn
\cdots &  & \cdots \nn
n_{\rho(A)}=k: &~~~&
\rho(A)=\{h_{i_1},h_{i_2},\ldots,h_{i_k}\}~~~,~~~2\leq
i_1<i_2<\cdots <\cdots < i_k\leq n-1~.~~~\nonumber \eea

\item With the knowledge of $\rho(A)$, we generate an ordered subset
${\cal O}(B)=\{h_2,\ldots,h_{n-1}\}/\rho(A)$. Then we choose the
standard {\sl ordered gauge} $2\prec 3\prec \cdots \prec n-1$,
and reconstruct the ordered splitting as the story in
\eref{recon-ordred-split}. Each ordered splitting contributes a
factor ${\cal C}_i$.

\item Collecting results for all possible ordered splitting, we get the
coefficient,
\bea \sum_{\rho(A)} (-)^{n_A} 2^{n-3}W_{(1\rho(A) n)}\times
\left\{ \sum_{\rm Ordered~Spliting({\cal O}(B))} {\cal
C}_i\right\}~.~~~\label{YM-BCJ-num}\eea
\end{itemize}

As a demonstration of above algorithm, let us compute the KK basis
of BCJ numerators for four-point Yang-Mills amplitude. For the
numerator $n_{(1|23|4)}$, we can generate the following terms,
\bea &&\rho(A)=\emptyset~~~,~~~ \mbox{Split}{\cal
O}[\{2,3\}]=\{\{2\},\{3\}\}~~ \Longrightarrow ~~+2 (\eps_1\cdot
\eps_4)(\eps_2\cdot k_1)(\eps_3\cdot k_{12})~,~~~ \nn
&&\rho(A)=\{2\}~~~,~~~  \mbox{Split}{\cal
O}[\{3\}]=\{3\}~~\Longrightarrow~~ -2 (\eps_1\cdot F_2\cdot \eps_4)
(\eps_3\cdot k_{12})~,~~~\nn
&&\rho(A)=\{3\}~~~,~~~  \mbox{Split}{\cal
O}[\{2\}]=\{2\}~~\Longrightarrow~~-2 (\eps_1\cdot F_3\cdot \eps_4)
(\eps_2\cdot k_1)~,~~~\nn
&&\rho(A)=\{2,3\}~~~,~~~  \mbox{Split}{\cal
O}[\emptyset]=\emptyset~~\Longrightarrow~~ +2 (\eps_1\cdot F_2\cdot
F_3\cdot \eps_4)~.~~~\nonumber \eea
Summing over all the results, we get
\bea n_{(1|23|4)} & = &2 (\eps_1\cdot \eps_4)(\eps_2\cdot
k_1)(\eps_3\cdot k_{12})-2 (\eps_1\cdot F_2\cdot \eps_4)
(\eps_3\cdot k_{12})\nonumber\\
&&~~~~~~~~~~~~~~~~~~~~~~-2 (\eps_1\cdot F_3\cdot \eps_4)
(\eps_2\cdot k_1)+2 (\eps_1\cdot F_2\cdot F_3\cdot
\eps_4)~.~~~\label{4p-n1234} \eea
Similarly, for $n_{(1|32|4)}$ we can reconstruct the following
terms,
\bea &&\rho(A)=\emptyset~~~,~~~\mbox{Split}{\cal
O}[\{3,2\}]=\{\{2\},\{3\}\}~~,~~\{3,2\}~~\Longrightarrow~~ +2
(\eps_1\cdot \eps_4)[(\eps_2\cdot k_1)(\eps_3\cdot
k_{1})+\eps_2\cdot F_3\cdot k_1]~,~~~ \nn
&&\rho(A)=\{2\}~~~,~~~\mbox{Split}{\cal
O}[\{3\}]=\{3\}~~\Longrightarrow~~-2 (\eps_1\cdot F_2\cdot \eps_4)
(\eps_3\cdot k_{1})~,~~~\nn
&&\rho(A)=\{3\}~~~,~~~\mbox{Split}{\cal
O}[\{2\}]=\{2\}~~\Longrightarrow~~ -2 (\eps_1\cdot F_3\cdot \eps_4)
(\eps_2\cdot k_{13})~,~~~\nn
&&\rho(A)=\{2,3\}~~~,~~~\mbox{Split}{\cal
O}[\emptyset]=\emptyset~~\Longrightarrow ~~+2 (\eps_1\cdot F_2\cdot
F_3\cdot \eps_4)~.~~~\nonumber \eea
Summing over all results, we get
\bea n_{(1|32|4)} & = & 2 (\eps_1\cdot \eps_4)[(\eps_2\cdot
k_1)(\eps_3\cdot k_{1})+\eps_2\cdot F_3\cdot k_1] \nonumber\\
&&~~~~ -2 (\eps_1\cdot F_2\cdot \eps_4) (\eps_3\cdot k_{1}) -2
(\eps_1\cdot F_3\cdot \eps_4) (\eps_2\cdot k_{13})+2 (\eps_1\cdot
F_2\cdot F_3\cdot \eps_4)~.~~~\label{4p-n1324}\eea

In paper \cite{Bjerrum-Bohr:2016axv}, the BCJ numerators for
four-point Yang-Mills are computed by directly manipulating the
${\rm Pf}'\Psi$ using cross-ratio identities, and the result is
given by
\bea n_{(1|23|4)} & =& - W_{(1234)}+ W_{(124)} (-(\eps_3\cdot k_4))
+ W_{(134)} (\eps_2\cdot k_1))\nn
& & + (\eps_1\cdot \eps_4)\left\{ t\left\{(\eps_2\cdot
\eps_3)(k_4\cdot k_3) -   (\eps_2\cdot k_4)( \eps_3\cdot
k_4)-(\eps_2\cdot k_3)(\eps_3\cdot k_4) \right\}+(1-t) (\eps_3\cdot
k_4)( \eps_2\cdot k_1)\right\}~,~~~\nn
n_{(1|32|4)} & =& - W_{(1324)}+ W_{(124)}(\eps_3\cdot k_1)-W_{(134)}
(\eps_2\cdot k_4)\nn
& & + (\eps_1\cdot \eps_4)\left\{ t(\eps_2\cdot k_4)( \eps_3\cdot
k_1)+(1-t) \left\{(\eps_2\cdot \eps_3)(k_4\cdot k_2) - (\eps_3\cdot
k_4)( \eps_2\cdot k_4)-(\eps_3\cdot k_2)(\eps_2\cdot k_4)
\right\}\right\}~,~~~\nonumber\eea
where $t$ is the gauge choice in cross-ratio identities. If taking
$t=1/2$, we get the relabeling symmetric expression between
$2\leftrightarrow 3$. While if taking $t=0$, we get
\bea n_{(1|23|4)} & =& - W_{(1234)}+ W_{(124)} (\eps_3\cdot k_{12})
+ W_{(134)} (\eps_2\cdot k_1))- (\eps_1\cdot \eps_4)\ (\eps_3\cdot
k_{12})( \eps_2\cdot k_1)~,~~~\nn
n_{(1|32|4)} & =& - W_{(1324)}+ W_{(124)}(\eps_3\cdot k_1)+W_{(134)}
(\eps_2\cdot k_{13})- (\eps_1\cdot \eps_4)[(\eps_2\cdot
k_1)(\eps_3\cdot k_{1})+\eps_2\cdot F_3\cdot k_1]~,~~~\nonumber\eea
which has a perfect agreement with results \eref{4p-n1234} and
\eref{4p-n1324} up to a factor $-2$.

{Although only four-point example is explicitly displayed, this algorithm can directly be applied to obtain BCJ numerators for tree level Yang-Mills amplitudes with more than four points.}


\section{Inspecting the amplitude relations through BCFW recursions}
\label{secBCFW}

Using gauge invariance, we have determined the non-trivial relations
between any EYM amplitudes and Yang-Mills amplitudes, with the
formula as shown in (\ref{generalEYM}) for generel EYM amplitudes. However, it would
be interesting to inspect it again by BCFW recursions. As mentioned
before, the relation between color-ordered EYM amplitudes and
color-ordered Yang-Mills amplitudes is the same as that between
color-ordered Yang-Mills-scalar amplitudes and pure scalar
amplitudes. So for simplicity, we would use BCFW recursion to
inspect the following two non-trivial relations,
\bea &&A^{\YMs}_{n,1}(1,\ldots,n;p)=\sum_{\shuffle}(\epsilon_p\cdot
Y_p)A^{\phi^3}_{n+1}(1,\{2,\ldots,n-1\}\shuffle\{p\},n)~,~~~\label{bcfw1gra}\\
&&A^{\YMs}_{n,2}(1,\ldots,n;p,q)=\sum_{\shuffle}(\epsilon_p\cdot
Y_p)A^{\YMs}_{n+1,1}(1,\{2,\ldots,n-1\}\shuffle\{p\},n;q)\nonumber\\
&&~~~~~~~~~~~~~~~~~~~~~~~~~~~~~~~~~+\sum_{\shuffle}(\epsilon_p\cdot
F_q\cdot
Y_q)A^{\phi^3}_{n+2}(1,\{2,\ldots,n-1\}\shuffle\{q,p\},n)~.~~~\label{bcfw2gra}\eea
For more complicated relations where $A^{\YMs}_{n,m}$ with $m>2$, of
course we can also investigate them by BCFW recursions. While $m\geq 2$, we need to consider both contribution from finite poles and boundary contribution, and although for $m>2$ cases the analysis would be more involved, the techniques for computing boundary contribution are no more than the $m=2$ case. Hence their analysis is similar to the relations of $A^{\YMs}_{n,2}$.

In fact, verify
the relation (\ref{bcfw1gra}) by BCFW recursion is trivial. We only
need to apply BCFW on both sides of (\ref{bcfw1gra}), while a
$\widehat{k}_1,\widehat{k}_n$ shifting is sufficient to detect all
the contributions for $A_{n,1}^{\YMs}$ and $A_{n+1}^{\phi^3}$. By
using the relation (\ref{bcfw1gra}) for $A_{n',1}^{\YMs}$ with
$n'<n$, we can prove it inductively. While the starting point of
induction, i.e., the relation for $A_{2,1}^{\YMs}$, can be verified
explicitly.

For the two-scalar one-gluon amplitude $A_{2,1}^{\YMs}(1,2;p)$, we
explicitly have
\bea A^{\YMs}_{2,1}(1,2;p)=f^{a_1'a_2'a_p'}\left(-{i\over
\sqrt{2}}\right)(k_1-k_2)\cdot\epsilon_p=(\epsilon_p\cdot
k_1)\left(i\sqrt{2}f^{a_1'a_p'a_2'}\right)=(\epsilon_p\cdot
k_1)A^{\phi^3}_3(1,p,2), \eea
where $A^{\phi^3}_3(1,p,2)$ is the three-point scalar amplitude with
cubic vertex defined as $f^{a_1'a_p'a_2'}$. We have absorbed the
factor $i\sqrt{2}$ into $\epsilon_p$. This operation just changes
the normalization factor and will not affect the following
discussions.

We will not repeat the trivial proof of (\ref{bcfw1gra}) here, but
jump to the more typical one (\ref{bcfw2gra}). While applying BCFW
deformation, for (\ref{bcfw1gra}) only residues of finite poles will
contribute, but for (\ref{bcfw2gra}) the boundary contribution is
un-avoidable. So we need to compare both sides of (\ref{bcfw2gra})
with the finite pole contributions as well as boundary contribution,
and the later will be computed by analyzing Feynman diagrams
\cite{Feng:2009ei,Feng:2010ku,Feng:2011twa}.

Now let us study the relation for $A_{n,2}^{\YMs}(1,\ldots,n;p,q)$
with two gluons by BCFW recursion. We shift the momenta of two
scalars $k_1$ and $k_n$, i.e.,
\bea k_1\to \WH k_1=k_1+zq~~~,~~~k_n\to \WH k_n=k_n-zq~,~~~\eea
where $q$ satisfies $k_1\cdot q=k_n\cdot q=q^2=0$. We shall
emphasize that the on-shell condition of scalar $k_n$ is not really
used in the following proof, thus this relation is also valid for
amplitudes with $k_n$ off-shell. As the standard BCFW recursion
arguments, under this deformation, amplitude $A$ as a rational
function of $z$ can be written as
\bea A=\Sl_{\text{Finite Poles}}\text{Res}{A(z)\over
z}+\text{Boundary terms}. \eea
To prove the relation (\ref{bcfw2gra}), we should confirm, (1) the sums
over finite poles for the LHS and RHS of (\ref{bcfw2gra}) match with
each other, (2) the boundary terms for the LHS and RHS of
(\ref{bcfw2gra}) match with each other.

\subsection{Contributions of finite poles}

Let us first treat the contributions from finite poles. According to
BCFW recursion, the contribution in the LHS of (\ref{bcfw2gra}) can
be written as
\bea A^{\YMs}_{n,2}(1,\ldots,n;p,q)&=&\Sl_{i=1}^{n-1}A\left(\WH
1,2,\ldots,i;p\Big|i+1,\ldots,\WH
n;q\right)+\Sl_{i=1}^{n-1}A\left(\WH
1,2,\ldots,i;q\Big|i+1,\ldots,\WH n;p\right)\nn
&&+\Sl_{i=1}^{n-2}A\left(\WH 1,2,\ldots,i;p,q\Big|i+1,\ldots,\WH
n\right)+\Sl_{i=2}^{n-1}A\left(\WH 1,2,\ldots,i\Big|i+1,\ldots,\WH
n;p,q\right)~,~~~\label{eq:BCFW2GrLHS} \eea
where $A(\alpha|\beta)$ is short for $A_L(\alpha,P_\alpha){i\over
P_\alpha^2}A_R(-P_\alpha,\beta)$, and $P_\alpha$ is the sum of all
momenta in set $\alpha$, which is the propagator in between the left
and right sub-amplitudes. The contribution from the finite poles in
the RHS of (\ref{bcfw2gra}) can be given by the sum of $\mathcal{F}_1, \mathcal{F}_2$
defined as
\bea \mathcal{F}_1&=&\Sl_{i=1}^{n-1}\Sl_{\shuffle}(\epsilon_p\cdot \WH
Y_p)A\left(\WH 1,\{2,\ldots,i\}\shuffle\{p\}\Big|i+1,\ldots,\WH
n;q\right)\label{bcfwT1-1}\\
&&+\Sl_{i=2}^{n-1}\Sl_{\shuffle}(\epsilon_p\cdot \WH
Y_p+\epsilon_p\cdot \WH P_{1\ldots i})A\left(\WH
1,2,\ldots,i\Big|\{i+1,\ldots,n-1\}\shuffle\{p\},\WH n;q\right)\label{bcfwT1-2}\\
&&+\Sl_{i=1}^{n-2}\Sl_{\shuffle}(\epsilon_p\cdot \WH Y_p)A\left(\WH
1,\{2,\ldots,i\}\shuffle\{p\};q\Big|i+1,\ldots,\WH n\right)\label{bcfwT1-3}\\
&&+\Sl_{i=1}^{n-1}\Sl_{\shuffle}(\epsilon_p\cdot \WH
Y_p+\epsilon_p\cdot \WH P_{1\ldots i})A\left(\WH
1,2,\ldots,i;q\Big|\{i+1,\ldots,n-1\}\shuffle \{p\},\WH
n\right)~,~~~\label{bcfwT1-4}
\eea
and
\bea \mathcal{F}_2&=&\Sl_{i=1}^{n-2}\Sl_{\shuffle}(\epsilon_p\cdot F_q\cdot
\WH Y_q)A\left(\WH 1,\{2,\ldots,i\}\shuffle
\{q,p\}\Big|i+1,\ldots,\WH n\right)\label{bcfwT2-1}\\
&&+\Sl_{i=2}^{n-1}\Sl_{\shuffle}(\epsilon_p\cdot F_q\cdot (\WH
Y_q+\WH P_{1\ldots i}))A\left(\WH
1,2,\ldots,i\Big|\{i+1,\ldots,n-1\}\shuffle\{q,p\},\WH n\right)\label{bcfwT2-2}\\
&&+\Sl_{i=1}^{n-1}\Sl_{\shuffle_1,\shuffle_2}(\epsilon_p\cdot
F_q\cdot \WH Y_q)A\left(\WH
1,\{2,\ldots,i\}\shuffle_1\{q\}\Big|\{i+1,\ldots,n-1\}\shuffle_2
\{p\},\WH n\right)~,~~~\label{bcfwT2-3}
\eea
where remind again $Y_p$ is defined as the sum of the momenta at the LHS of
$p$ in its corresponding sub-amplitude which has scalar origin in
$A^{\YMs}_{n,2}$. While $p$ is the the right sub-amplitude $A_R$,
the intermediate propagator $P_\alpha$ appears as the first leg of
$A_R$ which is at the LHS of $p$, and we write it explicitly out of $\WH
Y_p$ to avoid ambiguity.

Let us now compare the result in (\ref{eq:BCFW2GrLHS}) and
$\mathcal{F}_1,\mathcal{F}_2$. From inductive assumption, we know $A^{\YMs}_{n,1}$ for
any $n$ and the lower-point amplitude $A^{\YMs}_{n',2}$ with $n'<n$
satisfy (\ref{bcfw1gra}), (\ref{bcfw2gra}). So we have,
\begin{enumerate}
  \item The sum of
$A_L$ in (\ref{bcfwT1-1}) is equal to the $A_L$ in the first
term of (\ref{eq:BCFW2GrLHS}) by relations of YMs amplitude with
one gluon.
  \item In the (\ref{bcfwT2-3}), $(\epsilon_p\cdot F_q\cdot \WH Y_q)=(\epsilon_p\cdot q)(\epsilon_q\cdot \WH Y_q)-(\epsilon_p\cdot \epsilon_q)(q\cdot \WH
  Y_q)$, while for the latter factor, $$\sum_{\shuffle_1}(q\cdot
  \WH Y_q)A_L(\WH 1, \{2,\ldots,i\}\shuffle_1\{q\},\WH
  P_{1\ldots i,q})=0$$ due to BCJ relation. For the former
  factor,
 \bea  &&\sum_{\shuffle_1}(\epsilon_p\cdot q)(\epsilon_q\cdot
  \WH Y_q)A\left(\WH
1,\{2,\ldots,i\}\shuffle_1\{q\}\Big|\{i+1,\ldots,n-1\}\shuffle_2
\{p\},\WH n\right)\nonumber\\
&&=(\epsilon_p\cdot q)A\left(\WH
1,2,\ldots,i;q\Big|\{i+1,\ldots,n-1\}\shuffle_2 \{p\},\WH
n\right) ~.~~~\nonumber\eea
Noting that this term plus (\ref{bcfwT1-4}), and that $\epsilon_p\cdot
\WH P_{1\ldots i}+\epsilon_p\cdot q=\epsilon_p\cdot \WH
P_{1\ldots i,q}$, we get
\bea &&\Sl_{\shuffle}(\epsilon_p\cdot Y_p+\epsilon_p\cdot \WH
P_{1\ldots i},q)A_L(\WH 1,2,\ldots,i,-\WH P_{1\ldots
i,q};q){i\over P^2_{1\ldots i,q}}A_R(\WH P_{1\ldots
i,q},\{i+1,\ldots,n-1\}\shuffle \{p\},\WH n)\nonumber~,~~~\eea
which is equal to the second term of (\ref{eq:BCFW2GrLHS}) by
relations of YMs amplitude with one gluon for $A_R$.
  \item The $A_L$ in (\ref{bcfwT1-3}) and (\ref{bcfwT2-1}) sums to,
\bea &&\Sl_{\shuffle}(\epsilon_p\cdot \WH Y_p)A_L(\WH
1,\{2,\ldots,i\}\shuffle\{p\},-\WH P_{1\ldots i,p,q};q){i\over
P^2_{1\ldots i,p,q}}A_R(\WH P_{1\ldots i,p,q},i+1,\ldots,\WH
n)\nonumber\\
&&+\Sl_{\shuffle}(\epsilon_p\cdot F_q\cdot \WH Y_q)A_L(\WH
1,\{2,\ldots,i\}\shuffle \{q,p\},-\WH P_{1\ldots i,p,q}){i\over
\WH P^2_{1\ldots i,p,q}}A_R(\WH P_{1\ldots i,p,q},i+1,\ldots,\WH
n)~,~~~\nonumber\eea
which equals to the third term of (\ref{eq:BCFW2GrLHS}) by
relations of lower-point YMs amplitudes with two gluons.
  \item The $A_R$ in (\ref{bcfwT1-2}) and (\ref{bcfwT2-2}) sums to
  \bea &&\Sl_{\shuffle}(\epsilon_p\cdot (\WH Y_p+ \WH P_{1\ldots
i}))A_L(\WH 1,2,\ldots,i,-\WH P_{1\ldots i}){i\over P^2_{1\ldots
i}}A_R(\WH P_{1\ldots i},\{i+1,\ldots,n-1\}\shuffle\{p\},\WH
n;q)\nonumber\\
&&+\Sl_{\shuffle}(\epsilon_p\cdot F_q\cdot (\WH Y_q+\WH
P_{1\ldots i}))A(\WH 1,2,\ldots,i,-\WH P_{1\ldots i}){i\over \WH
P^2_{1\ldots i}}A_R(\WH P_{1\ldots
i},\{i+1,\ldots,n-1\}\shuffle\{q,p\},\WH n)~,~~~\nonumber \eea
  which equals to the fourth term of (\ref{eq:BCFW2GrLHS}) by
  relations of lower-point YMs amplitude with two gluons.
\end{enumerate}
Hence, all contributions of finite poles in the LHS and RHS of
(\ref{bcfw2gra}) under $(k_1,k_n)$-shifting match with each other.

\subsection{The boundary contributions}

%
\begin{figure}
\centering
\includegraphics[width=6in]{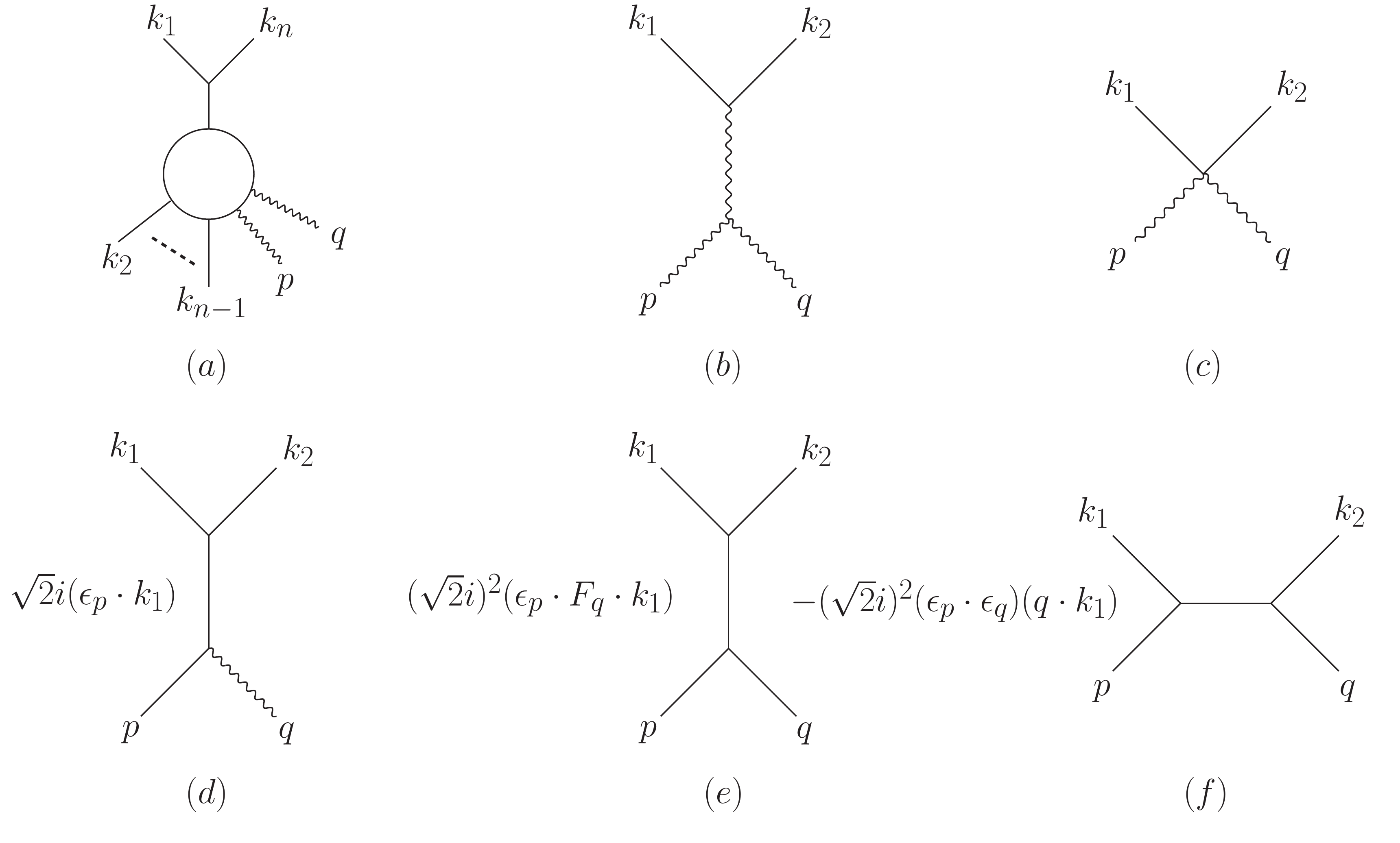}
\caption{Boundary of the left hand side of relation for amplitudes with more than two scalars and two gluons.}\label{Fig:boundary}
\end{figure}

Now let us discuss the boundary contributions. We would consider the
situations with $n>2$ and $n=2$ separately, since there are more
subtleties in the latter situation.

\noindent{\bf The case with $n>2$:} in this case, the boundary term
of the BCFW recursion under $(k_1,k_n)$-deformation comes from the
diagram as shown in Fig.(\ref{Fig:boundary}.a).  In the LHS of
(\ref{bcfw2gra}), it is given by
\bea f^{a'_na'_1e}{i\over s_{1n}}\W
A^{\YMs}_{n-1,2}(2,\ldots,n-1,P_{1,n}^e;p,q)~,~~~\eea
where $\W A^{\YMs}_{n-1,2}(P_{1,n}^e,2,\ldots,n-1;p,q)$ denotes sum
of all possible diagrams with one off-shell scalar line $P_{1,n}^e$,
$(n-2)$ on-shell scalars $k_2,\ldots,k_{n-1}$ as well as two
on-shell gluons $p$, $q$. In the RHS of (\ref{bcfw2gra}), it is
given by
\bea
&&\Sl_{\shuffle}(\epsilon_p\cdot Y_p)f^{a'_na'_1e}{i\over s_{1n}}\W A^{\YMs}_{n,1}(\{2,\ldots,n-1\}\shuffle \{p\},P_{1,n}^e;q)\nonumber\\
&&+\Sl_{\shuffle}(\epsilon_p\cdot F_q\cdot Y_q)f^{a'_na'_1e}{i\over
s_{1n}}\W A^{\phi^3}_{n+1}(\{2,\ldots,n-1\}\shuffle
\{q,p\},P_{1,n}^e)~.~~~
\eea
By the use of generalized $U(1)$-decoupling identity \cite{Du:2011js}
\bea \Sl_{\shuffle}\W
A_{n+m,k}^{\YMs}(\{1,\ldots,n-1\}\shuffle\{\alpha_1,\ldots,\alpha_m\},-P_{1\ldots
n-1,\alpha_1\ldots \alpha_m,h_1\ldots
h_k};h_1,\ldots,h_k)=0~,~~~\eea
we re-express the boundary term in the RHS of (\ref{bcfw2gra}) as
\bea
&&\Sl_{\shuffle}(\epsilon_p\cdot Y_p)f^{a'_na'_1e}{i\over s_{1n}}\W A^{\YMs}_{n,1}(2,\{3,\ldots,n-1\}\shuffle \{p\},P_{1,n}^e;q)\nonumber\\
&&+\Sl_{\shuffle}(\epsilon_p\cdot F_q\cdot Y_q)f^{a'_na'_1e}{i\over
s_{1n}}\W A^{\phi^3}_{n+1}(2,\{3,\ldots,n-1\}\shuffle
\{q,p\},P_{1,n}^e)~.~~~\label{bcfwBoundaryT1}
\eea
Remind that our proof of (\ref{bcfw2gra}) does not rely on the
on-shell condition of the right-most scalar $k_n$, hence
(\ref{bcfw2gra}) is also valid for amplitudes with off-shell $k_n$.
Assuming the validation of (\ref{bcfw2gra}) for YMs amplitude with
$n'<n$ gluons, we simply get the sum (\ref{bcfwBoundaryT1}) as
$f^{a'_na'_1e}{i\over s_{1n}}\W
A^{YMs}_{n-1,2}(2,\ldots,n-1,P_{1,n}^e;p,q)$, which is identical to
the boundary contribution in the LHS of (\ref{bcfw2gra}).

\noindent{\bf The case with $n=2$:} this case is much more subtle.
The boundary contributions in the LHS of (\ref{bcfw2gra}) come from
the diagrams as shown in Fig.(\ref{Fig:boundary}.b),
Fig.(\ref{Fig:boundary}.c), while the boundary contributions in the
RHS of (\ref{bcfw2gra}) come from the diagrams as shown in
Fig.(\ref{Fig:boundary}.d), Fig.(\ref{Fig:boundary}.e),
Fig.(\ref{Fig:boundary}.f).

According to the Feynman rules for Yang-Mills-scalar amplitudes, we
can compute the three terms for the RHS of (\ref{bcfw2gra}) as
\bea
\mbox{Fig.(\ref{Fig:boundary}.d)}&=&f^{a'_2a'_1e}f^{ea'_pa'_q}\,{i\over
s_{12}}\,(\sqrt{2}i)^2\,(\epsilon_q\cdot p)\,(\epsilon_p\cdot
k_1)~,~~~\\
\mbox{Fig.(\ref{Fig:boundary}.e)}&=&f^{a'_2a'_1e}f^{ea'_qa'_p}\,{i\over
s_{12}}\,(\sqrt{2}i)^2(\epsilon_p\cdot F_q \cdot k_1)~,~~~\\
\mbox{Fig.(\ref{Fig:boundary}.f)}&=&-f^{a'_1a'_qe}f^{ea'_pa'_2}\,(\sqrt{2}i)^2\,{i\over
2}(\epsilon_p\cdot\epsilon_q)~.~~~ \eea
On the other hand, we can compute the two terms for LHS of
(\ref{bcfw2gra}) as
\bea
\mbox{Fig.(\ref{Fig:boundary}.b)}&=&f^{a'_2a'_1e}f^{ea'_pa'_q}\,{i\over
s_{12}}\,(\sqrt{2}i)^2\,\left[(\epsilon_q\cdot p)(\epsilon_p\cdot
k_1)-(\epsilon_p\cdot q)(\epsilon_q\cdot
k_1)+(\epsilon_p\cdot\epsilon_q)(q\cdot k_1)\right]\nn
&&+(-i)f^{a'_2a'_1e}f^{ea'_pa'_q}\,{1\over
2}\,(\epsilon_p\cdot\epsilon_q)~.~~~\nn
\mbox{Fig.(\ref{Fig:boundary}.c)}&=&f^{a'_1a'_pe}f^{ea'_qa'_2}\,{i\over
2}\,(\epsilon_p\cdot\epsilon_q)+f^{a'_1a'_qe}f^{ea'_pa'_2}\,{i\over
2}\,(\epsilon_p\cdot\epsilon_q)~.~~~ \eea
If we re-write the second line in the result of
Fig.(\ref{Fig:boundary}.b) by Jacobi identity
$$f^{a'_2a'_1e}f^{ea'_pa'_q}=f^{a'_1a'_pe}f^{ea'_qa'_2}-f^{a'_1a'_qe}f^{ea'_pa'_2}~,~~~$$
then the matching of boundary contribution in both sides of
(\ref{bcfw2gra}) can be easily checked.

With above discussions, we have confirmed the non-trivial relations
between YMs amplitude and pure scalar amplitudes (hence the EYM
amplitude and Yang-Mills amplitudes) by BCFW recursion relations.
The proof of relations for YMs amplitude with more than two gluons
requires more labors, but the strategy is similar, which includes
comparing the contributions from finite poles and boundary
contributions. We will not discuss it further.

\section{Inspecting the amplitude relations through KLT relation}
\label{secKLT}

In the following discussions we will demonstrate that, at least in
the first few simplest scenarios, the newly discovered
multi-graviton relations
\cite{Nandan:2016pya,Stieberger:2016lng,delaCruz:2016gnm} can be
readily understood from the perspective of KLT relations. It was
demonstrated in \cite{Bern:1999bx} that the KLT relation provides a
much more perturbation-friendly construction of the EYM amplitudes,
which would be otherwise difficult to calculate in viewing of the
infinite vertices that constitute the linearized gravity Feynman
rules. In this setting, EYM amplitude factorizes into a copy of pure
gluon amplitude and a copy that gluon interacts with scalars,
through which the color dependence is introduced. To have simpler
expression, we will use the $(n-2)!$ symmetric KLT relation first
introduced in
\cite{BjerrumBohr:2010ta,BjerrumBohr:2010zb,BjerrumBohr:2010yc},
\begin{eqnarray}
A^{\EYM}(1,2,\ldots,n) & = & \lim_{k_n^2\to 0}{1\over k_n^2}\sum_{\alpha,\beta\in \mathcal{S}_{n-2}}{{A}^{\YM}(n,\alpha,1)S[\alpha|\beta]A^{\YMs}(1,\beta,n)}\label{eq:klt-eym}\\
 & = & \sum_{\alpha\in S_{n-2}}{A}^{\YM}(1,\alpha,n)\,n(1,\alpha,n)~.~~~\nonumber
\end{eqnarray}
where the numerator in the expression defined using gluon scalar
currents
\begin{equation}
n(1,\alpha,n)=\sum_{\beta\in S_{n-2}}\mathcal{S}[\alpha|\beta]J^{\YMs}(1,\beta,n)~~~~\label{eq:klt-insp-numerator}
\end{equation}
carries both kinematic and color factors. The formula defined in
(\ref{eq:klt-insp-numerator}) has provided a way of evaluating the
numerator $n(1,\alpha,n)$. However, it is obvious that, directly
calculating all currents and then making the sum is not an efficient
method. There are two alternative methods to compute the
coefficients $n(1,\alpha,n)$. The first is to carry out the
summation step by step as was done in \cite{Du:2011js,Du:2016tbc}.
The idea is to divide the full $S_{n-2}$ permutation sums appearing
in (\ref{eq:klt-insp-numerator}) into $(n-2)$ blocks of $S_{n-3}$
permutation sums, such that in each block we can pull out a format
of BCJ sums. Then one can use the Fundamental BCJ relation for
currents to simplify the expression and arrive at a similar sum as
the one given in (\ref{eq:klt-insp-numerator}) but with only
$S_{n-3}$ permutation sums. Iterating this procedure several times,
we can finally compute the coefficients. Establishing the
Fundamental BCJ relation for currents is a crucial point for this
method, and we will show how to do this in the later sections. The
second method is, however, less straightforward. When expanding the
amplitude into KK basis with the formulation given in the second
line of (\ref{eq:klt-eym}), it is shown in \cite{Cachazo:2013iea,
Fu:2012uy, Du:2013sha, Fu:2014pya} that, the coefficients
$n(1,\alpha,n)$ are nothing but the numerators of Del
Duca-Dixon-Maltoni (DDM) basis provided we write the whole
$A^{\YMs}(1,\beta,n)$ amplitude into BCJ form (i.e., numerators
satisfying the Jacobi relations). Using this aspect, the problem is
translated to computing the BCJ numerators of DDM basis by any
conventional methods.

The purpose of this section is to show that, the newly discovered EYM amplitude relations can also be fitted in the framework of KLT relations. The methods that developed in the computation of BCJ numerators in various theories \cite{Du:2011js,Du:2016tbc} are also well-suited in the analysis of EYM amplitude expansion, with only a few modification. This connects the problem of EYM amplitude expansion with many other theories. In the following discussions, we will use both methods developed years ago for computing the BCJ numerators to address
the problem of constructing the expansion coefficients
$n(1,\alpha,n)$.

\subsection{The case with single gluon}

For the purpose of being self-contained we list the color-ordered
Feynman rules for gluon-scalar interaction presented in
\cite{Bern:1999bx} in the Appendix \ref{Appendix:feyn-rules}.
Consider first the scalar Yang-Mills amplitudes when there is only
one gluon. Note that a (color-stripped) gluon propagator does not
transmit the color/flavor of scalars attached to its two ends, so
that for single trace part of the partial amplitude, gluon lines
cannot be internal or the color factors carried by the scalars at
its two ends factorize. A consequence is that all single gluon
amplitudes are consisting of cubic graphs. For example at four
points when, say leg $3$, is the gluon line there are only three
cubic graphs in the KK sector, up to anti-symmetry of the
three-vertices,
\bea \begin{array}{c}
\includegraphics[width=1.62cm]{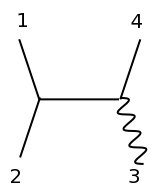}~~~~~~~
\includegraphics[width=1.62cm]{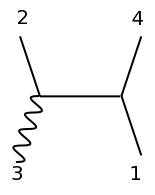}~~~~~~
\includegraphics[width=1.62cm]{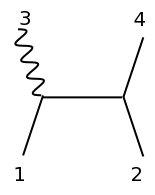}\end{array}~.~~~\nonumber\eea
It is very important to notice that  the color-kinematics duality is
ensured by the vanishing of the sum of their numerators (even at the
off-shell level) 
\begin{equation}
f^{a_{1}a_{2}a_{4}}(k_{1}+k_{2}+k_{4})\cdot\epsilon_{3}\sim k_{3}\cdot\epsilon_{3}=0~.~~~
\end{equation}
This observation (i.e., only cubic vertex is allowed and the gauge
invariance), when generalized to higher points, indicates that the
Feynman diagrams provide the desired BCJ form. In particular, an
$n$-point DDM half-ladder numerator $n(1,2,3,\cdots
i,p_{g},i+1,\cdots,n)$ is therefore given by the corresponding
Feynman diagrams as 
\begin{eqnarray}
k_{12}^{2}k_{123}^{2}\cdots\,
\begin{minipage}{4.09cm} \includegraphics[width=4.09cm]{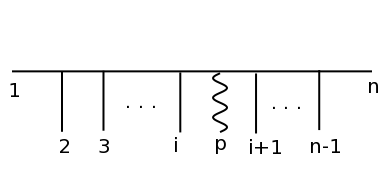}  \end{minipage}
 & = & \left[f^{1,2,*}f^{*,3,*}\dots f^{*,i,a}\times\delta^{ab}\times f^{b,i+1,*}\cdots f^{*,n-1,n}\right]\nonumber \\
 &  &~~~~~~~~~~~~~~~~~~ \times \sqrt{2}^{n-2}(-1)^{n+1}\,i\, \epsilon_{p}\cdot\left(k_{1}+\cdots+k_{i}\right)~.~~~\label{Fig.1.5}
\end{eqnarray}
So what are the allowed DDM numerators? The Yang-Mills-scalar theory
has a gauge group and a flavor group. The flavor ordering fixes the
ordering of scalars, thus the only allowed freedom is the location
of gluon leg along the DDM-chain. In other words, the desired
expansion coefficients in (\ref{eq:klt-eym}) are nothing but the one
given in (\ref{Fig.1.5}) with all possible insertions of gluon legs,
\begin{equation}
A^{\EYM}_{n,1}(1,2,\cdots,n;p)= \sqrt{2}^{n-2}(-1)^{n+1}\,i\, \sum_{\shuffle}(\epsilon_{p}\cdot X_{p}){A}^{\YM}(1,\{2,\ldots,n-1\}\shuffle\{p\},n)~,~~~
\end{equation}
and we find agreement with the new single graviton relation (up to
an overall factor).

\subsection{The four-point gluon-scalar amplitude involving two gluons}

Next we consider Yang-Mills-scalar amplitudes involving two gluons.
For this case, since Feynman diagrams will involve the four-point
vertex, the BCJ form will not be manifest for Feynman diagrams and
the computation will be more complicated. Thus in this subsection,
we will follow the method of summing over the color-ordered KK
basis.

At four points we have the following two KK basis amplitudes, 
\begin{eqnarray}
A^{\YMs}_{2,2}(1,2_{g},3_{g},4) & = & \frac{n_{s}}{s_{12}}-\frac{n_{t}}{s_{23}}+n_{4} =
 \begin{minipage}{1.62cm}  \includegraphics[width=1.62cm]{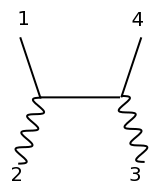} \end{minipage}
 +
 \begin{minipage}{1.62cm}  \includegraphics[width=1.62cm]{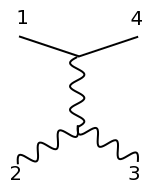} \end{minipage}
 +
 \begin{minipage}{1.62cm} \includegraphics[width=1.62cm]{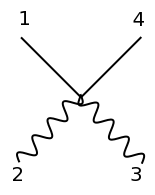}  \end{minipage}~,~~~ \\
A^{\YMs}_{2,2}(1,3_{g},2_{g},4) & = & -\frac{n_{u}}{s_{13}}+\frac{n_{t}}{s_{23}}+n_{4} =
 \begin{minipage}{1.62cm} \includegraphics[width=1.62cm]{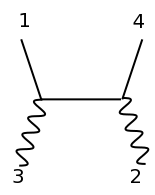}  \end{minipage}
  +
  \begin{minipage}{1.62cm}  \includegraphics[width=1.62cm]{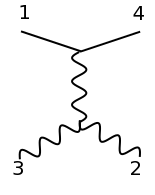}  \end{minipage}
  +
  \begin{minipage}{1.62cm} \includegraphics[width=1.62cm]{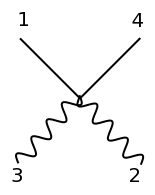}  \end{minipage}~,~~~
\end{eqnarray}
where $n_{s}$, $n_{t}$, $n_{u}$ and $n_{4}$ denote the factors
\begin{flalign} n_{s} & = k_{12}^{2}\,
\begin{minipage}{1.62cm}  \includegraphics[width=1.62cm]{ngg-s} \end{minipage}
  =  \Big(((k_{3}+k_{4})-k_{1})\cdot\epsilon_{2}\Big)\Big((k_{4}-(k_{1}+k_{2}))\cdot\epsilon_{3}\Big)\times\delta^{a_{1}a_{4}}\left(\frac{-i}{\sqrt{2}}\right)^{2}
  \,i~,~~~&&\label{eq:4pt-ns}\end{flalign}
\begin{flalign} n_{t} & =  k_{23}^{2}\,
\begin{minipage}{1.62cm}  \includegraphics[width=1.62cm]{ngg-t} \end{minipage}
  =  \Big(\big(\epsilon_{2}\cdot\epsilon_{3}\big)\big(
 (k_{3}-k_{2}) \cdot (k_{4}-k_{1})\big)
 +\big(\left(k_{4}-k_{1}\right)\cdot\epsilon_{3}\big)\big(\left((k_{1}+k_{4})-k_{3}\right)\cdot\epsilon_{2}\big)\nonumber &&\\
 & ~~~~~~~~~~~~~~~~~~~~~~~~~~~~~~~~~+\big(\left(k_{4}-k_{1}\right)\cdot\epsilon_{2}\big)\big(\left(k_{2}-(k_{1}+k_{4})\right)\cdot\epsilon_{3}\big)\Big)\times\delta^{a_{1}a_{4}}\left(\frac{i}{\sqrt{2}}\right)\left(\frac{-i}{\sqrt{2}}\right)
 \,
 i~,~~~&&\label{eq:4pt-nt}\end{flalign}
\begin{flalign}
n_{u} & =  (-1)k_{13}^{2}\,
\begin{minipage}{1.62cm}  \includegraphics[width=1.62cm]{ngg-u-prime} \end{minipage}  =  (-1)\big(\left((k_{2}+k_{4})-k_{1}\right)\cdot\epsilon_{3}\big)\big(\left(k_{4}-(k_{1}+k_{3})\right)\cdot\epsilon_{2}\big)\times\delta^{a_{1}a_{4}}\left(\frac{-i}{\sqrt{2}}\right)^{2} \,
i~,~~~\label{eq:4pt-nu}&&
\end{flalign}
and 
\bea n_{4}  =
\begin{minipage}{1.62cm} \includegraphics[width=1.62cm]{ngg-4}  \end{minipage}
 =
 \frac{i}{2}\delta^{a_{1}a_{4}}(\epsilon_{2}\cdot\epsilon_{3})~.~~~\label{eq:4pt-n4}
\eea 
Now that with quartic graph present, the original Jacobi identity
inevitably needs to be modified if color-kinematics duality is to
remain holding. To better keep track of how this is done we write
the BCJ sum of the two KK basis amplitudes in terms of the factors
just introduced so that every term appears in the sum has a clear
graphical interpretation. Also for future reference we analytically
continue one of the scalar legs, say leg $4$, and write 
\begin{eqnarray}
 &&s_{21}A^{\YMs}_{2,2}(1,2_g,3_g,4)+(s_{21}+s_{23})A^{\YMs}_{2,2}(1,3_g,2_g,4)\label{eq:4pt-off-shell-bcj} \\
 &&=s_{12}\left(\frac{n_{s}}{s_{12}}-\frac{n_{t}}{s_{23}}+n_{4}\right)+(s_{21}+s_{23})\left(-\frac{n_{u}}{s_{13}}+\frac{n_{t}}{s_{23}}+n_{4}\right)\nonumber\\
 &&=\left(n_{s}+n_{t}+n_{u}+(s_{12}-s_{13})n_{4}\right)+k_{4}^{2}\left(- \, \frac{n_{u}}{s_{13}}+n_{4}\right)~.~~~\nonumber
\end{eqnarray}
The fact that BCJ sum vanishes in the on-shell limit suggests that
the Jacobi identity is modified as
$n_{s}+n_{t}+n_{u}+(s_{12}-s_{13})n_{4}=0$ up to terms proportional
to $k_{4}^{2}$. A careful inspection shows that these terms actually
cancel completely. Plugging equations (\ref{eq:4pt-ns}) to
(\ref{eq:4pt-n4}) into the left hand side of this modified Jacobi
sum, we see that (neglecting an overall factor
$(-i)\delta^{a_{1}a_{4}}/2$), 
\begin{eqnarray}
(\epsilon_{2}\cdot\epsilon_{3})~~\text{ part }&:&~  -\,\left(k_{3}-k_{2}\right)\cdot\left(k_{4}-k_{1}\right)+(s_{12}-s_{13})=0~,~~\label{eq:jacobi-ee-part}\\
(\epsilon_{2}\cdot k)(\epsilon_{3}\cdot k)~\text{ part }&:&~ 2^{2}\,\big(k_{1}\cdot\epsilon_{2}\big)\,\big((k_{1}+k_{2})\cdot\epsilon_{3}\big)~,~~ \leftarrow\text{contribution from }n_{s}~,~~~
\nonumber \\
 && +2^{2}\,(-1)\big((k_{1}+k_{3})\cdot\epsilon_{2}\big)\,\big(k_{1}\cdot\epsilon_{3}\big)~,~~~  \leftarrow\text{contribution from }n_{u}~,~~~\nonumber \\
 && +2\Big(\left(k_{3}\cdot\epsilon_{2}\right)\left((2k_{1}+k_{2})\cdot\epsilon_{3}\right)-\left((2k_{1}+k_{3})\cdot\epsilon_{2}\right)\left(k_{2}\cdot\epsilon_{3}\right)\Big)~,~~ \leftarrow\text{contribution from }n_{t}~,  \nonumber\\
 & &=0~.~~~\label{eq:jacobi-ek-part}\end{eqnarray}
We obtain the numerator by feeding the off-shell continued BCJ sum
just computed into the KLT inspired prescription
(\ref{eq:klt-insp-numerator}), taking the modified Jacobi identity
into account, yielding 
\begin{eqnarray*}
n(13_g2_g4) & = & \frac{1}{k_{4}^{2}}s_{31}\left[s_{21}A^{\YMs}_{2,2}(1,2_g,3_g,4)+(s_{21}+s_{23})A^{\YMs}_{2,2}(1,3_g,2_g,4)\right]\\
 & = & - \, n_{u}+s_{13}\,n_{4}=
 \begin{minipage}{2.14cm} \includegraphics[width=2.14cm]{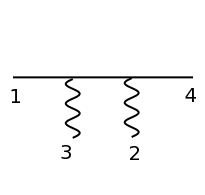}  \end{minipage}
 + s_{13}\,
 \begin{minipage}{2.14cm}  \includegraphics[width=2.14cm]{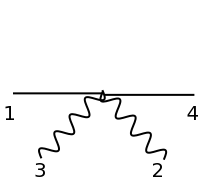} \end{minipage}~,~~~
\end{eqnarray*}
where by an abuse of notation we neglected factors of inverse
propagators so that the graphs appear in the equation above should
be understood as representing the corresponding numerators rather
than the original Feynman graphs. In the following discussions we
shall not distinguish numerators from Feynman graphs unless it is
not apparent from the context. The other two gluon numerator at four
points can be readily obtained by swapping labels
$(2\leftrightarrow3)$. Inserting the half-ladder numerators back
into KLT relation and we find agreement with the two graviton
relation (equation (4) in \cite{Nandan:2016pya}). 
\begin{eqnarray}
A^{\EYM}_{2,2}(1,4;h_2,h_3) & = & (-2i)\Big((\epsilon_{2}\cdot X_{2})(\epsilon_{3}\cdot X_{3})-\frac{1}{4}(\epsilon_{2}\cdot\epsilon_{3})s_{21}\Big){A}^{\YM}_4(1,2,3,4)+(2\leftrightarrow3)~.~~~
\end{eqnarray}
Note that the above relation is not exactly the same as
(\ref{bcfw1gra}), but equivalent to it after using certain BCJ
relations, and note particularly that the new
$\epsilon_2\cdot\epsilon_3$ term came from the quartic graph
contribution.

\subsubsection*{Off-shell continued Jacobi identity}

The key point of the above calculation is the modified Jacobi
identity when some of the legs becoming off-shell, e.g.,
$n_{s}+n_{t}+n_{u}+(s_{12}-s_{13})n_{4}=0$. This modification will
lead to modified fundamental BCJ relations, to be discussed later.
When considering situations for higher points, one note that the
color dependency will factorize when the scalars are connected by an
internal gluon line, thus the single trace part of a two-gluon
partial amplitude can only contain graphs derivable from those
appearing at four-point case by welding pure scalar currents to
their two scalar legs. Therefore we only need to consider
analytically continuing the two scalar lines of the modified Jacobi
identity when two gluons are present. Careful inspection of
(\ref{eq:jacobi-ek-part}) shows that the $(\epsilon\cdot
k)(\epsilon\cdot k)$ part of the Jacobi sum is a pair-wise
cancelation, up to terms proportional to $(\epsilon_{2}\cdot k_{2})$
or $(\epsilon_{3}\cdot k_{3})$, and therefore remains valid even
when scalars become off-shell. The only modification comes from the
$(\epsilon_{2}\cdot\epsilon_{3})$ part. To completely cancel the
$(k_{2}-k_{3})\cdot(k_{1}-k_{4})$ factor produced by $n_{t}$, we see
that the quartic graph needs to be multiplied by the same factor.
The off-shell continued identity we need for all two gluon
amplitudes is then 
\begin{equation}
\begin{minipage}{1.62cm} \includegraphics[width=1.62cm]{ngg-s}  \end{minipage}
+
\begin{minipage}{1.62cm}   \includegraphics[width=1.62cm]{ngg-t} \end{minipage}
-
\begin{minipage}{1.62cm}  \includegraphics[width=1.62cm]{ngg-u-prime} \end{minipage}
+ \left(k_{2}-k_{3}\right)\cdot\left(k_{1}-k_{4}\right) \,
\begin{minipage}{1.62cm}  \includegraphics[width=1.62cm]{ngg-4} \end{minipage}
=0~,~~~
\end{equation}
and we will be using this identity in the following discussions.

\subsection{The five-point YMs amplitudes with two gluons}

Having presented the example of four points with two gluons, we
further show an example of five-point amplitude with two gluons.
Again, we will use the method of summing over color-ordered KK
basis.

At five points the number of graphs increases considerably. Recall
from \cite{Bern:2008qj} that there are $15$ different graphs in
total in the KK sector at five points if the amplitudes are to be
described by cubic graphs only, $6$ of them are independent when
Jacobi identities are taken into account. Similarly we label the
cubic graphs as $n_{1}$, $n_{2}$, $\ldots$, $n_{15}$, and we regard
the quartic graphs as additional corrections $n_{16}$, $n_{17}$,
$n_{18}$. The amplitudes are given by 
\begin{eqnarray}
A^{\YMs}_{3,2}(1,2_{g},3_{g},4,5) & = & \frac{n_{1}}{s_{12}s_{45}}+\frac{n_{2}}{s_{23}s_{15}}+\frac{n_{3}}{s_{34}s_{12}}+\frac{n_{4}}{s_{45}s_{23}}+\frac{n_{5}}{s_{15}s_{34}}+\frac{n_{16}}{s_{45}}+\frac{n_{17}}{s_{15}}~,~~~\\
A^{\YMs}_{3,2}(1,4,3_{g},2_{g},5) & = & \frac{n_{6}}{s_{14}s_{25}}+\frac{n_{5}}{s_{34}s_{15}}+\frac{n_{7}}{s_{23}s_{14}}+\frac{n_{8}}{s_{25}s_{34}}+\frac{n_{2}}{s_{15}s_{23}}+\frac{n_{18}}{s_{14}}+\frac{n_{17}}{s_{15}}~,~~~\\
A^{\YMs}_{3,2}(1,3_{g},4,2_{g},5) & = & \frac{n_{9}}{s_{13}s_{25}}-\frac{n_{5}}{s_{34}s_{15}}+\frac{n_{10}}{s_{24}s_{13}}-\frac{n_{8}}{s_{25}s_{34}}+\frac{n_{11}}{s_{15}s_{24}}-2\frac{n_{17}}{s_{15}}~,~~~\\
A^{\YMs}_{3,2}(1,2_{g},4,3_{g},5) & = & \frac{n_{12}}{s_{12}s_{35}}+\frac{n_{11}}{s_{24}s_{15}}-\frac{n_{3}}{s_{34}s_{12}}+\frac{n_{13}}{s_{35}s_{24}}-\frac{n_{5}}{s_{15}s_{34}}-2\frac{n_{17}}{s_{15}}~,~~~\\
A^{\YMs}_{3,2}(1,4,2_{g},3_{g},5) & = & \frac{n_{14}}{s_{14}s_{35}}-\frac{n_{11}}{s_{24}s_{15}}-\frac{n_{7}}{s_{23}s_{14}}-\frac{n_{13}}{s_{35}s_{24}}-\frac{n_{2}}{s_{15}s_{23}}+\frac{n_{18}}{s_{14}}+\frac{n_{17}}{s_{15}}~,~~~\\
A^{\YMs}_{3,2}(1,3_{g},2_{g},4,5) & = & \frac{n_{15}}{s_{13}s_{45}}-\frac{n_{2}}{s_{23}s_{15}}-\frac{n_{10}}{s_{24}s_{13}}-\frac{n_{4}}{s_{45}s_{23}}-\frac{n_{11}}{s_{15}s_{24}}+\frac{n_{16}}{s_{45}}+\frac{n_{17}}{s_{15}}~.~~~
\end{eqnarray}
Together there are $15$ cubic graphs and $3$ quartic graphs in the
two gluon scalar Yang-Mills amplitudes at five points, which we list
below, 
\begin{equation}
\begin{array}{ccccc}
n_{1}= \begin{minipage}{1.62cm} \includegraphics[width=1.62cm]{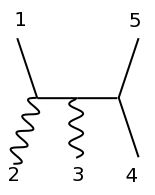}  \end{minipage}~,
& n_{2}= \begin{minipage}{1.62cm} \includegraphics[width=1.62cm]{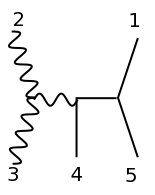}  \end{minipage}~,
& n_{3}=\begin{minipage}{1.62cm}  \includegraphics[width=1.62cm]{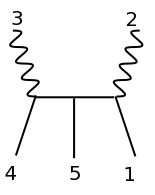} \end{minipage}~,
& n_{4}=\begin{minipage}{1.62cm}    \includegraphics[width=1.62cm]{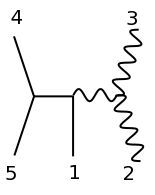}\end{minipage}~,
&n_{5}=\begin{minipage}{1.62cm}  \includegraphics[width=1.62cm]{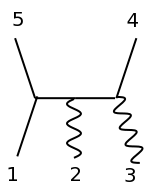} \end{minipage}~, \\
 n_{6}= \begin{minipage}{1.62cm}  \includegraphics[width=1.62cm]{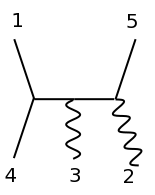} \end{minipage}~,
& n_{7}= \begin{minipage}{1.62cm} \includegraphics[width=1.62cm]{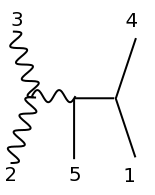}  \end{minipage}~,
& n_{8}= \begin{minipage}{1.62cm}   \includegraphics[width=1.62cm]{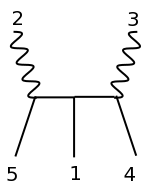} \end{minipage}~,
&n_{9}= \begin{minipage}{1.62cm}   \includegraphics[width=1.62cm]{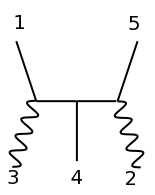} \end{minipage}~,
& n_{10}= \begin{minipage}{1.62cm}   \includegraphics[width=1.62cm]{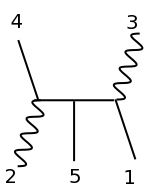} \end{minipage}~,\\
n_{11}= \begin{minipage}{1.62cm}  \includegraphics[width=1.62cm]{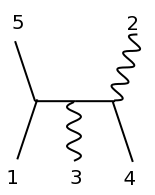}  \end{minipage}~,
& n_{12}= \begin{minipage}{1.62cm} \includegraphics[width=1.62cm]{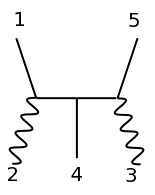}  \end{minipage}~,
&n_{13}= \begin{minipage}{1.62cm}   \includegraphics[width=1.62cm]{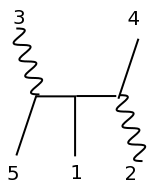} \end{minipage}~,
& n_{14}= \begin{minipage}{1.62cm}   \includegraphics[width=1.62cm]{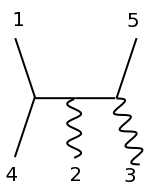} \end{minipage}~,
& n_{15}= \begin{minipage}{1.62cm}   \includegraphics[width=1.62cm]{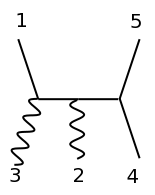} \end{minipage}~,\\
n_{16}= \begin{minipage}{1.62cm}   \includegraphics[width=1.62cm]{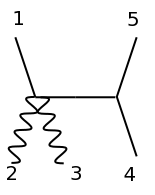} \end{minipage}~,
& n_{17}= \begin{minipage}{1.62cm}  \includegraphics[width=1.62cm]{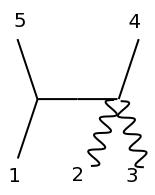} \end{minipage}~,
& n_{18}= \begin{minipage}{1.62cm}  \includegraphics[width=1.62cm]{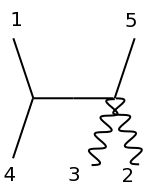} \end{minipage}~.~~~
\end{array}
\end{equation}
As in the case of a cubic theory, not all graphs are independent.
Together there are seven Jacobi identities derived from cyclic
permutations involving one gluon, one of the identities obtainable
as a linear combination of the others, 
\begin{eqnarray}
&&n_{3}-n_{5}+n_{8} = 0~~~,~~~n_{3}-n_{1}+n_{12} = 0~~~~,~~~
n_{4}-n_{2}+n_{7} = 0~,~~~\nonumber\\
&&n_{8}-n_{6}+n_{9}= 0~~~,~~~
n_{10}-n_{9}+n_{15}  = 0~~~,~~~
n_{10}-n_{11}+n_{13} = 0~~~~,~~~\Big(n_{13}-n_{12}+n_{14} =0 \Big)~.~~~\nonumber
\end{eqnarray}
In addition there are three more modified Jacobi identities where
two gluons participate the permutations, and therefore contains
quartic graphs, 
\begin{eqnarray}
&&n_{4}-n_{1}+n_{15}-(s_{21}-s_{31})n_{16} = 0~,~~~\\
&&n_{5}-n_{2}+n_{11}+(s_{34}-s_{24})n_{17} = 0~~~,~~~
n_{7}-n_{6}+n_{14}+(s_{35}-s_{25})n_{18} = 0~.~~~
\end{eqnarray}
The above constraints allows us to trade $n_{7}$, $n_{8}$, $\ldots$,
$n_{15}$ in terms of the first six independent cubic graphs plus the
three quartic graphs, 
\begin{eqnarray}
&&n_{7}  =  n_{2}-n_{4}~~~,~~~n_{8}  =  -n_{3}+n_{5}~~~,~~~n_{9}  =  n_{3}-n_{5}+n_{6}~,~~~\\
&&n_{10}  =  -n_{1}+n_{3}+n_{4}-n_{5}+n_{6}-(s_{21}-s_{31})n_{16}~,~~~\\
&&n_{11}  =  n_{2}-n_{5}-(s_{34}-s_{24})n_{17}~~~,~~~n_{12}  =  n_{1}-n_{3}~,~~~\\
&&n_{13}  =  n_{1}+n_{2}-n_{3}-n_{4}-n_{6}-(s_{34}-s_{24})n_{17}+(s_{21}-s_{31})n_{16}~,~~~\\
&&n_{14} =  -n_{2}+n_{4}+n_{6}-(s_{35}-s_{25})n_{18}~~~,~~~n_{15}  =  n_{1}-n_{4}+(s_{21}-s_{31})n_{16}~.~~~
\end{eqnarray}
Furthermore we note that all three quartic graphs actually
contribute the same value, 
\begin{equation}
n_{18}=n_{17}=n_{16}=(\epsilon_{2}\cdot\epsilon_{3})\times (\frac{-1}{2}) f^{a_{1}a_{4}a_{5}}~.~~~
\end{equation}
Bearing all these in mind we calculate the five-point numerator
$n(12_{g}3_{g}45)$ from KLT relation by summing over KK basis and
get
\begin{eqnarray}
n(12_{g}3_{g}45) & = & \sum_{\sigma\in  S_{3}}\mathcal{S}[234|\sigma]J^{\YMs}(1,\sigma,5)\label{eq:5pt-n12345}\\
 & = & n_{1}+s_{12}\,n_{16} =
 \begin{minipage}{2.72cm} \includegraphics[width=2.72cm]{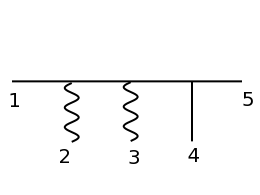} \end{minipage}
 +s_{12} \,
  \begin{minipage}{2.72cm}  \includegraphics[width=2.72cm]{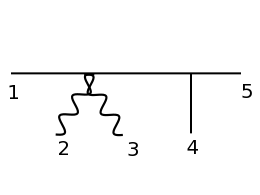} \end{minipage}~.~~~
  \nonumber
\end{eqnarray}
All other numerators follow the same derivation, and we obtain 
\begin{eqnarray}
n(12_{g}43_{g}5) & = &
\begin{minipage}{2.72cm} \includegraphics[width=2.72cm]{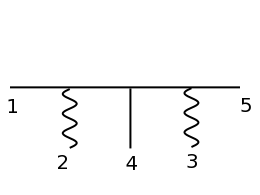}   \end{minipage}
 +s_{21} \,
\begin{minipage}{2.72cm} \includegraphics[width=2.72cm]{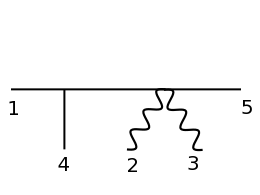}   \end{minipage}~,~~~
\\
n(13_{g}2_{g}45) & = &
\begin{minipage}{2.72cm}  \includegraphics[width=2.72cm]{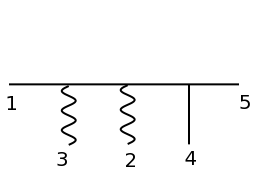}  \end{minipage}
+s_{31} \,
\begin{minipage}{2.72cm}   \includegraphics[width=2.72cm]{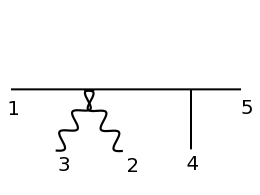}  \end{minipage}~,~~~
\\
n(13_{g}42_{g}5) & = &
\begin{minipage}{2.72cm}  \includegraphics[width=2.72cm]{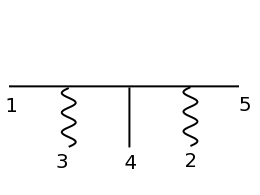}  \end{minipage}
+s_{31} \,
\begin{minipage}{2.72cm}    \includegraphics[width=2.72cm]{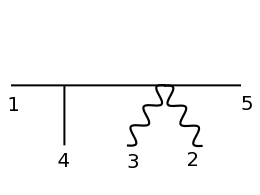} \end{minipage}~,~~~
\\
n(142_{g}3_{g}5) & = &
\begin{minipage}{2.72cm}   \includegraphics[width=2.72cm]{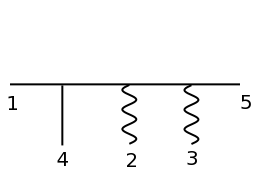} \end{minipage}
+(s_{21}+s_{24})\,
\begin{minipage}{2.72cm}   \includegraphics[width=2.72cm]{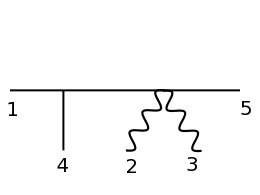} \end{minipage}~,~~~
\\
n(143_{g}2_{g}5) & = &
\begin{minipage}{2.72cm}  \includegraphics[width=2.72cm]{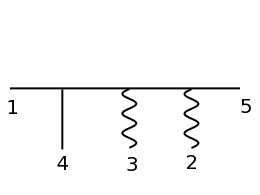}  \end{minipage}
+(s_{31}+s_{34})\,
\begin{minipage}{2.72cm}   \includegraphics[width=2.72cm]{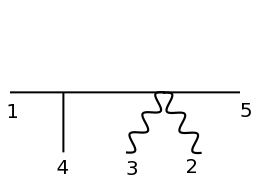}  \end{minipage}~.~~~
\label{eq:5pt-n14325}
\end{eqnarray}
Plugging the above results into DDM expression yields the two
graviton EYM amplitude at five points, 
\begin{eqnarray}
A^{\EYM}_{3,2}(1,4,5;h_2,h_3) & = & \sqrt{2}^{3}\,i\, \Big((\epsilon_{2}\cdot X_{2})(\epsilon_{3}\cdot X_{3})-\frac{1}{4}(\epsilon_{2}\cdot\epsilon_{3})s_{21}\Big){A}^{\YM}_5(1,2,3,4,5)\nonumber\\
 &  & + \sqrt{2}^{3}\,i\, \Big((\epsilon_{2}\cdot X_{2})(\epsilon_{3}\cdot X_{3})-\frac{1}{4}(\epsilon_{2}\cdot\epsilon_{3})s_{21}\Big){A}^{\YM}_5(1,2,4,3,5)+\cdots
\end{eqnarray}
%

\subsection{The five and higher point amplitude involving two gravitons}
\label{sec:5pt-and-higher} 

Having witness that KLT relation successfully explains the new EYM
amplitude expression for two graviton scattering at four and five
points, perhaps it is not much of a surprise that the explanation
generalizes to higher points. Indeed, one can actually read off the
$n$-point two gluon numerator, and the two graviton EYM amplitude is
determined by the corresponding DDM expression. We shall use the
algorithm introduced originally for the pure scalar scenario in
\cite{Du:2011js} to systematically calculate the numerator (i.e., to
systematically sum over KK basis). As we shall see, in the case when
only two gluons $(p,q)$ are involved, the numerators remain fairly
simple, 
\begin{eqnarray}
n(12\cdots i~p_{g}\cdots j~q_{g}\cdots n) & = &
\begin{minipage}{3.67cm}  \includegraphics[width=3.67cm]{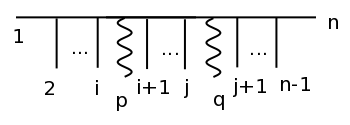}  \end{minipage}
 +2(p\cdot Y_p)
 \begin{minipage}{3.67cm}   \includegraphics[width=3.67cm]{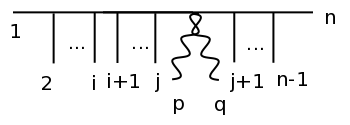}  \end{minipage}
 \label{eq:2-gluon-numerator}\\
 & = &  \sqrt{2}^{n-2}(-1)^{n+1}\,i\,\left( (\epsilon_{p}\cdot x_{p})(\epsilon_{q}\cdot x_{q})
 - \frac{1}{2} (\epsilon_{p}\cdot\epsilon_{q})\,(p\cdot Y_p) \right)~.~~~ \nonumber
\end{eqnarray}

\noindent {\bf A brief review of the algorithm for numerators in the
scalar scenario:} For the purpose of being self-contained, we
briefly review the algorithm used by the authors in \cite{Du:2011js}
and \cite{Du:2016tbc} to calculate numerators. The idea is to divide
the full $S_{n-2}$ permutation sum appears in the numerator-current
relation $n(1~\alpha~ n)=\sum_{\beta\in
S_{n-2}}\mathcal{S}[\alpha^{T}|\beta]\,J^{\YMs}(1,\beta, n)$ into
BCJ sums, and proceed repeatedly if the Fundamental BCJ relation
between currents admits further simplifications. For example, it was
shown in \cite{Du:2011js} that the Fundamental BCJ relation between
$\phi^{3}$ currents yields another current, with the leg running
through all insertions in the BCJ sum fixed at the off-shell
continued line, 
\begin{equation}
s_{21}\,
\begin{minipage}{2.14cm} \includegraphics[width=2.14cm]{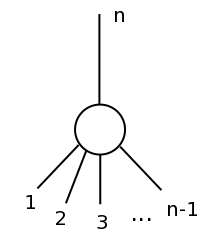}   \end{minipage}
+(s_{21}+s_{31}) \,
\begin{minipage}{2.14cm}  \includegraphics[width=2.14cm]{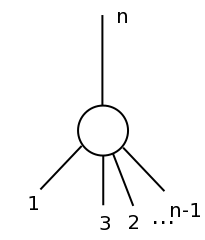}  \end{minipage}
+\cdots= k_{n}^{2}\,
\begin{minipage}{2.14cm}  \includegraphics[width=2.14cm]{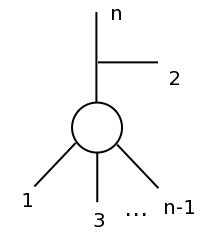}  \end{minipage}~,~~~\label{eq:off-shell-bcj-phi3}
\end{equation}
so that if we divide the full permutation sum $S_{3}$ in the
five-point numerator calculation into BCJ sums, after substituting
these summations using Fundamental BCJ relation
(\ref{eq:off-shell-bcj-phi3}), the collected result is yet another
BCJ sum, but only performed over permutations of the legs of
fewer-point sub-currents, 
\begin{eqnarray}
\sum_{\beta\in S_{3}}\mathcal{S}[432|\beta_{2}\beta_{3}\beta_{4}]\,J^{\phi^3}(1,\beta_{2},\beta_{3},\beta_{4},5) & = & s_{21}s_{31}\Bigl(s_{41}J^{\phi^3}(14325)+(s_{41}+s_{43})J^{\phi^3}(13425)+\ldots\Bigr)\nonumber\\
 &  & +s_{21}(s_{31}+s_{32})\Bigl(s_{41}J^{\phi^3}(14235)+(s_{41}+s_{42})J^{\phi^3}(12435)+\ldots\Bigr)\nonumber \\
 & = & k_{5}^{2}\,s_{21}\Bigl(s_{31}
 \begin{minipage}{2.14cm} \includegraphics[width=2.14cm]{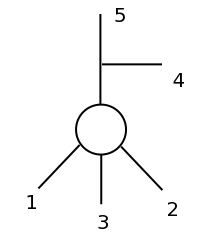}   \end{minipage}
 +(s_{31}+s_{32})
 \begin{minipage}{2.14cm}   \includegraphics[width=2.14cm]{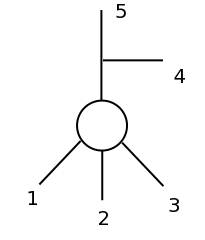}  \end{minipage}
 \Bigr)~,~~~\label{eq:5-pt-demon}
\end{eqnarray}
where we used (\ref{eq:off-shell-bcj-phi3}) to replace the first and
the second line of the equation above with the two graphs in
(\ref{eq:5-pt-demon}). The result is another BCJ sum over currents.
Repeat the substitution using Fundamental BCJ relation, and we
obtain the numerator
\begin{eqnarray}
 &  & k_{1234}^{2}\,k_{123}^{2}\,k_{12}^{2} \,
 \begin{minipage}{2.14cm}  \includegraphics[width=2.14cm]{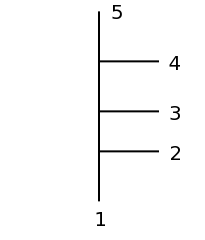}~~~~\end{minipage}~.~~~
\end{eqnarray}

\subsubsection*{The five-point scalar Yang-Mills numerators involving two gluons}

The calculation explained above only complicates slightly when few
gluons are present. As far as single trace contributions are
concerned, all amplitudes are consisted of Jacobi satisfying cubic
graphs when only one gluon participates the scattering, and the same
algorithm applies. It is straightforward to see that the numerator
is given by $n(12\cdots p_{g}\cdots n)=f^{1,2,*}f^{*,3,*}\cdots
f^{*,n-1,n}\,\epsilon_{p}\cdot x_{p}$, which when plugged into the
summing expression readily reproduces the new EYM formula. In other
words, (\ref{Fig.1.5}) can also be understood from this point of
view.

Things will become a little bit more complicated when two and more
gluons are involved, since quartic vertices start to come into play,
although they still remain quite manageable, in the sense that  the
modified  Fundamental BCJ relations brought by the quartic term also
permit repeated use of the relation when we carry out the summation.
Explicitly, at five points the two-gluon Fundamental BCJ relations
are modified as 
 \begin{eqnarray}
s_{21}J^{\YMs}(12_{g}3_{g}45)+s_{2,13}J^{\YMs}(13_{g}2_{g}45)+s_{2,134}J^{\YMs}(13_{g}42_{g}5)
& = & k_{5}^{2}
\begin{minipage}{2.14cm}  \includegraphics[width=2.14cm]{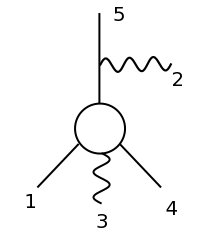}  \end{minipage}
\label{eqn:ofshell-bcj-1}~,~~~ \\
s_{21}J^{\YMs}(12_{g}34_{g}5)+s_{2,13}J^{\YMs}(132_{g}4_{g}5)+s_{2,134}J^{\YMs}(134_{g}2_{g}5) & = & k_{5}^{2}\Bigl(
\begin{minipage}{2.14cm}  \includegraphics[width=2.14cm]{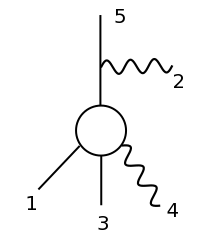}  \end{minipage}
+
\begin{minipage}{2.14cm}  \includegraphics[width=2.14cm]{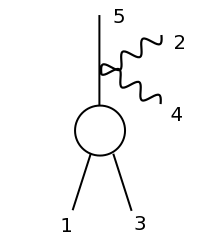}  \end{minipage}
\Bigr)~,~~~~~~~\label{eqn:ofshell-bcj-2}\\
 & \vdots & \nonumber
\end{eqnarray}
The rules to modification is as follows. Generically one only needs
to replace the appropriate scalar by gluon lines in the original
Fundamental BCJ relation between currents
(\ref{eq:off-shell-bcj-phi3}), and the right hand side of the
equation is a current with the running leg fixed at the off-shell
line. The only exception is when the running leg is gluonic, also
that either leg $1$ or leg $n-1$ (legs adjacent to the off-shell
line) is a gluon line. In these cases an additional current needs to
be added, where a quartic vertex resides on the off-shell line
connects the two gluons.

We leave the details of a proof to these relations at five-point to
Appendix \ref{Appendix:proof--offshell-bcj} because of its
complicated nature. The principles are however not much different
from the pure scalar scenario and is conceptually straightforward.
Basically we cancel graphs related by Jacobi identities among
Berends-Giele decomposed five-point current in the BCJ sum. The
result after cancelation is then collected and identified to be the
Berends-Giele decomposition of the right hand side of the equation.
The proof for generic $n$ points follows rather trivially from the
structure of the proof, since adding more scalar lines into
sub-currents at peripherals does not change Jacobi identities.

Assuming the Fundamental BCJ relations above, it is not difficult to
see that the numerator is genuinely given by the formula
(\ref{eq:2-gluon-numerator}) we claimed earlier. Consider for
example the derivation that leads to numerator $n(123_{g}4_{g}5)$,
\begin{eqnarray}
n(123_{g}4_{g}5) & = & s_{21}s_{31}\Bigl(s_{41}J^{\YMs}(14_{g}3_{g}25)+(s_{41}+s_{43})J^{\YMs}(13_{g}4_{g}25)+\ldots\Bigr)\nonumber\\
 &  & +s_{21}(s_{31}+s_{32})\Bigl(s_{41}J^{\YMs}(14_{g}23_{g}5)+(s_{41}+s_{42})J^{\YMs}(124_{g}3_{g}5)+\ldots\Bigr)\nonumber \\
 & = & k_{5}^{2}\,s_{21}\Bigl(s_{31}
 \begin{minipage}{2.14cm}  \includegraphics[width=2.14cm]{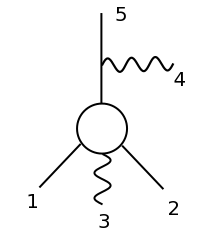}  \end{minipage}
 +(s_{31}+s_{32})\Bigl(
 \begin{minipage}{2.14cm}   \includegraphics[width=2.14cm]{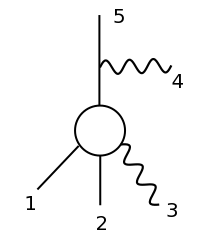}  \end{minipage}
 +
 \begin{minipage}{2.14cm}  \includegraphics[width=2.14cm]{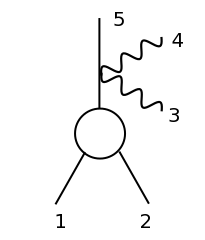}  \end{minipage}
 \Bigr)\Bigr)~.~~~\label{eq:2nd-line-34g-eqn}
\end{eqnarray}
As was explained earlier we obtain the numerator by first dividing
the full $S_{n-2}$ permutation sum appears in the KLT inspired
prescription (\ref{eq:klt-insp-numerator}) into BCJ sums, and then
use the Fundamental BCJ relation between currents to fix the $n-2$
legs one by one in descending order. For the most part, this
procedure is not different from the derivation of a pure scalar
numerator, and the result does contain a cubic half ladder graph.
The only modification occurs whenever the leg we attempt to fix is
gluonic, in which case an additional graph is included, where a
quartic vertex connecting both gluon lines emerges. The derivation
afterwards again follows that of a pure scalar numerator. In the
$n(123_{g}4_{g}5)$ example this leads to 
\begin{equation}
n(123_{g}4_{g}5)=k_{1234}^{2}\,k_{123}^{2}\,k_{12}^{2}
\begin{minipage}{2.72cm}  \includegraphics[width=2.72cm]{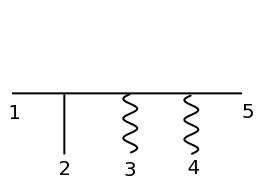}  \end{minipage}
+k_{1234}^{2}\,k_{12}^{2} (s_{31}+s_{32})
\begin{minipage}{2.72cm}   \includegraphics[width=2.72cm]{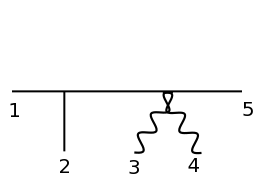}  \end{minipage}
~.~~~
\end{equation}
Note that the Mandelstam variables associated with the quartic graph
was furnished by momentum kernel. Careful inspection of the
derivation that leads to (\ref{eq:2nd-line-34g-eqn}) shows that they
should contain the inner products between gluon line carrying the
smaller label and all the scalar lines which precede it. As another
illustration we consider $n(12_{g}34_{g}5)$,
\begin{eqnarray}
n(12_{g}34_{g}5)
 & = & k_{5}^{2}\,s_{21}\Bigl(s_{31}\Bigl(
 \begin{minipage}{2.14cm}  \includegraphics[width=2.14cm]{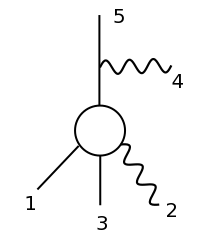}  \end{minipage}
 +
 \begin{minipage}{2.14cm}   \includegraphics[width=2.14cm]{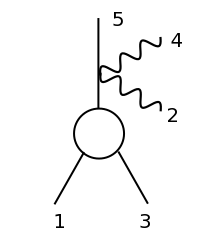}  \end{minipage}
 \Bigr)+(s_{31}+s_{32})
 \begin{minipage}{2.14cm}  \includegraphics[width=2.14cm]{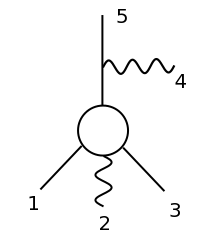}  \end{minipage}
 \Bigr)~.~~~
\end{eqnarray}
A repeated use of the Fundamental BCJ relation yields 
\begin{equation}
n(12_{g}34_{g}5)=k_{1234}^{2}\,k_{123}^{2}\,k_{12}^{2}
\begin{minipage}{2.72cm}  \includegraphics[width=2.72cm]{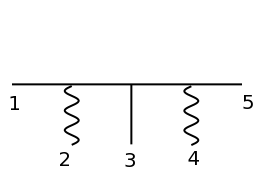}  \end{minipage}
+k_{1234}^{2}\,k_{13}^{2}\, s_{21}
\begin{minipage}{2.72cm}   \includegraphics[width=2.72cm]{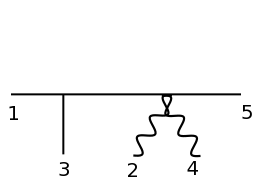}  \end{minipage}
~.~~~
\end{equation}
As a verification, note that applying the same rules to derive
numerators with all possible combinations of gluon positions yields
the same results as those listed from equation (\ref{eq:5pt-n12345})
to (\ref{eq:5pt-n14325})
 previously obtained exclusively for five points.

\section{Conclusion}
\label{SecConclusion}

In this paper, we studied the newly discovered EYM amplitude relation by gauge invariance principle, the BCFW recursion relation as well as the KLT relation respectively. It turns out that the problem of EYM amplitude expansion is also closely related to the problem of computing BCJ numerators and the boundary contribution of BCFW terms.

The major context of this paper is devoted to the principle of gauge
invariance applied to the determination of EYM amplitude relations. We propose a constructive algorithm by expanding any EYM amplitude $A^{\EYM}_{n,m}$ as a linear sum of $A^{\EYM}_{n+i,m-i}$ for $i=1,2,\ldots,m$ with given expansion coefficients, and the contributing terms of $A^{\EYM}_{n+i,m-i}$ are determined by $A^{\EYM}_{n+i-1,m-i+1}$. This means that any contributing terms can be recursively determined by the very first one $A^{\EYM}_{n+1,m-1}$, while keeping the gauge invariance in each step. This leads to a compact formula \eref{generalEYM} for general EYM amplitude
relations with arbitrary number of gravitons. Realizing that the
expansion of Einstein-Yang-Mills amplitude into Yang-Mills
amplitudes shares the same kinematic coefficient as the expansion of
Yang-Mills-scalar amplitude into cubic-scalar amplitudes, we copy
the EYM amplitude relation to YMs amplitude relation, and generalize
the later one to the expansion of pure Yang-Mills amplitude into
cubic-scalar amplitudes by the help of Pfaffian expansion. With the
Yang-Mills amplitude expanded recursively into the cubic graphs, we
further outline the strategy of rewriting the scalar amplitudes into
KK basis, manifesting the color-kinematics duality and computing the
BCJ numerators of Yang-Mills amplitude.

We also study the EYM amplitude relations in the S-matrix framework,
and present the proof of EYM amplitude relations with two gravitons
by BCFW recursion relations. In this case, any choice of deformed
momenta is not possible to avoid the boundary contributions, so we
need to compare the contributions of both sides in the relations
from finite poles and also the boundary. The matching of both
contributions also constraints the possible form of the non-trivial
relations. Besides, we examine the problem again from the
perspective of KLT relations. The expansion coefficients of EYM
amplitude relations are identical to the BCJ numerators of DDM
basis, and by computing the BCJ relations for currents we confirm
the validation of EYM amplitude relations.

Following our results, there are many interesting directions to
explore further. In our paper, one of the most important results is
the recursive construction \eref{generalEYM} of EYM amplitude
relation. We have claimed this expression by a few explicit examples
plus the guidance of gauge invariance principle. For the confirmation of the claim, a rigorous derivation by other methods is favorable. In an upcoming paper, we would explore the recursive construction directly from operations on the CHY-integrand level. Furthermore, we believe that, such recursive pattern can also find its hints in the BCFW recursion relation or KLT relation investigation of EYM amplitude expansion, which worth to work on with.

Another possible work would be that, in our recursive construction,
the gauge invariance is manifest for all gravitons at each step
except the first one that started the recursive algorithm. As shown
in
\cite{Boels:2016xhc,Arkani-Hamed:2016rak,Rodina:2016mbk,Rodina:2016jyz},
for Yang-Mills theory, the requirement of gange invariance for
$(n-1)$ points is sufficient to guarantee the correctness of the
full amplitude. This observation seems to be also true in the EYM
theory, thus finding an explicit proof along the same line as in
\cite{Arkani-Hamed:2016rak,Rodina:2016mbk,Rodina:2016jyz} would be a
thing worth to do.

A most interesting and important future direction would be the
systematic study of the CHY-integrand expansion. In \S\ref{secCHY},
we have laid down the general framework for the expansion, while in
the whole paper we are focus only on the expansion of (reduced)
Pfaffian. However, many CHY-integrands, such as $({\rm Pf}'(A))^2$
can be obtained from Pfaffian with proper reduction. Thus our
results could be easily generalized to many other theories.
Especially by similar calculations, we can check if the soft theorem
can be used to uniquely determine the amplitude for some theories,
such as NLSM as advertised in
\cite{Arkani-Hamed:2016rak,Rodina:2016mbk,Rodina:2016jyz}.

Finally, as a byproduct of the EYM expansion, we have outlined the strategy of computing BCJ numerators\footnote{The polynomial
expression of BCJ numerator of (reduced) Pfaffian has been applied
to the proof of vanishing double poles in a recent work
\cite{Huang:2017ydz}. } from the expansion relation for general EYM amplitudes. The four-point example shows the procedure of computing the BCJ numerators as polynomial of $(\epsilon\cdot \epsilon)$, $(\epsilon\cdot k)$ and $(k\cdot k)$, constructed neatly from the expansion coefficients of EYM amplitudes into Yang-Mills amplitudes. This
construction, when generalized to loop-level, would fascinate many
important calculations involving gravitons.

\section*{Acknowledgments}

We would like to thank Fei Teng for valuable discussions.
BF is supported by Qiu-Shi Funding and the National Natural Science
Foundation of China (NSFC) with Grant No.11575156, No.11135006, and No.11125523. YD would like to acknowledge NSFC under
Grant Nos.11105118, 111547310, as well as the support from 351 program of Wuhan
University. RH would like to acknowledge the supporting from NSFC
No.11575156 and the Chinese Postdoctoral Administrative Committee.

\appendix

\section{Scalar Yang-Mills Feynman rules}
\label{Appendix:feyn-rules}

For reference purposes we list below the color-ordered Feynman rules
for constructing scalar Yang-Mills amplitudes \cite{Bern:1999bx}.
The scalars and gluons are understood to be represented by straight
lines and wavy lines respectively,
\bea
\begin{minipage}{2.14cm}   \includegraphics[width=2.14cm]{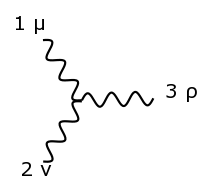} \end{minipage}
  =  \frac{i}{\sqrt{2}}\eta^{\mu\nu}(k_{1}-k_{2})^{\rho}+\text{cyclic}
& ~~,~~ &
 \begin{minipage}{2.14cm}  \includegraphics[width=2.14cm]{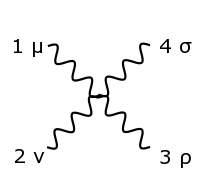}   \end{minipage}
  =
  i\,\eta^{\mu\rho}\eta^{\nu\sigma}-\frac{i}{2}(\eta^{\mu\nu}\eta^{\rho\sigma}+\eta^{\mu\sigma}\eta^{\nu\rho})~,~~~\eea
\bea
 \begin{minipage}{2.14cm}  \includegraphics[width=2.14cm]{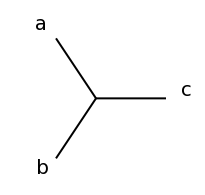}  \end{minipage}
  =  \sqrt{2}\,f^{abc}
&~~,~~&
 \begin{minipage}{2.14cm}  \includegraphics[width=2.14cm]{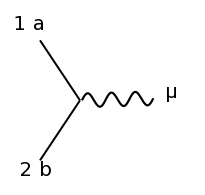}  \end{minipage}
  =
  -\,\frac{i}{\sqrt{2}}\,\delta^{ab}(k_{1}-k_{2})^{\mu}~,~~~\eea
\bea
 \begin{minipage}{2.14cm}  \includegraphics[width=2.14cm]{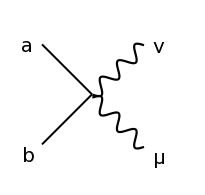}  \end{minipage}
  =  \frac{i}{2}\delta^{ab}\eta^{\mu\nu}
& ~~,~~ &
 \begin{minipage}{2.14cm}  \includegraphics[width=2.14cm]{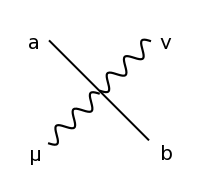}  \end{minipage}
  =  -i\,\delta^{ab}\eta^{\mu\nu}~.~~~
\eea
%


\section{Graphical proof of the two-gluon Fundamental BCJ relation between currents}
\label{Appendix:proof--offshell-bcj}

As a demonstration of the general idea, in this appendix we prove
two of the Fundamental BCJ relations at five points involving two
gluons, equations (\ref{eqn:ofshell-bcj-1}) and
(\ref{eqn:ofshell-bcj-2}), following the method used in
\cite{Du:2011js} (which was also briefly outlined earlier in \S
\ref{sec:5pt-and-higher}). We shall neglect repeating a similar
proof for generic $n$ points, as it can be readily derived by
induction and by attaching more external legs on the sub-currents
represented by blank circles in the graphs below.

\subsection*{Relations with no gluon adjacent to the off-shell leg}

Consider first the configuration where the leg running over all
possible insertions in the BCJ sum is a gluon, and the other gluon
is non-adjacent to the off-shell leg. We would like to prove that 
\begin{equation}
s_{21}J^{\YMs}(12_{g}3_{g}45)+(s_{2,13})J^{\YMs}(13_{g}2_{g}45)+(s_{2,134})J^{\YMs}(13_{g}42_{g}5)=
\begin{minipage}{2.2cm}  \includegraphics[width=2.2cm]{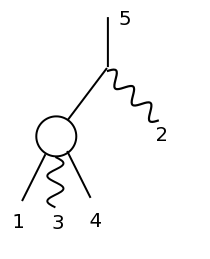}  \end{minipage}
~.~~~\label{eq:bcj-sum-nonadj}
\end{equation}
For this purpose we Berends-Giele decompose all three currents
appear in the BCJ sum, yielding altogether nine graphs,
\begin{flalign} s_{21}J^{\YMs}(12_{g}3_{g}45) & = s_{21}\underset{(a1)}{
\begin{minipage}{2.2cm} \includegraphics[width=2.2cm]{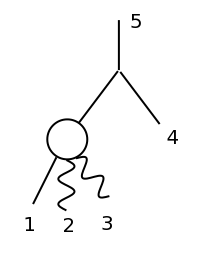}
\end{minipage} } +s_{21}\underset{(a2)}{ \begin{minipage}{2.2cm}
\includegraphics[width=2.2cm]{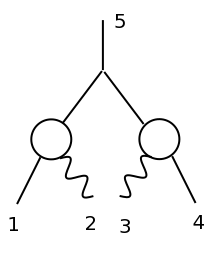}  \end{minipage} }
+s_{21}\underset{(a3)}{ \begin{minipage}{2.2cm}
\includegraphics[width=2.2cm]{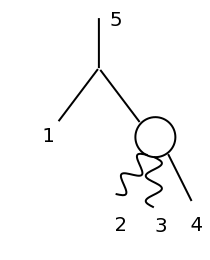}  \end{minipage} }~,~~~&&\end{flalign}
\begin{flalign} s_{2,13}J^{\YMs}(13_{g}2_{g}45) & =
s_{2,13}\underset{(b1)}{\begin{minipage}{2.2cm}
\includegraphics[width=2.2cm]{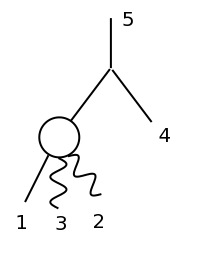} \end{minipage} }
+s_{2,13}\underset{(b2)}{\begin{minipage}{2.2cm}
\includegraphics[width=2.2cm]{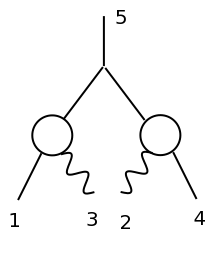}  \end{minipage} }+s_{2,13}\underset{(b3)}{\begin{minipage}{2.2cm}
\includegraphics[width=2.2cm]{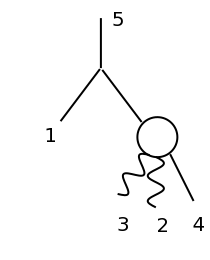} \end{minipage} }~,~~~ &&
\end{flalign}
\begin{flalign}
s_{2,134}J^{\YMs}(13_{g}42_{g}5) & = s_{2,134}\underset{(c1)}{
\begin{minipage}{2.2cm}
\includegraphics[width=2.2cm]{apdx1-c1}   \end{minipage} }
+s_{2,134}\underset{(c2)}{\begin{minipage}{2.2cm}
\includegraphics[width=2.2cm]{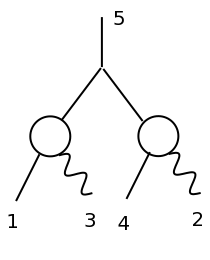}   \end{minipage}
}+s_{2,134}\underset{(c3)}{ \begin{minipage}{2.2cm}
\includegraphics[width=2.2cm]{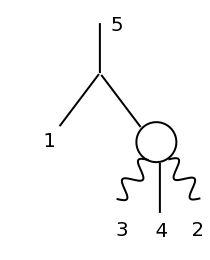}  \end{minipage}
}~,~~~&&\end{flalign}
and notice that, aside from $(c1)$, rest of the graphs can be
regrouped as BCJ sums of sub-currents. Indeed, graphs $(a1)$and
$(b1)$ together make up a BCJ sum of the sub-currents involving legs
$1$, $2$ and $3$, 
\begin{equation}
s_{21}\underset{(a1)}{\begin{minipage}{2.2cm} \includegraphics[width=2.2cm]{apdx1-a1}   \end{minipage} }
+s_{2,13}\underset{(b1)}{ \begin{minipage}{2.2cm}   \includegraphics[width=2.2cm]{apdx1-b1}  \end{minipage} }
=k_{123}^{2} \begin{minipage}{2.2cm} \includegraphics[width=2.2cm]{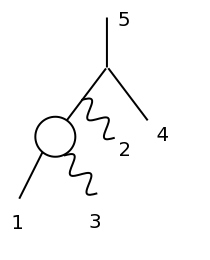}   \end{minipage}
+k_{123}^{2}\begin{minipage}{2.2cm}   \includegraphics[width=2.2cm]{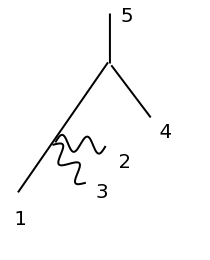}  \end{minipage}~,~~~
\label{eq:proof1-1}
\end{equation}
and graph $(a2)$ is by itself a (trivial) BCJ sum of three point
current. The combination of $(b2)$ and $(c2)$ is also a BCJ sum of
the three point current, after eliminating part of the sum that
carries an $s_{21}$ using $U(1)$ decoupling identity,
\begin{equation}
s_{21}\underset{(b2)}{ \begin{minipage}{2.2cm}  \includegraphics[width=2.2cm]{apdx1-b2}  \end{minipage} }
+s_{2,13}\underset{(c2)}{ \begin{minipage}{2.2cm}  \includegraphics[width=2.2cm]{apdx1-c2}   \end{minipage} }
=s_{23} \begin{minipage}{2.2cm} \includegraphics[width=2.2cm]{apdx1-c2}    \end{minipage}~,~~~\label{eq:proof1-2}
\end{equation}
and similarly $(a3)$, $(b3)$ and $(c3)$ combine to give, up to terms
vanishing under $U(1)$ decoupling identity,
\begin{eqnarray}
& & s_{21}\underset{(a3)}{ \begin{minipage}{2.2cm}  \includegraphics[width=2.2cm]{apdx1-a3}  \end{minipage} }
+s_{2,13}\underset{(b3)}{ \begin{minipage}{2.2cm} \includegraphics[width=2.2cm]{apdx1-b3}   \end{minipage} }
+s_{2,134}\underset{(c3)}{ \begin{minipage}{2.2cm}  \includegraphics[width=2.2cm]{apdx1-c3}  \end{minipage} } =k_{234}^{2} \begin{minipage}{2.2cm}  \includegraphics[width=2.2cm]{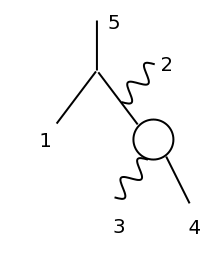}   \end{minipage}
-k_{234}^{2} \begin{minipage}{2.2cm}  \includegraphics[width=2.2cm]{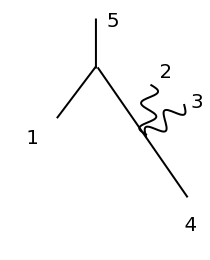}  \end{minipage}~.~~~~~
\label{eq:proof1-3}
\end{eqnarray}
In the equations above we have assumed the BCJ relations between
currents at four points. As for the remaining graph $(c1)$ that does
not regrouped with the others into a BCJ sum of sub-currents, we
rewrite the coefficient it carries using the kinematic identity
$s_{2,134}=k_{5}^{2}-s_{134}$ and then further Berends-Giele
decompose the part that carries a factor $s_{134}$, giving 
\begin{eqnarray}
s_{2,134}\underset{(c1)}{\begin{minipage}{2.2cm} \includegraphics[width=2.2cm]{apdx1-c1}   \end{minipage} }
 & = & k_{5}^{2} \begin{minipage}{2.2cm}  \includegraphics[width=2.2cm]{apdx1-c1}  \end{minipage}
 -s_{134} \begin{minipage}{2.2cm}   \includegraphics[width=2.2cm]{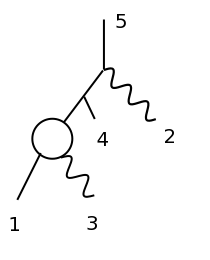}  \end{minipage}
 -s_{134} \begin{minipage}{2.2cm}  \includegraphics[width=2.2cm]{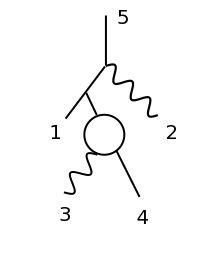}  \end{minipage}~.~~~
 \label{eq:proof1-4}
\end{eqnarray}
Because of the regrouping and the application of lower point BCJ
relation on sub-currents, the full BCJ sum (\ref{eq:bcj-sum-nonadj})
is now translated into the right hand side of equations
(\ref{eq:proof1-1}), graph $(a2)$, (\ref{eq:proof1-2}),
(\ref{eq:proof1-3}) and (\ref{eq:proof1-4}) combined. To see that
this combination is indeed identical to the right hand side of
equation (\ref{eq:bcj-sum-nonadj}) we must show that all other
graphs cancel, and this is true because of the Jacobi identities 
\begin{equation}
k_{123}^{2} \begin{minipage}{2.2cm}  \includegraphics[width=2.2cm]{apdx1-e3}  \end{minipage}
+s_{24} \begin{minipage}{2.2cm}  \includegraphics[width=2.2cm]{apdx1-c2}  \end{minipage}
-s_{134} \begin{minipage}{2.2cm} \includegraphics[width=2.2cm]{apdx1-d1}   \end{minipage} =0~,~~~
\end{equation}
and 
\begin{equation}
s_{12} \begin{minipage}{2.2cm}   \includegraphics[width=2.2cm]{apdx1-a2} \end{minipage}
+k_{234}^{2} \begin{minipage}{2.2cm}  \includegraphics[width=2.2cm]{apdx1-e4}  \end{minipage}
-s_{134} \begin{minipage}{2.2cm}  \includegraphics[width=2.2cm]{apdx1-d2}  \end{minipage} =0~,~~~
\end{equation}
and the fact that the following two graphs contribute the same, up
to a relative minus sign, 
\begin{eqnarray}
k_{123}^{2} \begin{minipage}{2.2cm}  \includegraphics[width=2.2cm]{apdx1-e1}  \end{minipage}
-k_{234}^{2} \begin{minipage}{2.2cm}  \includegraphics[width=2.2cm]{apdx1-e2}  \end{minipage}
& = & (\frac{i}{2}) f^{a_{1}a_{4}a_{5}}\epsilon_{2}\cdot\epsilon_{3} - \, (\frac{i}{2})f^{a_{1}a_{4}a_{5}}\epsilon_{2}\cdot\epsilon_{3}= 0~,~~~
\end{eqnarray}
therefore finishing our proof.

\subsection*{Relations with one gluon adjacent to the off-shell leg}

The proof when one gluon is adjacent to the off-shell leg follows
exactly the same derivation, except that now we have a few
additional quartic graphs. The relation we are aiming to prove is 
\begin{equation}
s_{21}J^{\YMs}(12_{g}34_{g}5)+s_{2,13}J^{\YMs}(132_{g}4_{g}5)+s_{2,134}J^{\YMs}(134_{g}2_{g}5)=
k_{5}^{2} \begin{minipage}{2.2cm} \includegraphics[width=2.2cm]{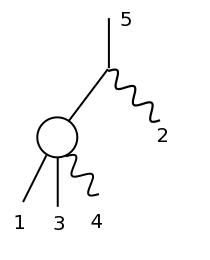}   \end{minipage}
+k_{5}^{2} \begin{minipage}{2.2cm}  \includegraphics[width=2.2cm]{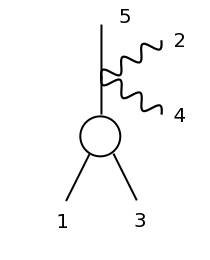}  \end{minipage}~.~~~
\label{eq:bcj-sum-adj}
\end{equation}
Currents that appear in the BCJ sum Berends-Giele decompose as 
\begin{flalign}
s_{21}J^{\YMs}(12_{g}34_{g}5) & = s_{21}\underset{(a1)}{
\begin{minipage}{2.2cm}  \includegraphics[width=2.2cm]{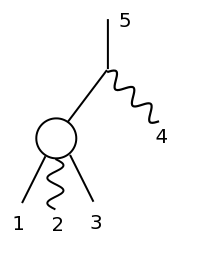}
\end{minipage} } +s_{21}\underset{(a2)}{ \begin{minipage}{2.2cm}
\includegraphics[width=2.2cm]{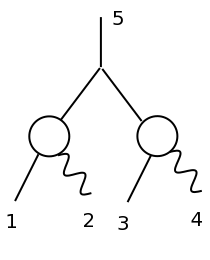}  \end{minipage} }
+s_{21}\underset{(a3)}{\begin{minipage}{2.2cm}
\includegraphics[width=2.2cm]{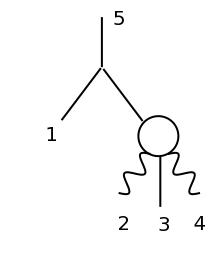}  \end{minipage} }~,~~~&&
\end{flalign}
\begin{flalign}
s_{2,13}J^{\YMs}(132_{g}4_{g}5) & = s_{2,13}\underset{(b1)}{
\begin{minipage}{2.2cm}
\includegraphics[width=2.2cm]{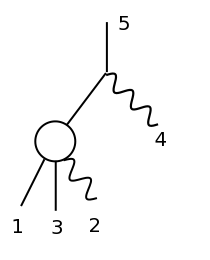}   \end{minipage} }
+s_{2,13}\underset{(b2)}{ \begin{minipage}{2.2cm}
\includegraphics[width=2.2cm]{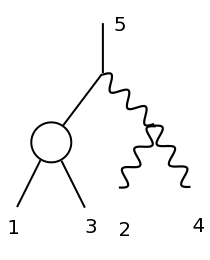}  \end{minipage}
}+s_{2,13}\underset{(b3)}{ \begin{minipage}{2.2cm}
\includegraphics[width=2.2cm]{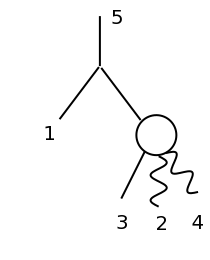}   \end{minipage} }
 +s_{2,13}\underset{(b4)}{ \begin{minipage}{2.2cm}  \includegraphics[width=2.2cm]{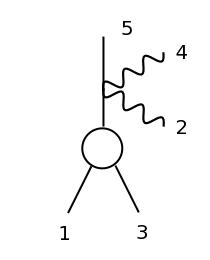}  \end{minipage} }~,~~~
\end{flalign}
\begin{flalign}
s_{2,134}J^{\YMs}(134_{g}2_{g}5) & = s_{2,134}\underset{(c1)}{
\begin{minipage}{2.2cm}
\includegraphics[width=2.2cm]{apdx2-c1}  \end{minipage} }
+s_{2,134}\underset{(c2)}{ \begin{minipage}{2.2cm}
\includegraphics[width=2.2cm]{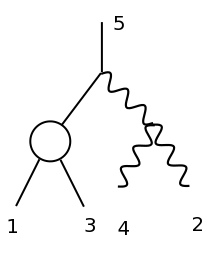}    \end{minipage} }
+s_{2,134}\underset{(c3)}{ \begin{minipage}{2.2cm}
\includegraphics[width=2.2cm]{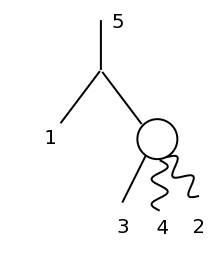}   \end{minipage} }
 +s_{2,134}\underset{(c4)}{ \begin{minipage}{2.2cm} \includegraphics[width=2.2cm]{apdx2-c4}   \end{minipage}
 }~.~~~
\end{flalign}
Note the presence of two new quartic graphs $(b4)$ and $(c4)$. As in
the previous example we regroup graphs into BCJ sums of
sub-currents. Graphs $(a1)$ and $(b1)$ make up a BCJ sum of the
sub-currents involving legs $1$, $2$ and $3$, 
\begin{equation}
s_{21}\underset{(a1)}{ \begin{minipage}{2.2cm} \includegraphics[width=2.2cm]{apdx2-a1}   \end{minipage} }
+s_{2,13}\underset{(b1)}{ \begin{minipage}{2.2cm}  \includegraphics[width=2.2cm]{apdx2-a2}  \end{minipage} }
=k_{123}^{2} \begin{minipage}{2.2cm}  \includegraphics[width=2.2cm]{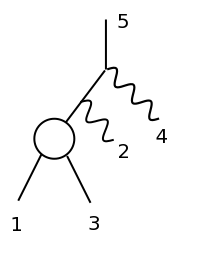}  \end{minipage}~,~~~
\label{eq:proof2-1}
\end{equation}
graph $(a2)$ forms a trivial BCJ sum of the three point current by
itself, graphs $(b2)$ and $(c2)$ add up to another BCJ sum after
eliminating terms using $U(1)$ decoupling, 
\begin{equation}
s_{21}\underset{(b2)}{ \begin{minipage}{2.2cm}  \includegraphics[width=2.2cm]{apdx2-b2}  \end{minipage} }
+s_{2,13}\underset{(c2)}{ \begin{minipage}{2.2cm}  \includegraphics[width=2.2cm]{apdx2-c2}  \end{minipage}}
=s_{23}\begin{minipage}{2.2cm}  \includegraphics[width=2.2cm]{apdx2-c2}   \end{minipage}~,~~~
\label{eq:proof2-2}
\end{equation}
and similarly for the sum of graphs $(a3)$, $(b3)$ and $(c3)$, 
\begin{eqnarray}
s_{21}\underset{(a3)}{ \begin{minipage}{2.2cm} \includegraphics[width=2.2cm]{apdx2-a3}   \end{minipage} }
+s_{2,13}\underset{(b3)}{ \begin{minipage}{2.2cm}   \includegraphics[width=2.2cm]{apdx2-b3}  \end{minipage} }
+s_{2,134}\underset{(c3)}{ \begin{minipage}{2.2cm} \includegraphics[width=2.2cm]{apdx2-c3}   \end{minipage}}=k_{234}^{2} \begin{minipage}{2.2cm}  \includegraphics[width=2.2cm]{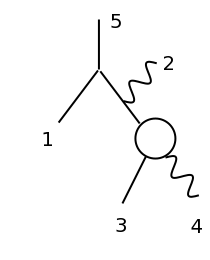}  \end{minipage}
+k_{234}^{2} \begin{minipage}{2.2cm}   \includegraphics[width=2.2cm]{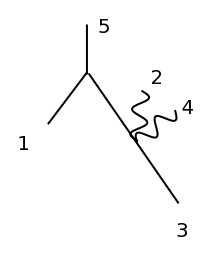} \end{minipage}~.~~~~\label{eq:proof2-3}
\end{eqnarray}
As for $(c1)$, we Berends-Giele decompose it as in the previous
example to give 
\begin{eqnarray}
s_{2,134}\underset{(c1)}{ \begin{minipage}{2.2cm} \includegraphics[width=2.2cm]{apdx2-c1}   \end{minipage}}
& = &
 k_{5}^{2} \begin{minipage}{2.2cm} \includegraphics[width=2.2cm]{apdx2-c1}   \end{minipage}
 -s_{134} \begin{minipage}{2.2cm}  \includegraphics[width=2.2cm]{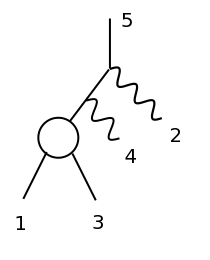}  \end{minipage}
 -s_{134} \begin{minipage}{2.2cm}  \includegraphics[width=2.2cm]{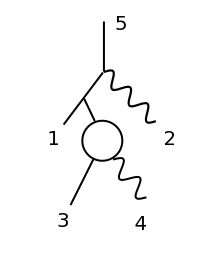}  \end{minipage}~.~~~
 \label{eq:proof2-4}
\end{eqnarray}
The full BCJ sum (\ref{eq:bcj-sum-adj}) is now translated into the
right hand side of equation (\ref{eq:proof2-1}), graph $(b2)$,
(\ref{eq:proof2-2}), (\ref{eq:proof2-3}), (\ref{eq:proof2-4}) plus
the additional graphs $(b4)$ and $(c4)$. To finish the proof we need
to further translate these graphs into one of those on the right
hand side of equation (\ref{eq:bcj-sum-adj}), and this is done by
using Jacobi identities 
\begin{equation}
k_{123}^{2} \begin{minipage}{2.2cm}  \includegraphics[width=2.2cm]{apdx2-e3}  \end{minipage}
+s_{24} \begin{minipage}{2.2cm}   \includegraphics[width=2.2cm]{apdx2-c2}  \end{minipage}
-s_{123} \begin{minipage}{2.2cm}   \includegraphics[width=2.2cm]{apdx2-d1} \end{minipage}
=(s_{4,13}-s_{2,13}) \begin{minipage}{2.2cm}   \includegraphics[width=2.2cm]{apdx2-c4}  \end{minipage}~,~~~
\end{equation}
\begin{equation}
s_{21} \begin{minipage}{2.2cm} \includegraphics[width=2.2cm]{apdx2-a2}   \end{minipage}
+k_{234}^{2} \begin{minipage}{2.2cm} \includegraphics[width=2.2cm]{apdx2-e4}   \end{minipage}
-s_{134} \begin{minipage}{2.2cm}  \includegraphics[width=2.2cm]{apdx2-d2}  \end{minipage}
=0~,~~~
\end{equation}
and the fact that the following two graphs contribute the same. 
\begin{eqnarray}
k_{234}^{2} \begin{minipage}{2.2cm}  \includegraphics[width=2.2cm]{apdx2-e2}  \end{minipage}
& = & s_{13} \begin{minipage}{2.2cm}  \includegraphics[width=2.2cm]{apdx2-c4}  \end{minipage} =(\frac{i}{2}) f^{a_{1}a_{3}a_{5}}\epsilon_{2}\cdot\epsilon_{4}~.~~~\end{eqnarray}
Collecting terms gives 
\begin{eqnarray}
k_{5}^{2} \begin{minipage}{2.2cm}  \includegraphics[width=2.2cm]{apdx2-c1}  \end{minipage}
 +\big((s_{4,13}-s_{2,13})+s_{2,13}+s_{2,134}+s_{13}\big)
 \begin{minipage}{2.2cm}  \includegraphics[width=2.2cm]{apdx2-c4}  \end{minipage} =
 k_{5}^{2} \begin{minipage}{2.2cm}  \includegraphics[width=2.2cm]{apdx2-c1}  \end{minipage}
 +k_{5}^{2} \begin{minipage}{2.2cm}  \includegraphics[width=2.2cm]{apdx2-c4}  \end{minipage}~,~~~\nonumber
\end{eqnarray}
which completes our proof.


\bibliographystyle{JHEP}
\bibliography{graviton}

\end{document}